\documentclass[prd,preprint,tightenlines,floatfix,showpacs,preprintnumbers,nofootinbib,eqsecnum,superscriptaddress]{revtex4}

\usepackage[dvips,final]{graphicx}
 \usepackage{amssymb}
  \usepackage{amsmath}
   \usepackage{amsfonts}
    \usepackage{epsfig}
     \usepackage{bm}% bold math
      \usepackage{multirow}

     \newcommand{\bp}{\mbox{\boldmath $p$}}
     
     \newcommand{\bk}{\mbox{\boldmath $k$}}
\begin{document}

\title{
\mbox{\boldmath $pp \to pp K^{+}K^{-}$} reaction at high energies}

\author{P. Lebiedowicz}
\email{piotr.lebiedowicz@ifj.edu.pl}
\affiliation{Institute of Nuclear Physics PAN, PL-31-342 Cracow, Poland}

\author{A. Szczurek}
\email{antoni.szczurek@ifj.edu.pl}
\affiliation{University of Rzesz\'ow, PL-35-959 Rzesz\'ow, Poland}
\affiliation{Institute of Nuclear Physics PAN, PL-31-342 Cracow, Poland}

\begin{abstract}
We evaluate differential distributions for the four-body
$p p \to p p K^{+} K^{-}$ reaction at high energies which constitutes 
an irreducible background to three-body processes $p p \to p p M$,
where $M=\phi$, $f_{2}(1275)$, $f_{0}(1500)$, $f_{2}'(1525)$, $\chi_{c0}$.
We consider central diffractive contribution
mediated by Pomeron and Reggeon exchanges
as well as completely new mechanism of emission of kaons from the proton lines.
We include absorption effects
due to proton-proton interaction and kaon-kaon rescattering.
We compare our results with measured cross sections for the CERN ISR experiment.
We make predictions for future experiments at RHIC, Tevatron and LHC.
Differential distributions in invariant two-kaon mass, 
kaon rapidities and transverse momenta of kaons are presented.
Two-dimensional distribution in $(y_{K^+}, y_{K^-})$ is particularly interesting.
The higher the incident energy, the higher preference for the same-hemisphere emission of kaons.
We find that the kaons from the new mechanism of emission directly
from proton lines are produced rather forward and backward 
but the corresponding cross section is rather small.
The processes considered here constitute a sizeable contribution to
the total proton-proton cross section as well as to kaon
inclusive cross section.

We consider a measurement of exclusive production of scalar
$\chi_{c0}$ meson in the proton-proton collisions via $\chi_{c0} \to K^{+}K^{-}$ decay.
The corresponding amplitude for exclusive
central diffractive $\chi_{c0}$ meson production
is calculated within the $k_{t}$-factorization approach.
The influence of kinematical cuts on the signal-to-background ratio is 
discussed. 

\end{abstract}


\maketitle

%--------------------------------------------------
\section{INTRODUCTION}
%--------------------------------------------------

The exclusive $p p \to pp K^{+} K^{-}$ reaction was studied
only at low energy \cite{Sibirtsev,Wilkin}.
Here the dominant mechanisms are exclusive $a_{0}(980)$ and $f_{0}(980)$ production \cite{Sibirtsev}
or excitation of nucleon and $\Lambda$ resonances \cite{Wilkin}.
The main aim of this paper is to discuss mechanisms of exclusive $K^{+} K^{-}$ production 
in hadron-hadron collisions at high energies.
Processes of central exclusive production
became recently a very active field of research (see e.g. Ref. \cite{ACF10} and references therein).
Although the attention is paid mainly to high-$p_{t}$ processes
that can be used for new physics searches (exclusive Higgs, $\gamma\gamma$ interactions, etc.),
measurements of low-$p_{t}$ signals are also very important as they can
help to constrain models of the backgrounds for the former ones.
The $p p \to pp K^{+} K^{-}$ reaction is a natural background for exclusive production
of resonances decaying into $K^{+} K^{-}$ channel, such as:
$\phi$, $f_{2}(1270)$, $f_{0}(1500)$, $f_{2}'(1525)$, $\chi_{c0}$.
The expected non-resonant background can be modeled using a
"non-perturbative" framework, mediated by Pomeron-Pomeron fusion
with an intermediate off-shell pion/kaon exchanged between the final-state particles.
The two-pion background to exclusive production of $f_{0}(1500)$ meson was discussed in Ref. \cite{SL09}.
In Refs.\cite{LSK09,LS10} we have studied production of $\pi^+ \pi^-$ pairs
for low and high energies.
Here we wish to present similar analysis for $K^{+}K^{-}$ production at high energies.
The dominant mechanism of the $pp \to pp \pi^+ \pi^-,pp \to pp K^+ K^-$ reactions at high energies
is relatively simple compared to that of the $pp \to nn \pi^+ \pi^+$ \cite{LS11}
or $pp \to pp \pi^0 \pi^0$ processes.
In Ref. \cite{SLTCS11} a possible measurement of the exclusive $\pi^+ \pi^-$ production
at the LHC with tagged forward protons has been studied.

A study of the centrally produced $\pi^{+}\pi^{-}$ and $K^{+}K^{-}$ channels
in $pp$ collisions has been performed experimentally 
at an incident beam momenta of 300 GeV/c  ($\sqrt{s} = 23.8$ GeV) \cite{WA76}
and 450 GeV/c ($\sqrt{s} = 29.1$ GeV) \cite{WA102}.
In the latter paper a study has been performed of resonance production rate
as a function of the difference in the transverse momentum vectors ($dP_{T}$)
between the particles exchanged from vertices.
An analysis of the  $dP_{T}$ dependence 
of the four-momentum transfer behavior
shows that the $\rho^{0}(770)$, $\phi(1020)$, $f_{2}(1270)$ and $f_{2}'(1525)$
are suppressed at small $dP_{T}$ in contrast to the $f_{0}(980)$, $f_{0}(1500)$ and $f_{0}(1710)$.
Different distributions are observed in the azimuthal angle
(defined as the angle between the $p_{t}$ vectors of the two outgoing protons)
for the different resonances (see \cite{WA102}).
The mass spectrum of the exclusive $K^{+}K^{-}$ system 
at the CERN Intersecting Storage Rings (ISR)
is shown e.g. in Ref.\cite{AFS85} at $\sqrt{s} = 63$ GeV
and in Ref.\cite{ABCDHW89} at $\sqrt{s} = 62$ GeV
(this is the highest energy at which normalized experimental data exist). 

Recently there was interest in central exclusive production of $P$-wave
quarkonia (Refs.~\cite{PST_chic0,PST_chic1,PST_chic2,LKRS10,LKRS11})
where the QCD mechanism is similar to the exclusive production of the 
Higgs boson. 
Furthermore, the $\chi_{c(0,2)}$ states are expected to annihilate via 
two-gluon processes into light mesons in particular into $K^+ K^-$.
Also some glueball candidates \cite{glueballs} can be searched for
in this channel.

The cross section for central exclusive production of $\chi_c$ mesons 
has been measured recently in proton-antiproton collisions at 
the Tevatron \cite{CDF_chic}.
In this experiment $\chi_c$ mesons are identified via decay to 
the $J/\psi + \gamma$ with $J/\psi \to \mu^{+}\mu^{-}$ channel. 
At the Tevatron the experimental invariant
mass resolution was not sufficient to distinguish between scalar,
axial and tensor $\chi_c$. While the branching fractions to this
channel for axial and tensor mesons are large \cite{PDG}
($\mathcal{B} = (34.4 \pm 1.5)\%$ and $\mathcal{B} = (19.5 \pm 0.8)\%$, respectively)
the branching fraction for the scalar meson is very small 
$\mathcal{B} = (1.16 \pm 0.08 )\%$ \cite{PDG}. 
Theoretical calculations have shown \cite{PST_chic2} that 
the cross section for exclusive $\chi_{c0}$ production obtained within the
$k_{t}$-factorization is much bigger than that for $\chi_{c1}$ and
$\chi_{c2}$. As a consequence, all $\chi_{c}$ mesons give similar
contributions to the $J/\psi + \gamma$ decay
channel. Clearly, the measurement via decay to the $J/\psi + \gamma$
channel at Tevatron cannot provide cross section for different species 
of $\chi_c$.

The scalar $\chi_{c0}$ meson decays
into several two-body (e.g. $\pi \pi$, $K^{+} K^{-}$, $p \bar{p}$)
and four-body final states (e.g. $\pi^{+} \pi^{-} \pi^{+} \pi^{-}$,
$\pi^{+} \pi^{-} K^{+} K^{-}$).
The observation of $\chi_{c0}$ CEP via two-body decay channels
is of special interest for studying the dynamics of heavy quarkonia.
The measurement of exclusive production of $\chi_{c0}$ meson
in proton-(anti)proton collisions via $\chi_{c0} \to \pi^{+} \pi^{-}$ decay 
has been already discussed in Ref. \cite{LPS11}.
In the present paper we analyze a possibility to measure $\chi_{c0}$ via
its decay to $K^+ K^-$ channel.
The branching fraction to this channel is relatively large 
$\mathcal{B}( \chi_{c0} \to K^{+} K^{-})  = (0.61 \pm 0.035)\%$ \cite{PDG}.
In addition, the axial $\chi_{c1}$ does not decay to the
$KK$ channel and the branching ratio for the $\chi_{c2}$ decay
into two kaons is smaller
$\mathcal{B}( \chi_{c2} \to K^{+} K^{-})  = (0.109 \pm 0.008)\%$ \cite{PDG}.
A much smaller cross section for $\chi_{c2}$ production as obtained
from theoretical calculation means that only $\chi_{c0}$ will contribute
to the signal.

Exclusive charmonium decays can be also studied in
$e^{+}e^{-}$ colliders. Here the $\chi_{cJ}$ states are
copiously produced in the radiative decays $\psi(2S) \to \gamma
\chi_{cJ}$ \cite{PDG}.
Recently the BESIII Collaboration performed a measurement of the
hadronic decays of the three $\chi_{cJ}$ states 
to $p \bar{p} K^{+} K^{-}$ 
($\bar{p}K^{+}\Lambda(1520)$, $\Lambda(1520) \bar{\Lambda}(1520)$ and 
$\phi p \bar{p}$) \cite{BES_ppKK}.
In the present paper we discuss a possibility to measure
$\chi_{c0}$ in the $K^+ K^-$ channel. 
Here, continuum backgrounds are expected to be larger than in the 
$e^+ e^-$ collisions.
This will discussed in the present paper.

%---------------------------------------------------------
\section{CENTRAL DIFFRACTIVE CONTRIBUTION}
\label{section:II}
%---------------------------------------------------------

%---------------------------------------------
\subsection{The $K N$ scattering}
\label{subsection:KN_elastic}
%---------------------------------------------

In order to fix parameters of our double Pomeron exchange (DPE) model we consider first elastic $KN$ scattering. 
The forward amplitudes $M_{KN} (s, t = 0)$ of the elastic scatterings
are written in terms of the Regge exchanges
%the simplified Regge-like forms:
\begin{eqnarray}
M_{K^{\pm} p \to K^{\pm} p}(s,0)= 
A_{I\!\!P}(s,0) + A_{f_{2}}(s,0) + A_{a_{2}}(s,0) \mp A_{\omega}(s,0) \mp A_{\rho}(s,0)\, , \nonumber\\
M_{K^{\pm} n \to K^{\pm} n}(s,0)=
A_{I\!\!P}(s,0) + A_{f_{2}}(s,0) - A_{a_{2}}(s,0) \mp A_{\omega}(s,0) \pm A_{\rho}(s,0).
\label{KN_amplitude_forward}
\end{eqnarray}
The optical theorem relates the total cross section for the scattering of a pair of hadrons
to the amplitude for elastic scattering: Im$M_{el}(s,t=0) \sim s \sigma_{tot}(s)$ .
When the centre-of-mass energy $\sqrt{s}$ is large
the elastic $KN$ scattering amplitude is a sum of the terms:
\begin{eqnarray}
A_{i}(s,t) = 
\eta_{i} \, s \, C_{i}^{K N}\left( \frac{s}{s_0} \right)^{\alpha_{i}(t)-1} 
\exp \left( \frac{B^{i}_{K N}}{2} t \right)\,,
\label{KN_amplitude_part}
\end{eqnarray}
where $i$ = $I\!\!P$, $f_{2}$, $a_{2}$, $\omega$ and $\rho$.
The energy scale $s_{0}$ is fixed at $s_{0} = 1$ GeV$^2$.
The values of coupling constants ($C_{i}^{K N}$)
are taken from the Donnachie-Landshoff analysis of the total
cross section in several hadronic reactions \cite{DL92}.
The parameters of Regge linear trajectories ($\alpha_{i}(t)=\alpha_{i}(0)+\alpha'_{i}t$) and
signature factors ($\eta_{i}$) used in the present calculations
are listed in Table \ref{tab:parameters}.
The slope of the elastic $KN$ scattering can be written as 
\begin{equation}
B(s) = B^{i}_{K N} + 2 \alpha'_{i}
 \ln \left( \frac{s}{s_0} \right)
\label{slope_NN}
\end{equation}
and only the $B^{i}_{K N}$ parameters are adjusted to the existing experimental data for the elastic $KN$ scattering. 

\begin{figure}[!h]    % 
\includegraphics[width=0.45\textwidth]{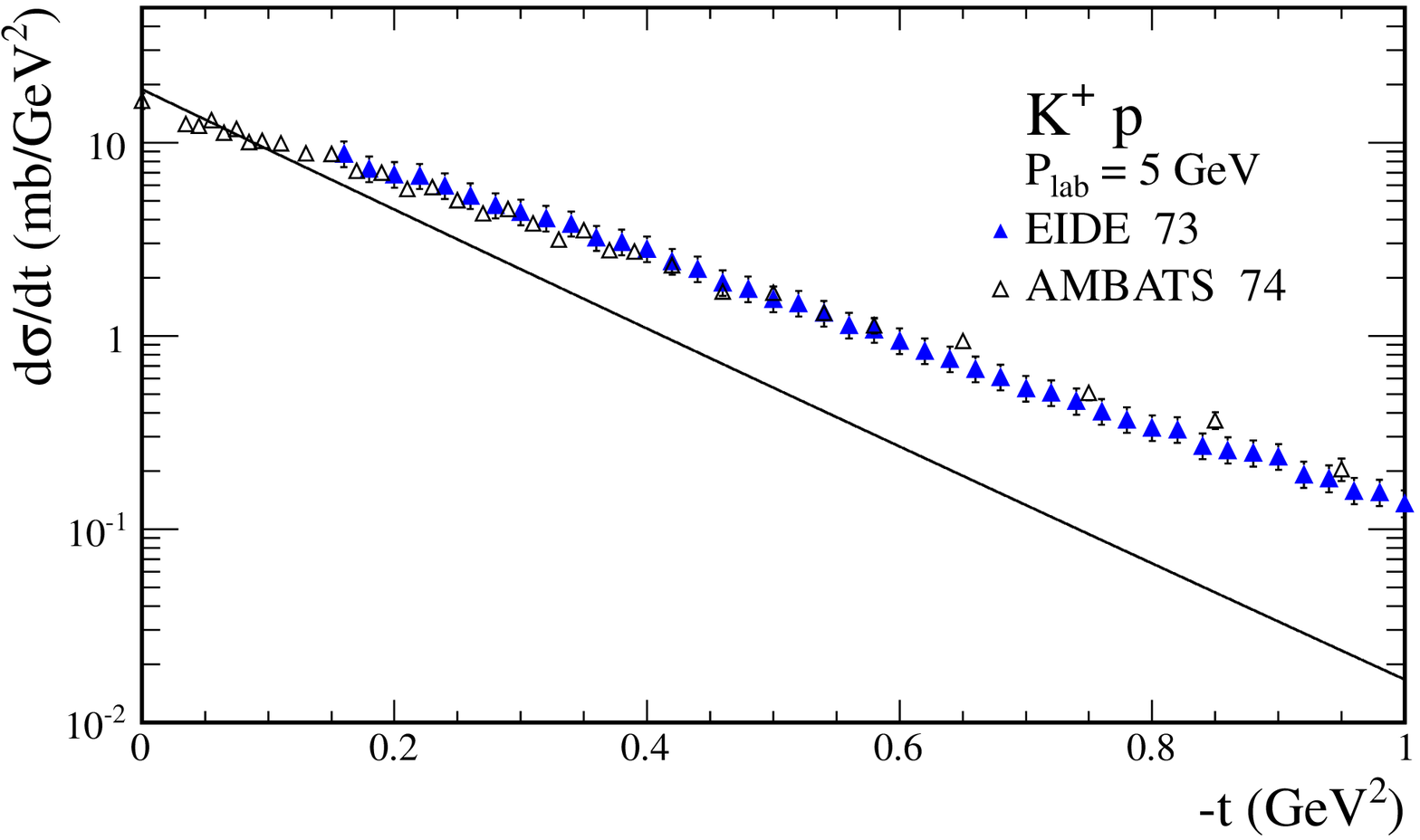}\qquad
\includegraphics[width=0.45\textwidth]{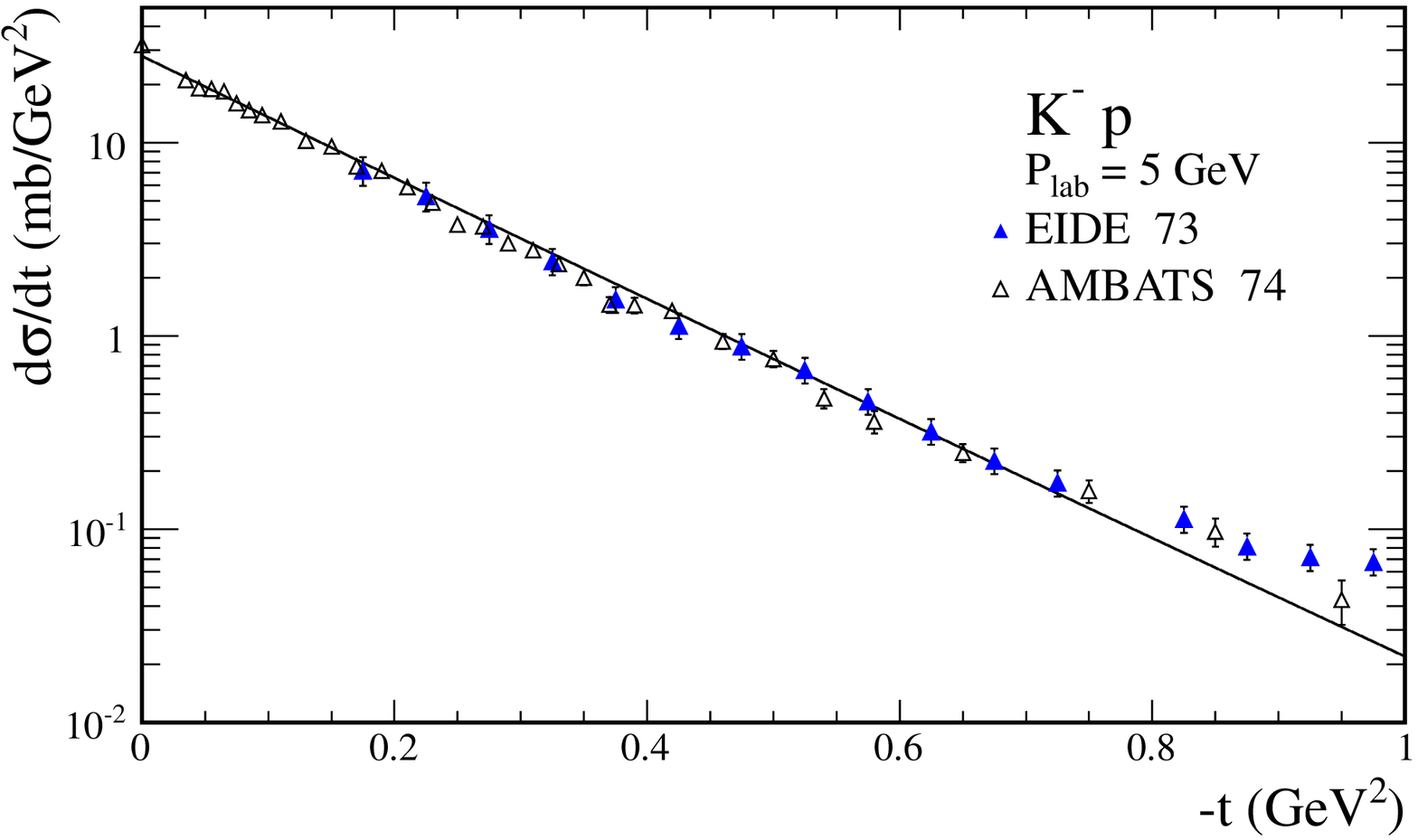}
\includegraphics[width=0.45\textwidth]{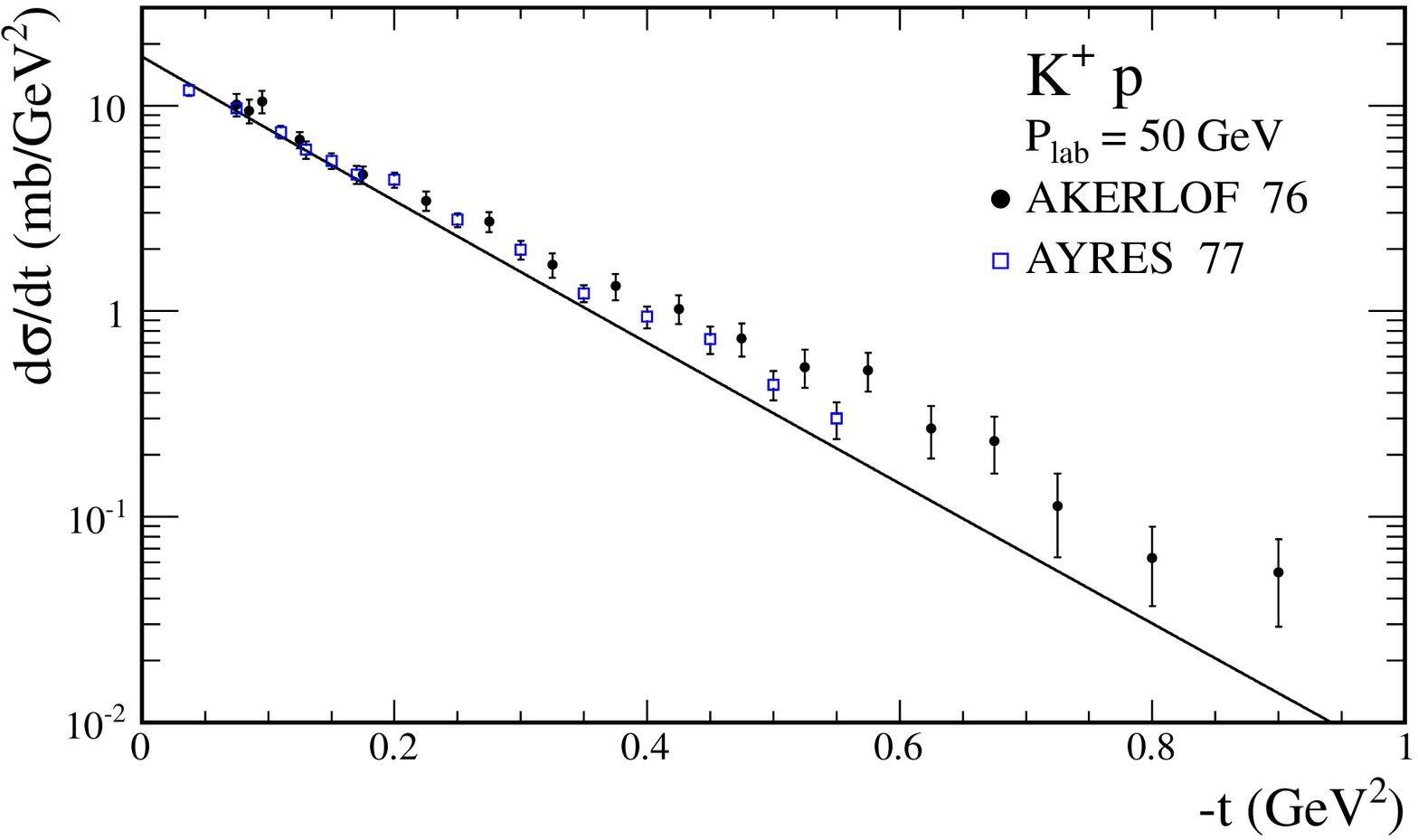}\qquad
\includegraphics[width=0.45\textwidth]{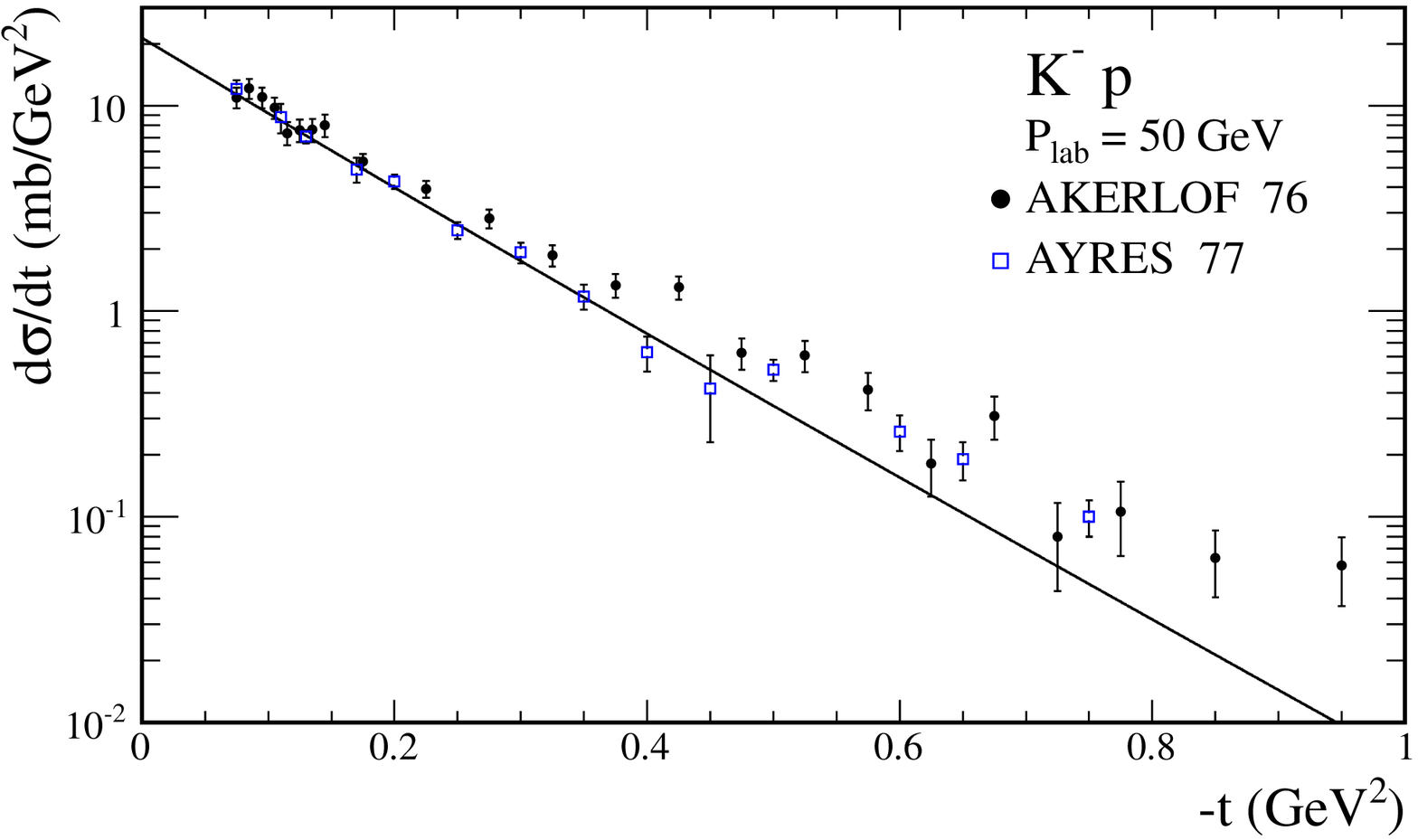}
\includegraphics[width=0.45\textwidth]{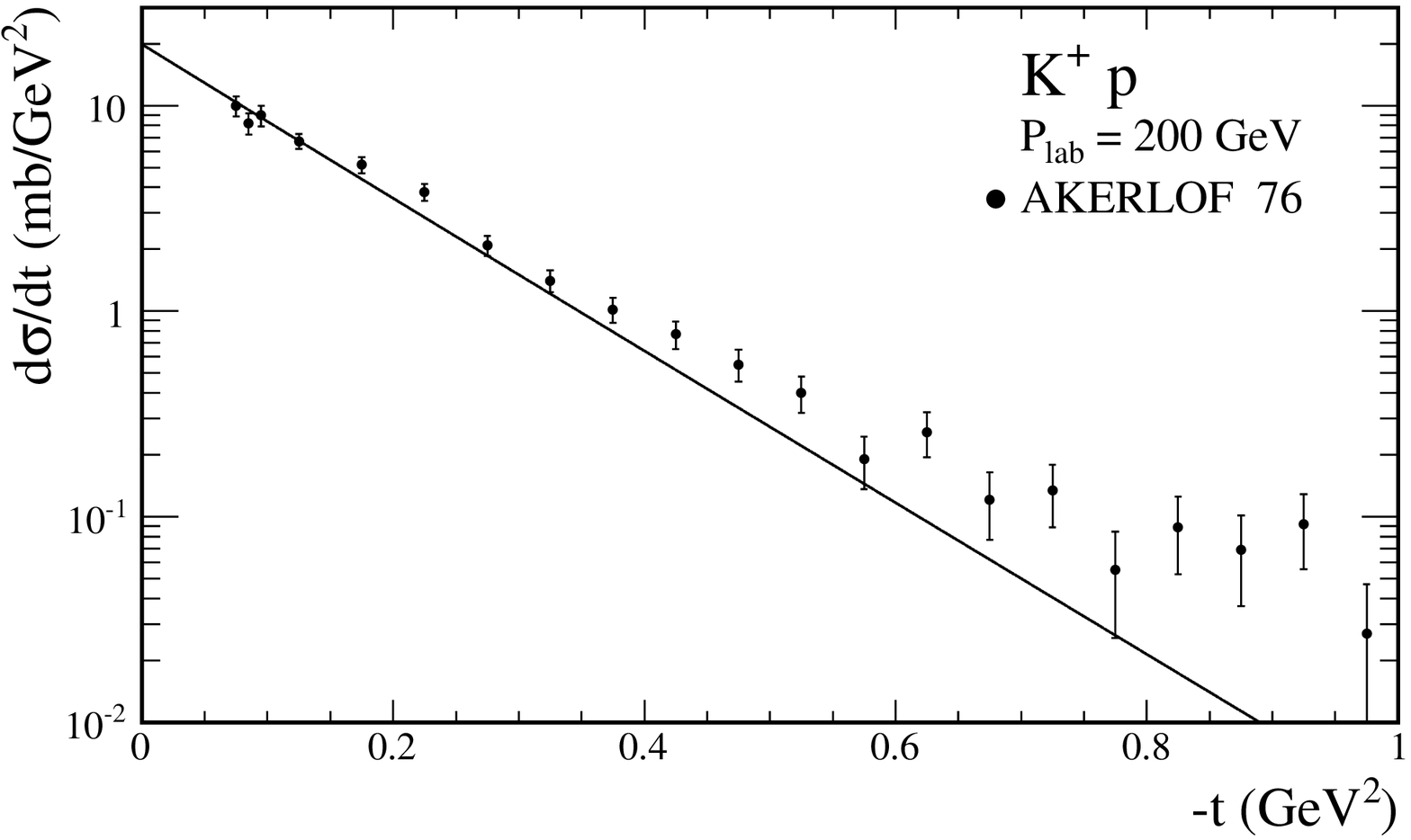}\qquad
\includegraphics[width=0.45\textwidth]{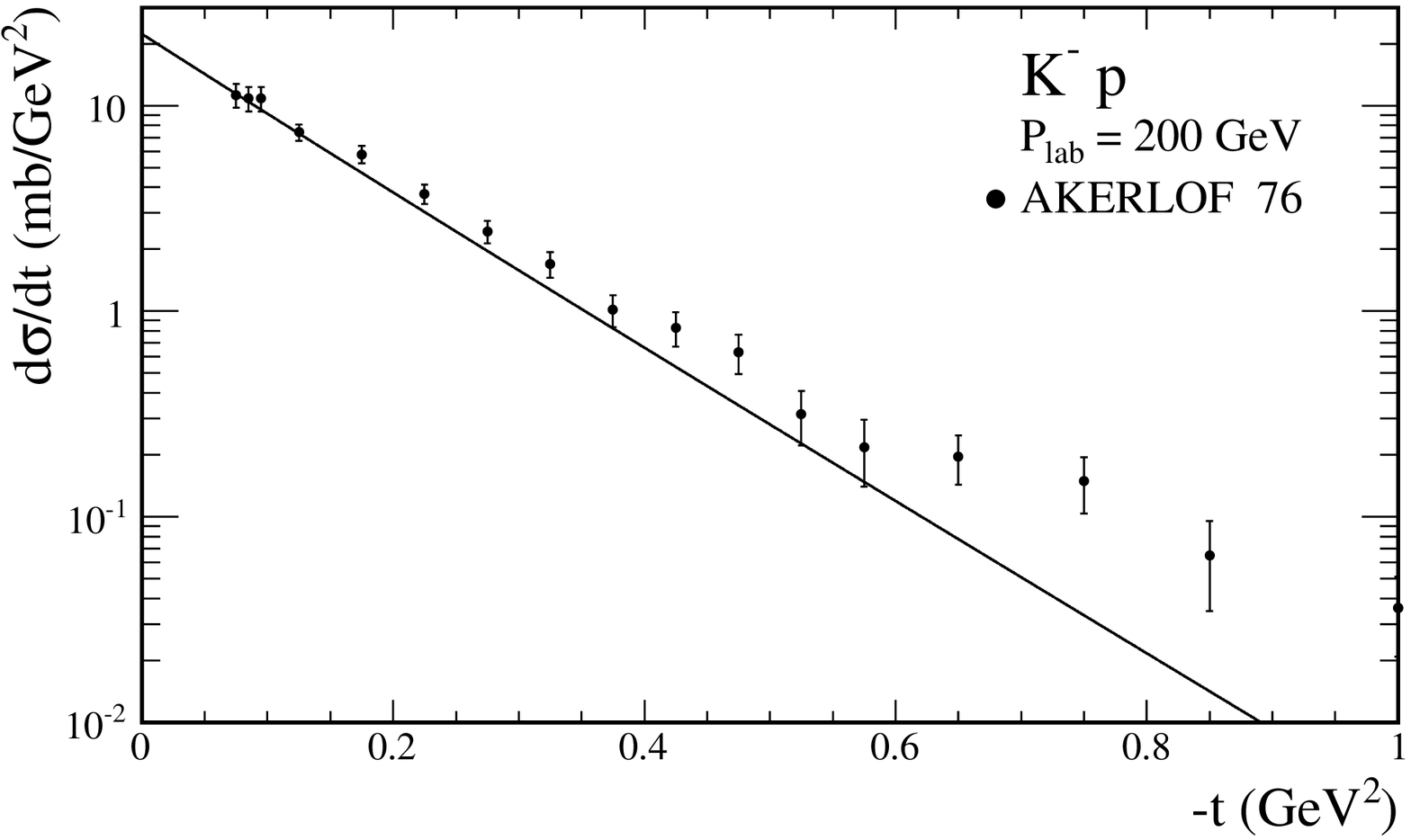}
   \caption{\label{fig:dsig_dt_KN}
   \small 
Differential distributions for $K^+ p$ (left) and $K^- p$ (right)
elastic scattering 
for three incident-beam momenta of $P_{lab}=$ 5, 50, 200 GeV.
The experimental data are taken from Refs \cite{dsig_dt}.
}
\end{figure}

The differential elastic cross section is expressed with the help
of the elastic scattering amplitude as usually:
\begin{equation}
\frac{d\sigma_{el}}{dt}=\frac{1}{16\pi s^{2}}|M_{KN}(s,t)|^{2}\; .
\label{dsigma_dt_elastic}
\end{equation}
The differential distributions $d\sigma_{el}/dt$
for both $K^+ p$ and $K^- p$ elastic scattering 
for three incident-beam momenta of $P_{lab}=$ 5 GeV, 
$P_{lab}=$ 50 GeV and $P_{lab}=$ 200 GeV are shown 
in Fig.\ref{fig:dsig_dt_KN}.
With the slope pareamters, as in Ref. \cite{LS10},
$B_{I\!\!P}^{K N}$ = $B_{I\!\!P}^{\pi N}$ = 5.5 GeV$^{-2}$,
$B_{I\!\!R}^{K N}$ = $B_{I\!\!R}^{\pi N}$ = 4 GeV$^{-2}$
for Pomeron and Reggeon exchanges,
a rather good description of experimental $d\sigma_{el}/dt$ is achieved.
The exception is the low energy $K^{+}p$ scattering.
Here $\Lambda$ baryon exchange is a possible mechanism in addition
to Pomeron and Reggeon exchanges.
%--------------------------------------------------------
%--------------------------------------------------------

\begin{figure}[!h]   
\includegraphics[width=0.45\textwidth]{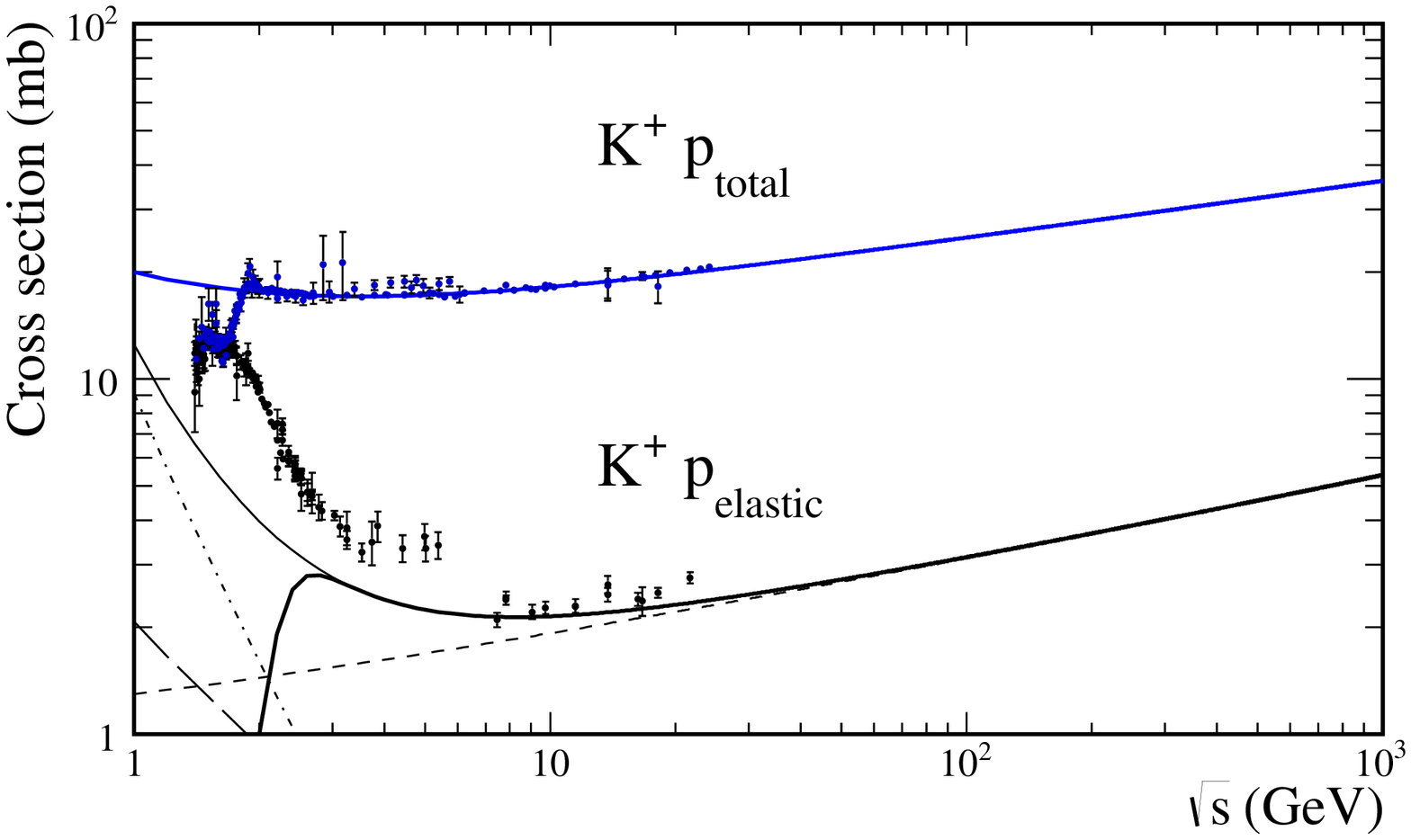}\qquad
\includegraphics[width=0.45\textwidth]{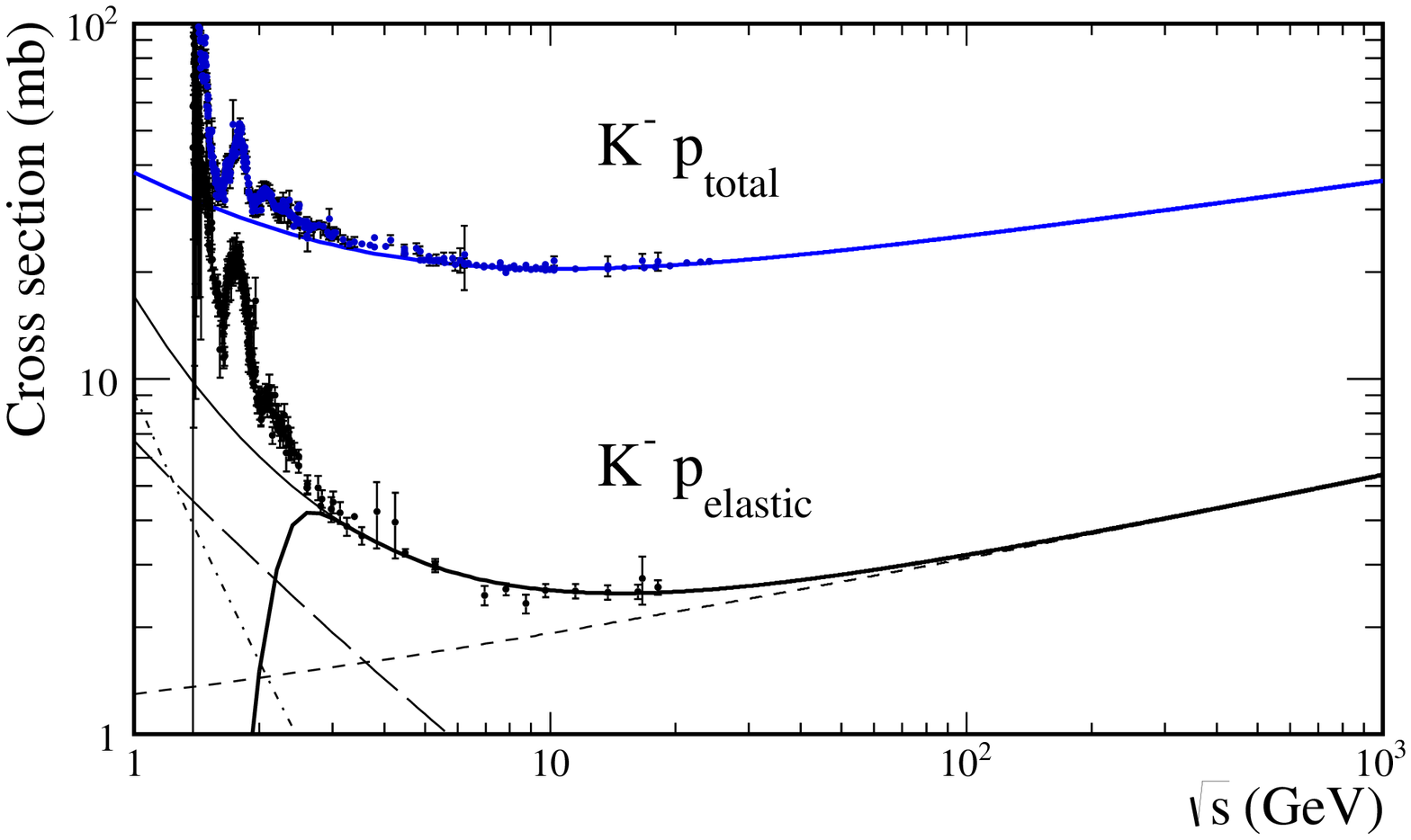}
\includegraphics[width=0.45\textwidth]{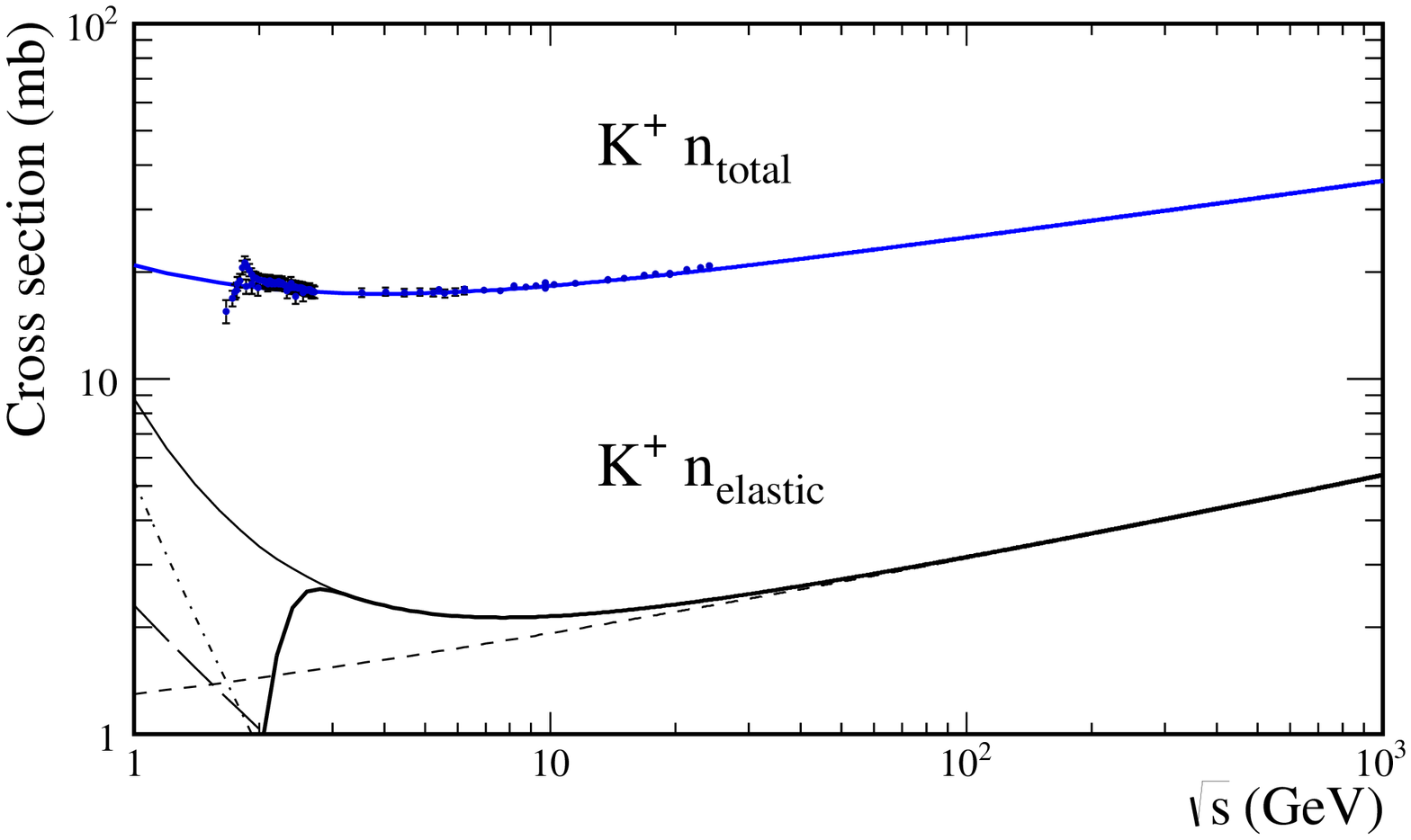}\qquad
\includegraphics[width=0.45\textwidth]{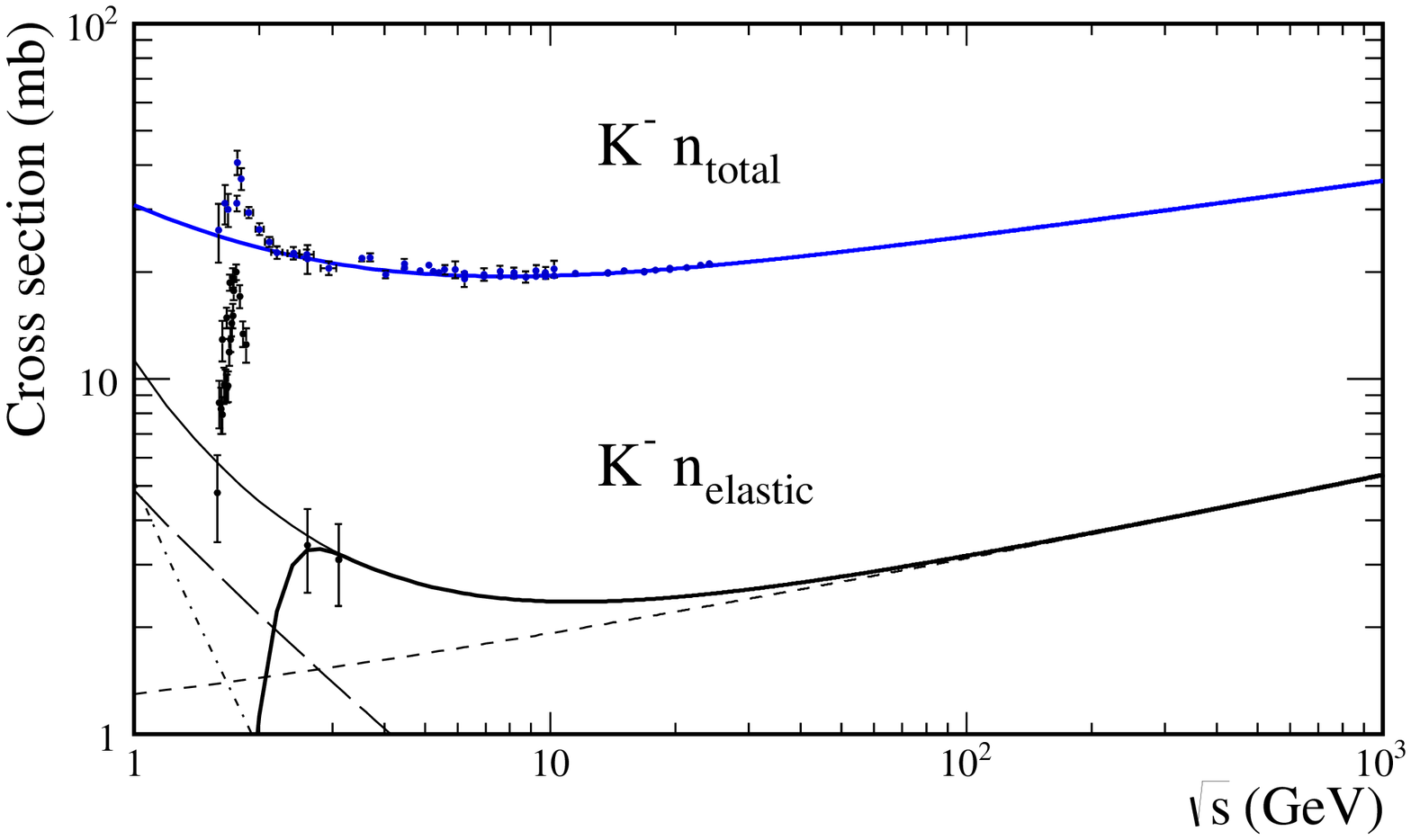}
   \caption{\label{fig:KN_elastic}
   \small 
The integrated cross section for the $K N$ total and elastic scattering.
The experimental data are taken from particle data book \cite{PDG}.
The lines are explained in the main text.
}
\end{figure}

We nicely describe the existing experimental data 
for elastic $KN$ scattering for $\sqrt{s} >$ 3 GeV.
as can be seen from Fig.\ref{fig:KN_elastic}.
In the Regge approach, high energy cross section is dominated 
by Pomeron exchange (dashed lines). The Reggeon exchanges dominate in the resonance
region (dash-dotted lines).
While the total cross section is just a sum of the Pomeron and Reggeon terms,
the elastic cross section have the interference term (long-dashed lines).
In order to exclude low energy regions the $M_{KN}(s,t)$ elastic scattering amplitudes
are corrected by purely phenomenological smooth cut-off correction factor (as in Ref. \cite{LS10}).
%
%--------------------------------------------------------
%--------------------------------------------------------

Our model sufficiently well describes the $K N$ data 
and includes absorption effects due to 
kaon-nucleon rescatterings in an effective way. 
This has a clear advantage for applications to the $p p \to p p K^{+}K^{-}$ reaction where the 
$K N$ absorption effects do not need to be included explicitly. 
Having fixed the parameters we can proceed to our four-body 
$p p \to p p K^+ K^-$ reaction.

%=======================================================================================
\begin{table}
\caption{Parameters of Pomeron and Reggeon exchanges 
determined from elastic and total cross sections
used in the present calculations.
}
\label{tab:parameters}
\begin{center}
\begin{tabular}{|c||c|c|c|c|c|}
\hline
       $i$ & $\eta_{i}$     & $\alpha_{i}(t)$   & $C_{i}^{N N}$ (mb)   & $C_{i}^{K N}$ (mb)   & $C_{i}^{KK}$ (mb) \\
\hline 
$I\!\!P$ & $i$             & 1.0808 + (0.25 GeV$^{-2}$) $t$ & 21.7    & 11.82                &  $\simeq$6.438    \\ 
$f_{2}$  & $(-0.860895+ i)$& 0.5475 + (0.93 GeV$^{-2}$) $t$ & 75.4875 & 15.67                &  $\simeq$3.253    \\ 
$\rho$   & $(-1.16158 - i)$& 0.5475 + (0.93 GeV$^{-2}$) $t$ &  1.0925 & 2.05                 &  $\simeq$3.847    \\ 
$a_{2}$  & $(-0.860895+ i)$& 0.5475 + (0.93 GeV$^{-2}$) $t$ &  1.7475 & 1.585                &  $\simeq$1.438    \\
$\omega$ & $(-1.16158 - i)$& 0.5475 + (0.93 GeV$^{-2}$) $t$ & 20.0625 & 7.055                &  $\simeq$2.481    \\
\hline
\end{tabular}
\end{center}
\end{table}
%=============================================================================================

%--------------------------------------
\subsection{Central diffractive production of \mbox{\boldmath $K^{+} K^{-}$}}
\label{subsection:background}
%--------------------------------------
%--------------------------------------------------------
\begin{figure}[!h]
\includegraphics[width=0.25\textwidth]{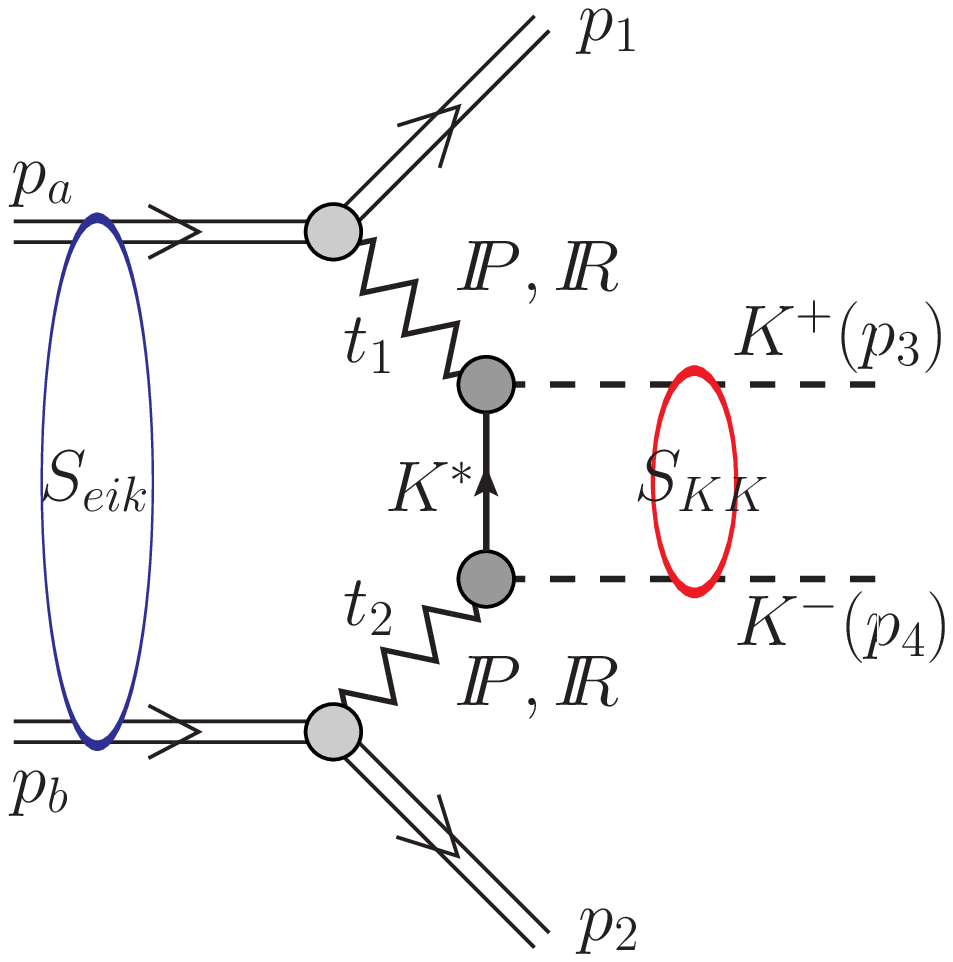}
\includegraphics[width=0.25\textwidth]{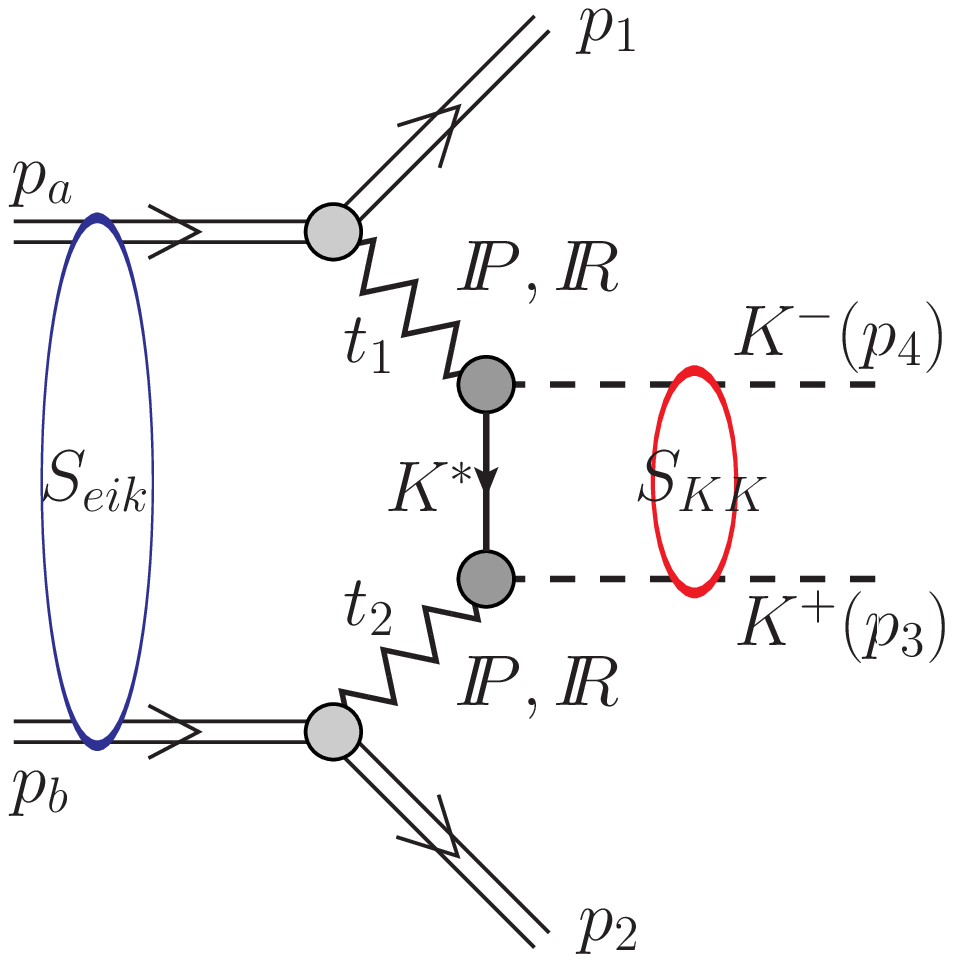}
  \caption{\label{fig:central_double_diffraction_diagrams_fsi}
  \small
The central diffractive mechanism of exclusive production of
$K^{+}K^{-}$ pairs including the absorptive corrections due to proton-proton interactions
as well as kaon-kaon rescattering.}
\end{figure}
%--------------------------------------------------------

The dominant mechanism of the exclusive production of
$K^{+}K^{-}$ pairs at high energies is sketched in
Fig.~\ref{fig:central_double_diffraction_diagrams_fsi}.
The formalism used in the calculation of the amplitude 
is explained in detail elsewhere for the $\pi^{+} \pi^{-}$ production \cite{LS10,LPS11}
and here only main aspects are discussed.
The full amplitude for the process $pp \to p K^{+} K^{-} p$
(with four-momenta $p_{a} + p_{b} \to p_{1} + p_{3} + p_{4} + p_{2}$, respectively)
is a sum of the Born and rescattering amplitudes
\begin{eqnarray}
\mathcal{M}^{full}_{pp\to ppKK}=
\mathcal{M}^{Born}+
\mathcal{M}^{pp-rescatt.}+
\mathcal{M}^{KK-rescatt.}.
\end{eqnarray}
The Born amplitude can be written as
\begin{eqnarray} \nonumber
\mathcal{M}^{Born}&=&
M_{13}(s_{13},t_1)F_{K}(\hat{t})\frac{1}{\hat{t}-m_{K}^{2}}F_{K}(\hat{t})M_{24}(s_{24},t_2)\\
&+&M_{14}(s_{14},t_1)F_{K}(\hat{u})\frac{1}{\hat{u}-m_{K}^{2}}F_{K}(\hat{u})M_{23}(s_{23},t_2)\, ,
\label{Regge_amplitude}
\end{eqnarray}
where $M_{ik}(s_{ik},t_{i}$ denotes "interaction" between forward proton ($i=1$)
or backward proton ($i=2$) and one of the two kaons 
($k=3$ for $K^{+}$, $k=4$ for $K^{-}$). 
The energy dependence of the
$K N$ elastic amplitudes is parameterized 
in terms of Pomeron and $f_{2}$, $a_{2}$, $\omega$ and $\rho$ Reggeon exchanges
as explained in section \ref{subsection:KN_elastic}.
The Donnachie-Landshoff parametrization is used only
above resonance regions
for the $K N$ subsystem energy $\sqrt{s_{ik}} > 2-3$ GeV.
%Below this energy resonance states are present in $KN$ subsystems. 
In order to exclude resonance regions the $M_{ik}$ terms are corrected
by a purely phenomenological smooth cut-off correction factors
which in practice modify the cross section only at large rapidities \cite{LS10}.

The kaon exchange as a meson exchange
is a correct description at rather low energies.
At higher energies a kaon reggezation is required \cite{LPS11}.
This is done by the following replacement:
\begin{eqnarray}
\frac{1}{\hat{t}/\hat{u} - m_{K}^2} \rightarrow
\beta_{M}(\hat{s}) \frac{1}{\hat{t}/\hat{u} - m_{K}^2} + \beta_{R}(\hat{s})
\mathcal{P}^{K}(\hat{t}/\hat{u},\hat{s}) \, ,
\label{generalized_kaon_propagator}
\end{eqnarray}
where we have introduced the kaon Regge propagator $\mathcal{P}^{K}(\hat{t}/\hat{u},\hat{s}) =
\mathcal{P}^{\pi}(\hat{t}/\hat{u},\hat{s})$ (see Ref.\cite{LPS11,pion_reggeization}).
Above we have written $\hat{s}, \hat{t}, \hat{u}$ to stress that these
are quantities for a subprocess rather than for a full reaction.
$\beta_M(\hat{s})$ and $\beta_R(\hat{s})$ are the phenomenological functions 
which role is to interpolate between meson and Reggeon exchange.
Here, as in Ref.\cite{LPS11}, we parametrize them as:
$\beta_M(\hat{s}) = \exp\left( -(\hat{s}- 4 m_{K}^2)
/\Lambda_{int}^2 \right)$, 
$\beta_R(\hat{s}) = 1 - \beta_M(\hat{s})$.
The parameter $\Lambda_{int}$ can be fitted to experimental data.
From our general experience in hadronic physics we expect it to be
about $\Lambda_{int} \sim 1-2$ GeV.

The form factors, $F(\hat{t}/\hat{u})$, correct for the off-shellness of the intermediate kaons
in the middle of the diagrams shown in
Fig.~\ref{fig:central_double_diffraction_diagrams_fsi}. In the following
they are parameterized as
\begin{equation}
F_{K}(\hat{t}/\hat{u})=\exp\left(\frac{\hat{t}/\hat{u}-m_{K}^{2}}{\Lambda^{2}_{off}}\right) \;,
\label{off-shell_form_factors}
\end{equation}
where the parameter $\Lambda_{off}$ is not known in general but, in
principle, could be fitted to the normalized experimental data.
How to extract $\Lambda_{off}$ will be discussed in the result section.

The absorptive corrections to the Born amplitude due to $pp$-interactions 
were taken into account in \cite{LPS11} as 
\begin{eqnarray}
{\cal M}^{pp-rescatt.}=
\frac{i}{8 \pi^{2} s}
\int d^{2} \bk_{t} A^{I\!\!P}_{pp \to pp}(s,k_{t}^{2})
{\cal M}^{Born}
(\bp^{\,*}_{a,t}-\bp_{1,t},\bp^{\,*}_{b,t}-\bp_{2,t}) \;,
\label{abs_correction}
\end{eqnarray}
where $p^{\,*}_{a} = p_{a} - k_{t}$, $p^{\,*}_{b} = p_{b} + k_{t}$
and $k_{t}$ is the transverse momentum exchanged in the blob.

The formula presented so far do not include $\pi \pi, KK \to KK$ rescatterings.
The pion-pion interaction at high energies was studied e.g. in Refs. \cite{SNS,SS}.
In full analogy to those works at the higher energies one can include
the $\pi \pi, KK \to KK$ rescattering for our four-body reaction by replacing
the normal (or reggeized) pion/kaon propagators (including vertex form factors).

The $KK \to KK$ subprocess amplitude for $t$ and $u$ diagrams
in Fig.~\ref{fig:central_double_diffraction_diagrams_fsi}
is written in the high-energy approximation 
\begin{eqnarray}
\frac{F_{K}^{2}(\hat{t})}{\hat{t} - m_{K}^2} &\rightarrow&
\frac{i}{16 \pi^2 \hat{s}}
\int d^2 \kappa  
\frac{F_{K}^{2}(\hat{t}_1)}{\hat{t}_1 - m_{K}^2}
M_{K^{+} K^{-} \to K^{+} K^{-}}(\hat{s},\hat{t}_{2}) \, , \nonumber \\
\frac{F_{K}^{2}(\hat{u})}{\hat{u} - m_{K}^2} &\rightarrow&
\frac{i}{16 \pi^2 \hat{s}}
\int d^2 \kappa  
\frac{F_{K}^{2}(\hat{u}_1)}{\hat{u}_1 - m_{K}^2}
M_{K^{-} K^{+} \to K^{-} K^{+}}(\hat{s},\hat{u}_{2}) \,.
\label{kk_resc}
\end{eqnarray}
Here the integration is over momentum in the loop (see \cite{SS}).
The quantities $\hat{t}_1$, $\hat{u}_1$ and $\hat{t}_2$, $\hat{u}_2$
are four-momenta squared of the exchanged objects
in the first and in the second step of the
rescattering process. Other details are explained in \cite{SNS}.

The elastic amplitudes in the $KK \to KK$ subprocesses are written as
\begin{eqnarray}
M_{KK \to KK}(\hat{s},\hat{t}_{2}/\hat{u}_{2})= 
  \beta'_M(\hat{s}) A_{KK \to KK}^{V-exch.}(\hat{t}_{2}/\hat{u}_{2})
+ \beta'_R(\hat{s}) A_{KK \to KK}^{Regge}(\hat{s},\hat{t}_{2}/\hat{u}_{2})\,,
\label{kk_amp}
\end{eqnarray}
for vector meson ($V = \rho,\omega,\phi$) exchanges and 
$\beta'_M(\hat{s}) = \exp( -(\hat{s}- 4 m_{K}^2)/\Delta \hat{s} )$,
$\beta'_R(\hat{s}) = 1 - \beta'_M(\hat{s})$, $\Delta \hat{s} = 9$ GeV$^{2}$.

The Regge-type interaction which
includes Pomeron and Reggeon ($f_{2}$, $a_{2}$, $\rho$ and $\omega$) exchanges
applies at higher energies:
\begin{eqnarray}
A_{K^{+} K^{-} \to K^{+} K^{-}}^{Regge}(\hat{s},\hat t_{2}) &=&
\eta_{i} \, \hat{s} \, C_{i}^{K K}\left( \frac{\hat{s}}{\hat{s}_0} \right)^{\alpha_{i}(\hat t_{2})-1} 
\exp \left( \frac{B^{i}_{K K}}{2} \hat{t_{2}} \right)
\, , \nonumber \\
A_{K^{-} K^{+} \to K^{-} K^{+}}^{Regge}(\hat{s},\hat u_{2})  &=&
\eta_{i} \, \hat{s} \, C_{i}^{K K}\left( \frac{\hat{s}}{\hat{s}_0} \right)^{\alpha_{i}(\hat u_{2})-1} 
\exp \left( \frac{B^{i}_{K K}}{2} \hat{u_{2}} \right),
\label{kk_kk_regge}
\end{eqnarray}
where the scale parameter $\hat{s_{0}}$ is taken as 1 GeV$^{2}$
and the $C_{i}^{KK}$ coupling constants can be evaluated 
assuming Regge factorization $C_{i}^{KK} = (C_{i}^{KN})^2 / C_{i}^{NN}$
and are listed in Table \ref{tab:parameters}.

At low energies the Regge type of interactions
is not realistic and rather $V = \rho,\omega,\phi$ meson exchanges
must be taken into account:
\begin{eqnarray}
A_{K^{+} K^{-} \to K^{+} K^{-}}^{V-exch.} (\hat{t}_{2})
&=&
g_{KKV} F_{KKV}({\hat t_{2}})
\frac{(p_3^{*\mu}+p_3^{\mu}) P_{\mu \nu} (p_4^{*\nu}+p_4^{\nu})}
{{\hat t_{2}} - m_{V}^2 + i  m_{V} \Gamma_{V}}
g_{KKV} F_{KKV}({\hat t_{2}})
\, , \nonumber \\
A_{K^{-} K^{+} \to K^{-} K^{+}}^{V-exch.} (\hat{u}_{2})
&=&
g_{KKV} F_{KKV}({\hat u_{2}})
\frac{(p_3^{*\mu}+p_4^{\mu}) P_{\mu \nu} (p_4^{*\nu}+p_3^{\nu})}
{{\hat u_{2}} - m_{V}^2 + i  m_{V} \Gamma_{V}}
g_{KKV} F_{KKV}({\hat u_{2}}) \,,
\label{kk_kk_mes}
\end{eqnarray}
where $P_{\mu \nu}(k) = -g_{\mu \nu} +  k_{\mu} k_{\nu} / m_{V}^2$
and the $KKV$ coupling constants $g_{KKV}$ are given from SU(3) symmetry relations
$2\, g_{KK \omega } = \sqrt{2} \,g_{KK \phi } = 2\, g_{KK \rho } = g_{\rho \pi \pi} = 6.04$ \cite{KRS97},
where the value of $g_{\rho \pi \pi}$ is determined by the decay width of the $\rho$ meson.
%
%--------------------------------------------------------
\begin{figure}[!h]
\includegraphics[width=0.2\textwidth]{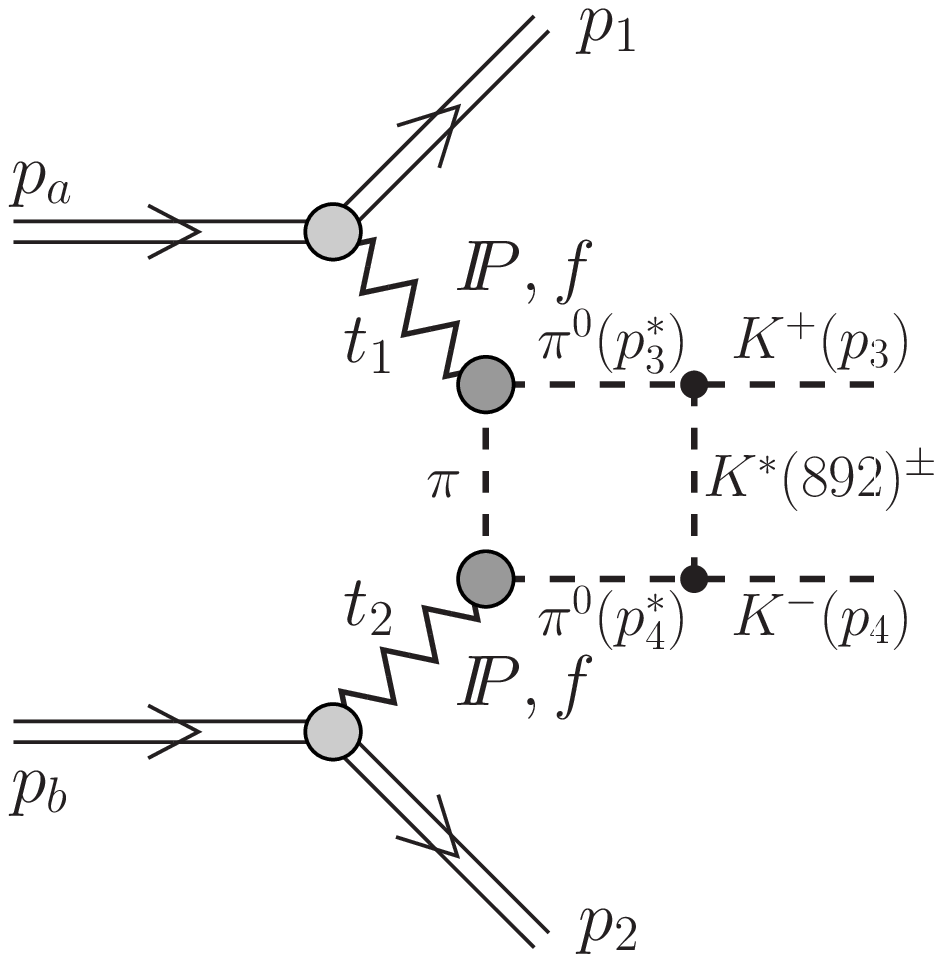}
\includegraphics[width=0.2\textwidth]{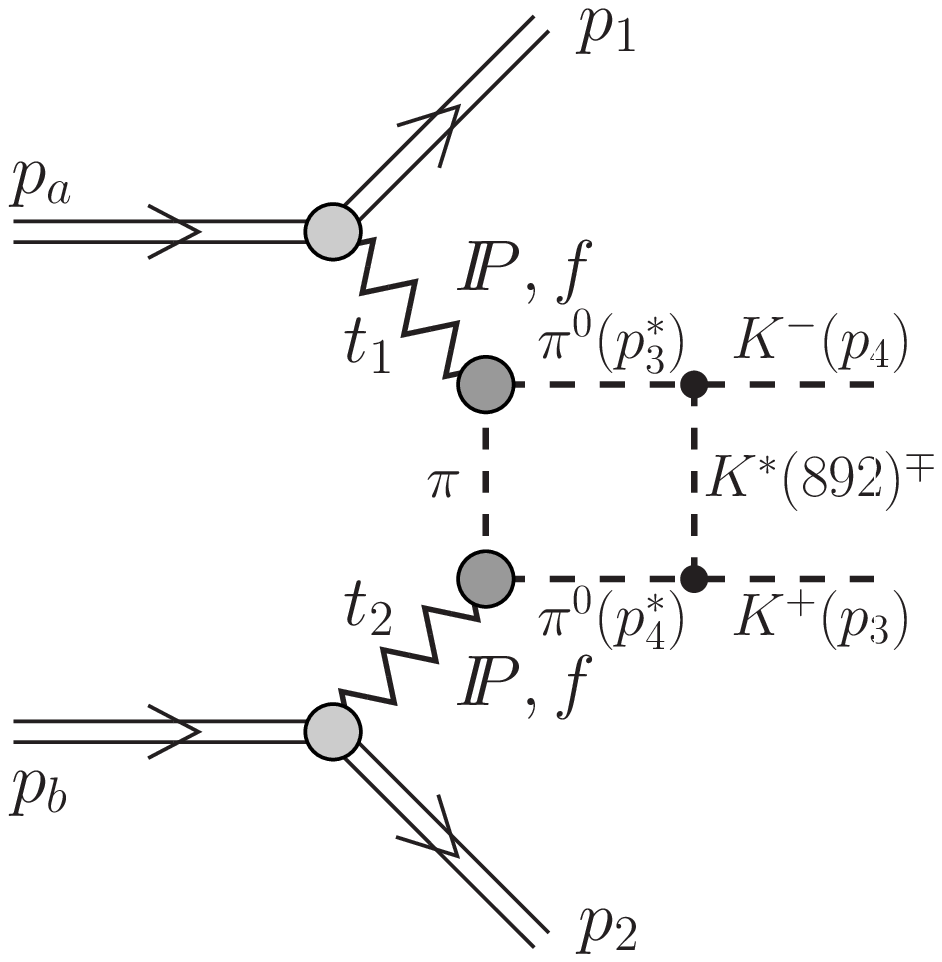}
\includegraphics[width=0.2\textwidth]{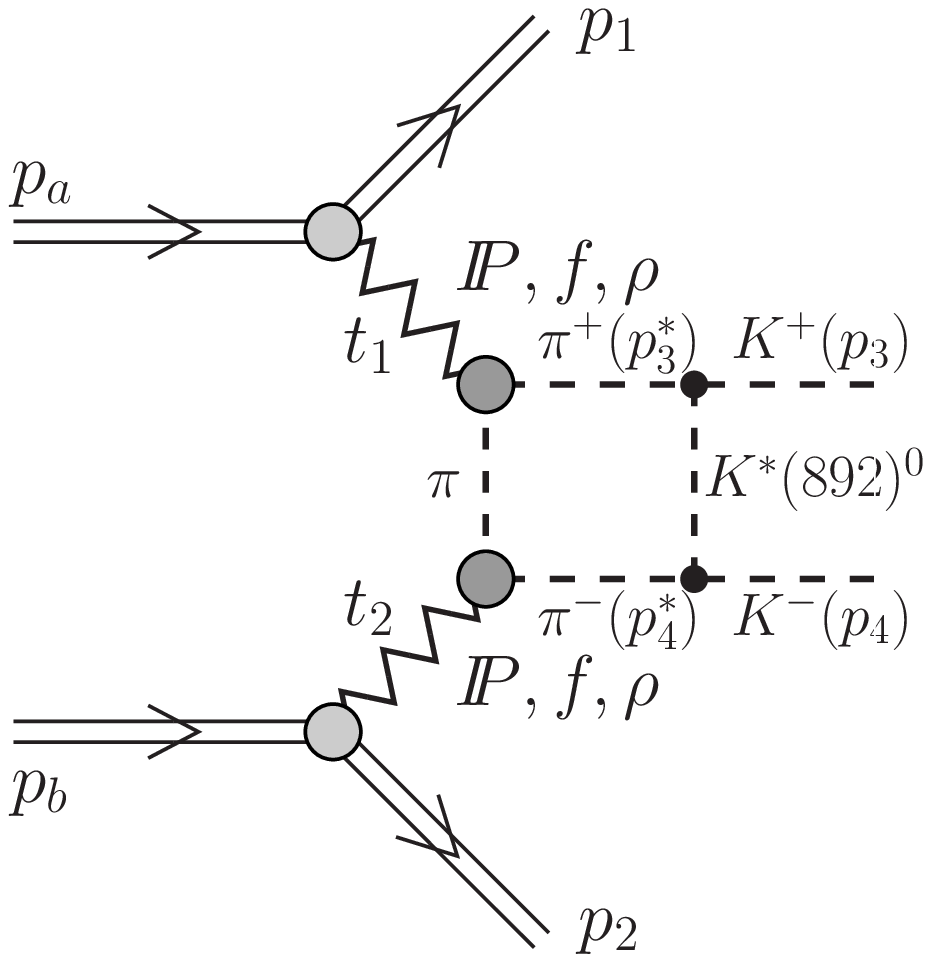}
\includegraphics[width=0.2\textwidth]{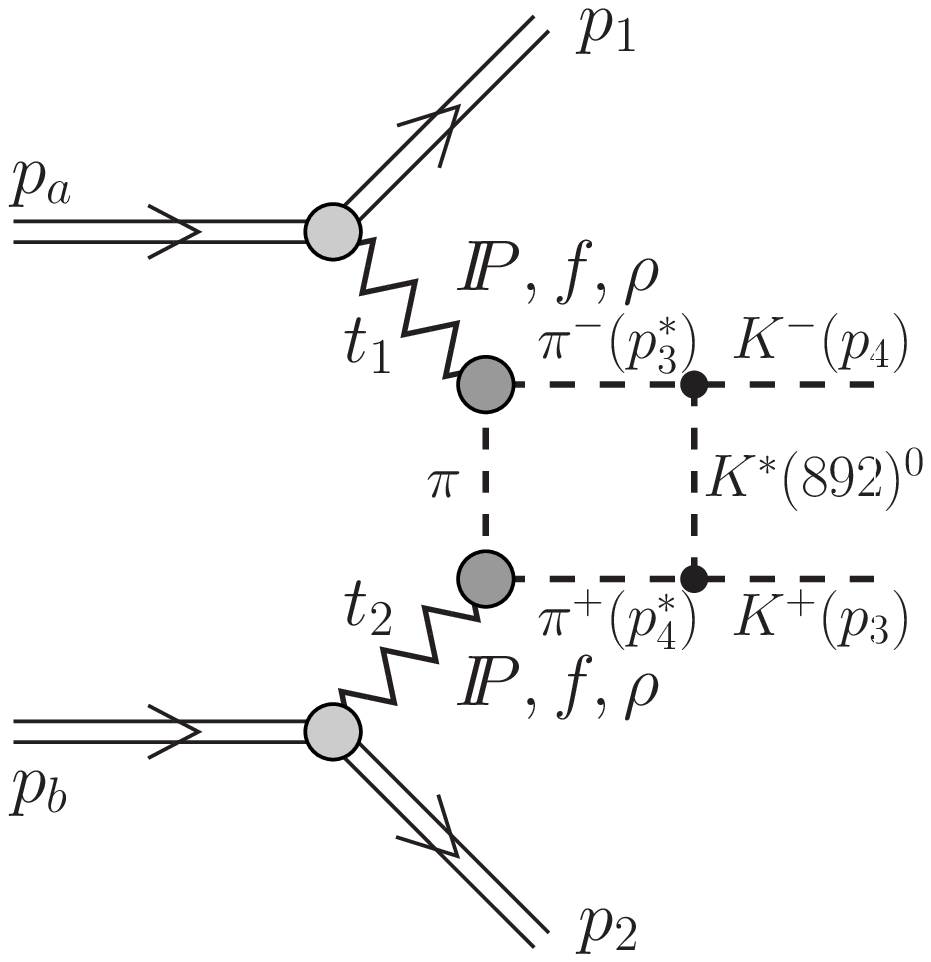}
  \caption{\label{fig:kstar_diagrams_fsi}
  \small
The central diffractive mechanism of exclusive production of
$K^{+}K^{-}$ pairs via the $K^{*}(892)$ meson exchanges.}
\end{figure}
%--------------------------------------------------------

Again the $\pi\pi \to KK$ subprocess amplitude is written in the high-energy approximation as
\begin{eqnarray}
\frac{F_{\pi}^{2}(\hat{t})}{\hat{t} - m_{\pi}^2} &\rightarrow&
\frac{i}{16 \pi^2 \hat{s}}
\int d^2 \kappa  
\frac{F_{\pi}^{2}(\hat{t}_1)}{\hat{t}_1 - m_{\pi}^2}
M_{\pi \pi \to K^{+} K^{-}}^{K^{*}-exch.}(\hat{t}_{2}) \, , \nonumber \\
\frac{F_{\pi}^{2}(\hat{u})}{\hat{u} - m_{\pi}^2} &\rightarrow&
\frac{i}{16 \pi^2 \hat{s}}
\int d^2 \kappa  
\frac{F_{\pi}^{2}(\hat{u}_1)}{\hat{u}_1 - m_{\pi}^2}
M_{\pi \pi \to K^{-} K^{+}}^{K^{*}-exch.} (\hat{u}_{2}) \,,
\label{pipi_kk_resc}
\end{eqnarray}
with
\begin{eqnarray}
M_{\pi \pi \to K^{+} K^{-}}^{K^{*}-exch.} (\hat{t}_{2})
&=&
g_{\pi K K^{*}} F_{\pi K K^{*}}({\hat t_{2}})
\frac{(p_3^{*\mu}+p_3^{\mu}) P_{\mu \nu} (p_4^{*\nu}+p_4^{\nu})}
{{\hat t_{2}} - m_{K^{*}}^2 + i  m_{K^{*}} \Gamma_{K^{*}}}
g_{\pi K K^{*}} F_{\pi K K^{*}}({\hat t_{2}})
\, , \nonumber \\
M_{\pi \pi \to K^{-} K^{+}}^{K^{*}-exch.} (\hat{u}_{2})
&=&
g_{\pi K K^{*}} F_{\pi K K^{*}}({\hat u_{2}})
\frac{(p_3^{*\mu}+p_4^{\mu}) P_{\mu \nu} (p_4^{*\nu}+p_3^{\nu})}
{{\hat u_{2}} - m_{K^{*}}^2 + i  m_{K^{*}} \Gamma_{K^{*}}}
g_{\pi K K^{*}} F_{\pi K K^{*}}({\hat u_{2}}) \,,
\label{pipi_kk_amp}
\end{eqnarray}
where now $P_{\mu \nu}(k) = -g_{\mu \nu} +  k_{\mu} k_{\nu} / m_{K^{*}}^2$ and 
we take $g_{\pi K K^{*}}= -\frac{1}{2}\, g_{\rho \pi \pi}$ \cite{KRS97}.

The quantities $F(k^2)$ in Eqs (\ref{kk_kk_mes}, \ref{pipi_kk_amp}) 
describe couplings of extended $\omega$ and $K^{*}$ mesons, respectively, and are parameterized in the exponential form:
\begin{equation}
F(k^2) = \exp\left(\frac{B_{V}}{4}(k^2-m_{V}^2)\right) \; .
\label{piKKstar_formfactors}
\end{equation}
Consistent with the definition of the coupling constant
the form factors are normalized to unity when $\omega$ or $K^{*}$ meson is on-mass-shell. 
We take $B_{V}$ = 4 GeV$^{-2}$.

The amplitudes given by formula (\ref{pipi_kk_amp}) are corrected by the factors
$( \hat{s}/ \hat{s}_{0} )^{\alpha_{K^{*}}(k^2)-1}$
to reproduce the high-energy Regge dependence.
We take $K^{*}$ meson trajectory as
$\alpha_{K^{*}}(k^2)=0.25 + \alpha'_{K^{*}} \,k^2$, with $\alpha'_{K^{*}}=0.83$ GeV$^{-2}$ \cite{pion_reggeization}.

The cross section is obtained by integration over the four-body phase space,
which is reduced to 8-dimensions and performed numerically
\begin{eqnarray}
\sigma =\int \frac{1}{2s} \overline{ |{\cal M}|^2} (2 \pi)^4
\delta^4 (p_a + p_b - p_1 - p_2 - p_3 - p_4)
\frac{d^3 p_1}{(2 \pi)^3 2 E_1}
\frac{d^3 p_2}{(2 \pi)^3 2 E_2}
\frac{d^3 p_3}{(2 \pi)^3 2 E_3}
\frac{d^3 p_4}{(2 \pi)^3 2 E_4}. \nonumber \\
\;
\label{dsigma_for_2to4}
\end{eqnarray}
The details how to conveniently reduce the number of kinematical integration variables
are given e.g. in \cite{LS10}.

%---------------------------------------------------------
\section{OTHER DIFFRACTIVE PROCESSES}
\label{section:IV}
%---------------------------------------------------------

\begin{figure}[!h]  
a)\includegraphics[width=0.3\textwidth]{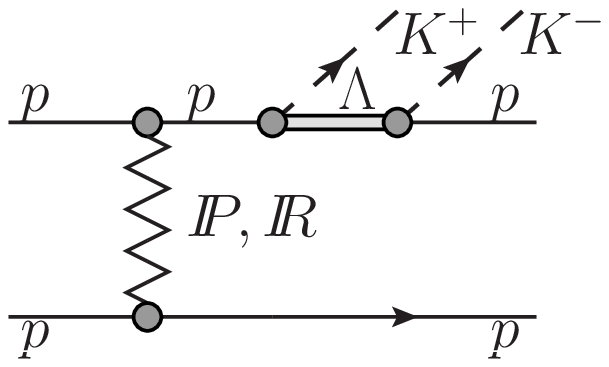}
b)\includegraphics[width=0.3\textwidth]{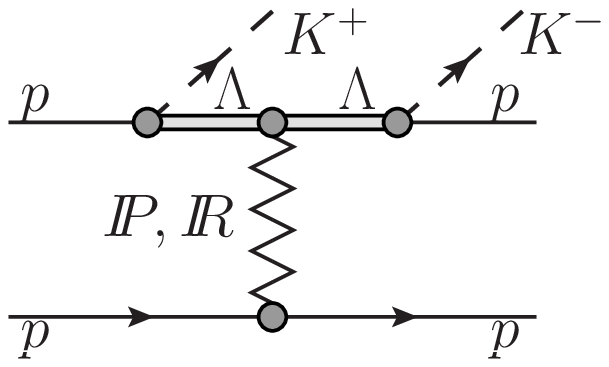}
c)\includegraphics[width=0.3\textwidth]{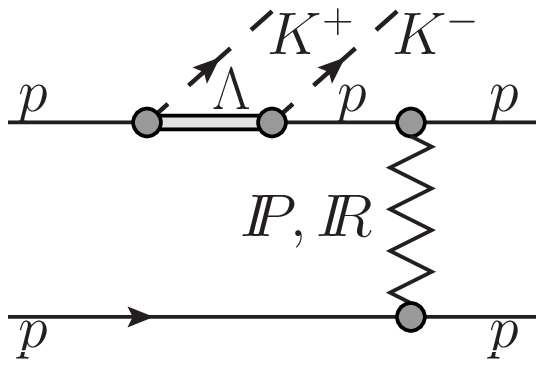}
d)\includegraphics[width=0.26\textwidth]{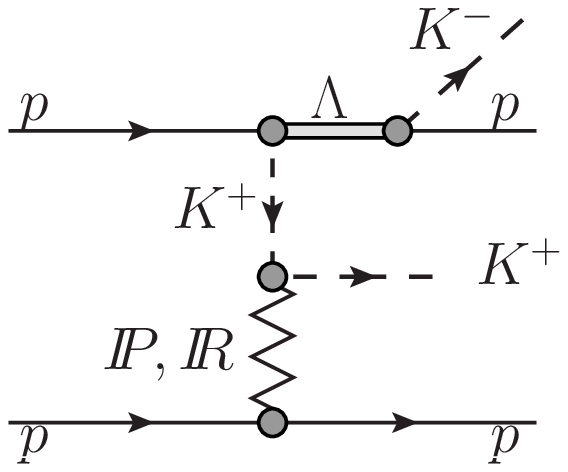}
e)\includegraphics[width=0.26\textwidth]{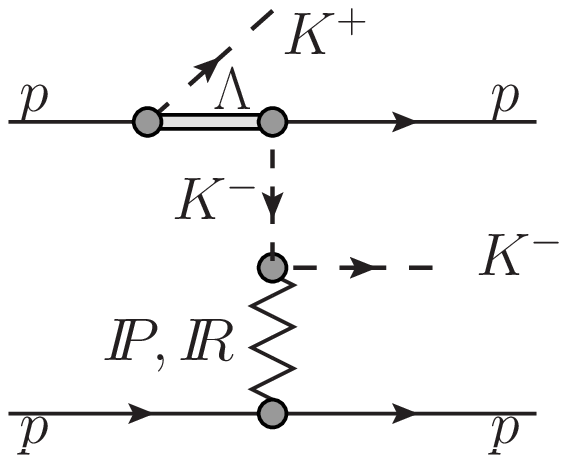}
   \caption{\label{fig:other_diagrams}
   \small
Other diffractive contributions leading to the $p p \to p p K^+ K^-$ channel.
}
\end{figure}

Up to now we have discussed only central diffractive 
contribution to the $pp \to ppK^{+}K^{-}$ reaction.
In general, there are also contributions with
other diffractive processes shown in Fig.\ref{fig:other_diagrams},
not e.valuated so far in the literature.
%--------------------------------------------------------
%--------------------------------------------------------
It is straightforward to evaluate the new diffractive contributions
of diagrams a) - e) and the Born amplitudes are given below:
\begin{eqnarray}
{\cal M}^{(a)}_{\lambda_{a}\lambda_{b} \to \lambda_{1}\lambda_{2}} &=&
\bar{u}(p_{1},\lambda_{1}) i \gamma_{5} S_{\Lambda}(p_{1fl}^{2}) i \gamma_{5} S_{p}(p_{1fp}^{2}) u(p_{a},\lambda_{a})\,
g_{\Lambda K N}^{2}\, 
F_{p}^{2}(p_{1fp}^{2}) \, F_{\Lambda}^{2}(p_{1fl}^{2}) \nonumber \\
&\times &
i s C_{I\!\!P}^{NN} \left( \frac{s}{s_{0}}\right)^{\alpha_{I\!\!P}(t_{2})-1} 
\exp\left(\frac{B_{I\!\!P}^{NN} t_{2}}{2}\right)\,
\delta_{\lambda_{2}\lambda_{b}},
\label{diagram_a}
\end{eqnarray}
\begin{eqnarray}
{\cal M}^{(b)}_{\lambda_{a}\lambda_{b} \to \lambda_{1}\lambda_{2}} &=&
\bar{u}(p_{1},\lambda_{1}) i \gamma_{5} S_{\Lambda}(p_{1fl}^{2}) S_{\Lambda}(p_{1il}^{2}) i \gamma_{5} u(p_{a},\lambda_{a})\,
g_{\Lambda K N}^{2}\, 
F_{\Lambda}^{2}(p_{1il}^{2}) \, F_{\Lambda}^{2}(p_{1fl}^{2}) \nonumber \\
&\times &
i s_{124} C_{I\!\!P}^{\Lambda N} \left( \frac{s_{124}}{s_{0}}\right)^{\alpha_{I\!\!P}(t_{2})-1} 
\left( \frac{s_{134}}{s_{th}^{pKK}}\right)^{\alpha_{\Lambda}(p_{1il}^{2})-1/2}\, %poprawka
\exp\left(\frac{B_{I\!\!P}^{\Lambda N} t_{2}}{2}\right)\,
\delta_{\lambda_{2}\lambda_{b}},
\label{diagram_b}
\end{eqnarray}
\begin{eqnarray}
{\cal M}^{(c)}_{\lambda_{a}\lambda_{b} \to \lambda_{1}\lambda_{2}} &=&
\bar{u}(p_{1},\lambda_{1}) S_{p}(p_{1ip}^{2}) i \gamma_{5} S_{\Lambda}(p_{1il}^{2}) i \gamma_{5} u(p_{a},\lambda_{a})\,
g_{\Lambda K N}^{2}\, 
F_{\Lambda}^{2}(p_{1il}^{2}) \, F_{p}^{2}(p_{1il}^{2}) \nonumber \\
&\times &
i s_{12} C_{I\!\!P}^{NN} \left( \frac{s_{12}}{s_{0}}\right)^{\alpha_{I\!\!P}(t_{2})-1} 
\left( \frac{s_{14}}{s_{th}^{pK}}\right)^{\alpha_{N}(p_{1ip}^{2})-1/2}\, %poprawka
\left( \frac{s_{34}}{s_{th}^{KK}}\right)^{\alpha_{\Lambda}(p_{1il}^{2})-1/2}\, %poprawka
\exp\left(\frac{B_{I\!\!P}^{NN} t_{2}}{2}\right)\,
\delta_{\lambda_{2}\lambda_{b}},\nonumber \\
\label{diagram_c}
\end{eqnarray}
\begin{eqnarray}
{\cal M}^{(d)}_{\lambda_{a}\lambda_{b} \to \lambda_{1}\lambda_{2}} &=&
\bar{u}(p_{1},\lambda_{1}) i \gamma_{5} S_{\Lambda}(p_{1fl}^{2}) i \gamma_{5} u(p_{a},\lambda_{a})\, S_{K}(p_{1fk}^{2})\,
g_{\Lambda K N}^{2}\, 
F_{\Lambda}^{2}(p_{1fl}^{2}) \, F_{K}^{2}(p_{1fk}^{2}) \nonumber \\
&\times &
i s_{23} C_{I\!\!P}^{KN} \left( \frac{s_{23}}{s_{0}}\right)^{\alpha_{I\!\!P}(t_{2})-1} 
\left( \frac{s_{134}}{s_{th}^{pKK}}\right)^{\alpha_{K}(p_{1fk}^{2})-1}\, %poprawka
\exp\left(\frac{B_{I\!\!P}^{KN} t_{2}}{2}\right)\,
\delta_{\lambda_{2}\lambda_{b}},
\label{diagram_d}
\end{eqnarray}
\begin{eqnarray}
{\cal M}^{(e)}_{\lambda_{a}\lambda_{b} \to \lambda_{1}\lambda_{2}} &=&
\bar{u}(p_{1},\lambda_{1}) i \gamma_{5} S_{\Lambda}(p_{1il}^{2}) i \gamma_{5} u(p_{a},\lambda_{a})\, S_{K}(p_{1ik}^{2})\,
g_{\Lambda K N}^{2}\, 
F_{\Lambda}^{2}(p_{1il}^{2}) \, F_{K}^{2}(p_{1ik}^{2}) \nonumber \\
&\times &
i s_{24} C_{I\!\!P}^{KN} \left( \frac{s_{24}}{s_{0}}\right)^{\alpha_{I\!\!P}(t_{2})-1} 
\left( \frac{s_{14}}{s_{th}^{pK}}\right)^{\alpha_{K}(p_{1ik}^{2})-1}\, %poprawka
\left( \frac{s_{13}}{s_{th}^{pK}}\right)^{\alpha_{\Lambda}(p_{1il}^{2})-1/2}\, %poprawka
\exp\left(\frac{B_{I\!\!P}^{KN} t_{2}}{2}\right)\,
\delta_{\lambda_{2}\lambda_{b}}, \nonumber \\
\label{diagram_e}
\end{eqnarray}
where $s_{0} = 1$ GeV$^2$ and $s_{th}^{pK} = (m_N + m_{K})^{2}$, $s_{th}^{pKK} = (m_N + 2 m_{K})^{2}$.
In the above equations
$u(p_{i},\lambda_{i})$, $\bar{u}(p_{f},\lambda_{f})=u^{\dagger}(p_{f},\lambda_{f})\gamma^{0}$
are the Dirac spinors (normalized as $\bar{u}(p) u(p) = 2 m_{N}$) of 
the initial and outgoing protons with the four-momenta $p$ and 
the helicities $\lambda$.
Here $s_{ij}=(p_{i}+p_{j})^{2}$, $s_{ijk}=(p_{i}+p_{j}+p_{k})^{2}$
are squared invariant masses of the $(i,j)$ and $(i,j,k)$ systems.
The four-momenta squared of the virtual particles are:
$p_{1il, 2il}^{2} = (p_{a,b} - p_{3})^{2}$,
$p_{1fl, 2fl}^{2} = (p_{1,2} + p_{4})^{2} = s_{14, 24}$,
$p_{1ik, 2ik}^{2} = (p_{1il, 2il} - p_{1, 2})^{2}$,
$p_{1fk, 2fk}^{2} = (p_{a,b} - p_{1il, 2il})^{2}$,
$p_{1ip, 2ip}^{2} = (p_{1il, 2il} - p_{4})^{2}$,
$p_{1fp, 2fp}^{2} = (p_{1fl, 2fl} + p_{3})^{2} = s_{134, 234}$.
While the four-momenta squared of transferred kaons and protons are $<0$, it is not the case for
transferred $\Lambda$'s where $p_{1il, 2il}^{2} < m_{\Lambda}^{2}$.
The propagators for the intermediate particles are respectively
\begin{eqnarray}
S_{K}(k^{2}) &=& {\frac{i}{k^{2} - m_{K}^{2}}}\,,\nonumber \\
S_{p}(k^{2}) &=& {\frac{i(k_{\nu} \gamma^{\nu} + m_{N})}{k^{2} - m_{N}^{2}}}\,,\nonumber \\
S_{\Lambda}(k^{2}) &=& {\frac{i(k_{\nu} \gamma^{\nu} + m_{\Lambda})}{k^{2} - m_{\Lambda}^{2}
%+ i m_{\Lambda} \Gamma_{\Lambda}
}}\,.
\label{propagators}
\end{eqnarray}
The form factors, $F_{i}(k^{2})$, correct for the off-shellness of the virtual particles
and are parameterised as
\begin{eqnarray}
F_{i}(k^{2})&=&\exp\left(\frac{-|k^{2}-m_{i}^{2}|}{\Lambda_{off}^{2}}\right)\,,
\label{form_factors}
\end{eqnarray}
where the parameter $\Lambda_{off}$ = 1 GeV is taken in practical calculations.
In our calculation the $\Lambda K N$ coupling constant is taken as $g^{2}_{\Lambda K N} = 14$ \cite{Laget}.

The Regge parameters in diagram (b) in Fig.\ref{fig:other_diagrams} (see Eq.(\ref{diagram_b})) 
are not known precisely and are assumed to be
$C_{I\!\!P}^{\Lambda N} \approx C_{I\!\!P}^{NN}$ (see Table \ref{tab:parameters})
and $B_{I\!\!P}^{\Lambda N} \approx B_{I\!\!P}^{NN}$ = 9 GeV$^{-2}$.
To reproduce the high-energy Regge dependence
the amplitudes given in Eqs (\ref{diagram_b} - \ref{diagram_e}) are corrected,
e.g. the amplitude of (\ref{diagram_d}) is multiplied by a factor 
$(s_{134}/ s_{th}^{pKK} )^{\alpha_{K}(p_{1fk}^2)-1}$.
The parameters of the Regge trajectories used in the calculation are given as 
$\alpha_{K}(k^2)= 0.7 (k^{2}-m_{K}^{2})$,
$\alpha_{p}(k^2)= -0.3 + 0.9 k^{2}$,
$\alpha_{\Lambda}(k^2)= -0.6 + 0.9 k^{2}$ 
for the kaon, proton and $\Lambda$ exchanges, respectively.

%--------------------------------------------------------
\begin{figure}[!h]  
\includegraphics[width=0.25\textwidth]{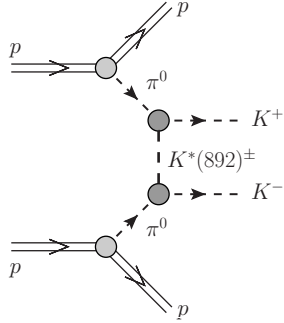}
   \caption{\label{fig:pipikk_diagram}
   \small
The $\pi\pi \to KK$ subprocess leading to the $p p \to p p K^+ K^-$ reaction.
}
\end{figure}
%--------------------------------------------------------
Finally we consider the $\pi\pi \to KK$ rescattering mechanism 
shown in Fig.\ref{fig:pipikk_diagram} which is
particularly important rather at lower energies,
e.g. for experiment PANDA to be built at GSI Darmstadt.
We write the Born amplitude according to Feynman rules as
\begin{eqnarray}
{\cal M}^{\pi\pi-KK}_{\lambda_{a}\lambda_{b} \to \lambda_{1}\lambda_{2}}(\hat{t},\hat{u}) &=&
\bar{u}(p_{1},\lambda_{1}) i \gamma_{5} u(p_{a},\lambda_{a}) S_{\pi}(t_{1}) \,
g_{\pi N N} F_{\pi NN}(t_{1}) F_{\pi K^{*}K}(t_{1})  \nonumber\\
&&
\left(M_{\pi \pi \to K^{+} K^{-}}^{K^{*}-exch.} (\hat{t}) + M_{\pi \pi \to K^{-} K^{+}}^{K^{*}-exch.} (\hat{u})\right)
\nonumber\\
&& \bar{u}(p_{2},\lambda_{2}) i \gamma_{5} u(p_{b},\lambda_{b}) S_{\pi}(t_{2}) \, 
g_{\pi N N} F_{\pi NN}(t_{2}) F_{\pi K^{*}K}(t_{2}),
\label{diagram_pipiKK}
\end{eqnarray}
where $g^2_{\pi N N}/4\pi$ = 13.5 value is taken 
and the $M_{\pi \pi \to KK}^{K^{*}-exch.}$ amplitudes are given by Eq.(\ref{pipi_kk_amp}).
%--------------------
\section{RESULTS}
\label{section:III}
\begin{figure}[!h]
\includegraphics[width = 0.45\textwidth]{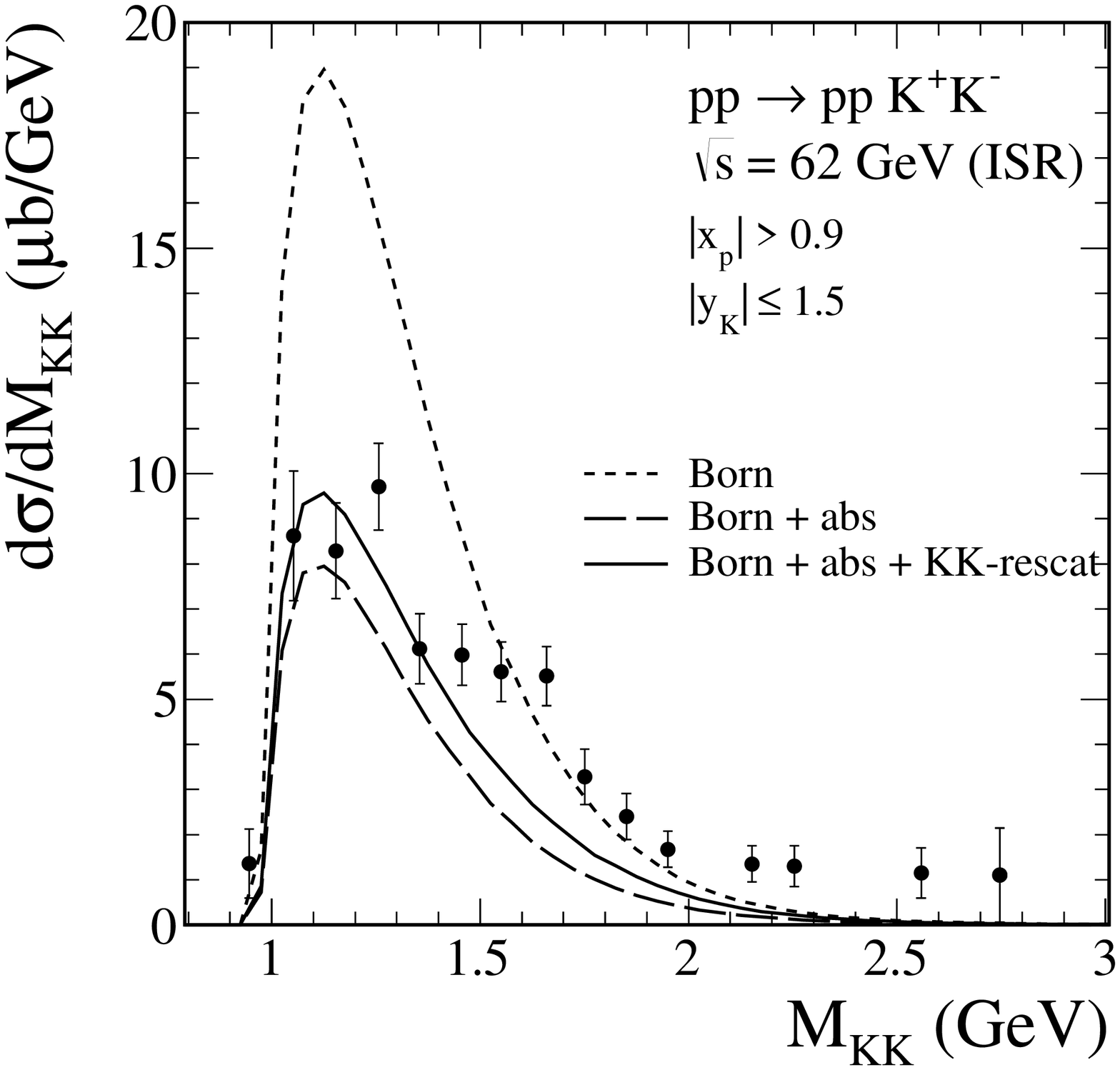}
\includegraphics[width = 0.45\textwidth]{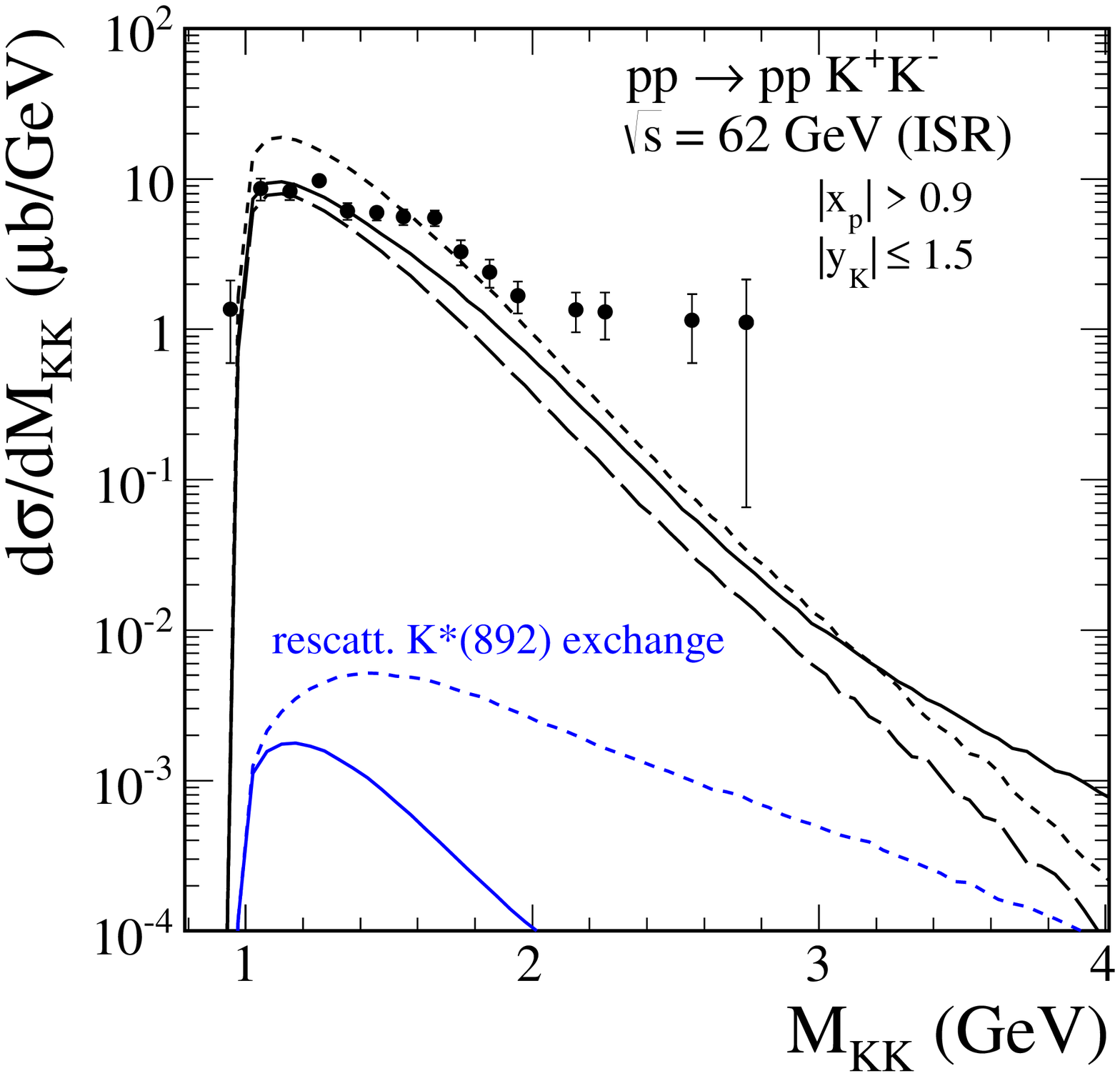}
 \caption{\label{fig:dsig_dmkk_ISR}
 \small
Differential cross section $d\sigma/dM_{KK}$
for the $pp \to pp K^{+} K^{-}$ reaction
at $\sqrt{s} = 62$ GeV with experimental
cuts relevant for the CERN ISR experimental data from Ref.~\cite{ABCDHW89}.
Right panel shows the same in logarithmic scale.
Results without (dotted line) and
with (solid line) absorption effects are shown.
Here $\Lambda_{off}^{2}=2$ GeV$^{2}$ and $\Lambda_{int} = 2$ GeV.
}
\end{figure}

%--------------------
Now we wish to show results and predictions for existing
and future experiments. We start with DPE mechanism which dominates at midrapidities.
In Fig.~\ref{fig:dsig_dmkk_ISR} we show the two-kaon invariant mass
distribution at the center-of-mass energy of the CERN ISR $\sqrt{s} = 62$ GeV \cite{ABCDHW89}.
In this calculation the experimental cuts on the rapidity
of both kaons and on longitudinal momentum fractions
(Feynman-$x$, $x_{F} = 2p_{\parallel}/\sqrt{s}$)
of both outgoing protons are included.
The experimental data show some small peaks above our flat model continuum.
They correspond to the $K^+ K^-$ resonances (e.g. $f_{2}(1270)$, $f_{2}'(1525)$)
which are not included explicitly in our calculation.
In the present analysis we are interested mostly
what happens above the region $M_{KK} > 2-3$ GeV (see right panel). 
The results depend on the value of the nonperturbative, 
a priori unknown parameter of the form factor
responsible for off-shell effects (see Eq.~(\ref{off-shell_form_factors})).
Our model with $\Lambda_{off}^{2}=2$ GeV$^{2}$ cut-off parameter fitted to the data provides an educated
extrapolation to the unmeasured region.
%We show results with the cut-off parameters $\Lambda_{off}^{2}=2$ GeV$^{2}$.
We compare results without (dotted lines) and
with absorption corrections including the $KK$-rescattering effect (solid line).
At the $\chi_{c0}$ mass the $KK$-rescattering leads
to an enhancement of the cross section compared to the calculation without $KK$-rescattering.
Below we shall use also this background predictions when analyzing
the signal ($\chi_{c0}$) to background ratio.
%--------------------------------------------------------
%--------------------------------------------------------

\begin{figure}[!h]
\includegraphics[width = 0.32\textwidth]{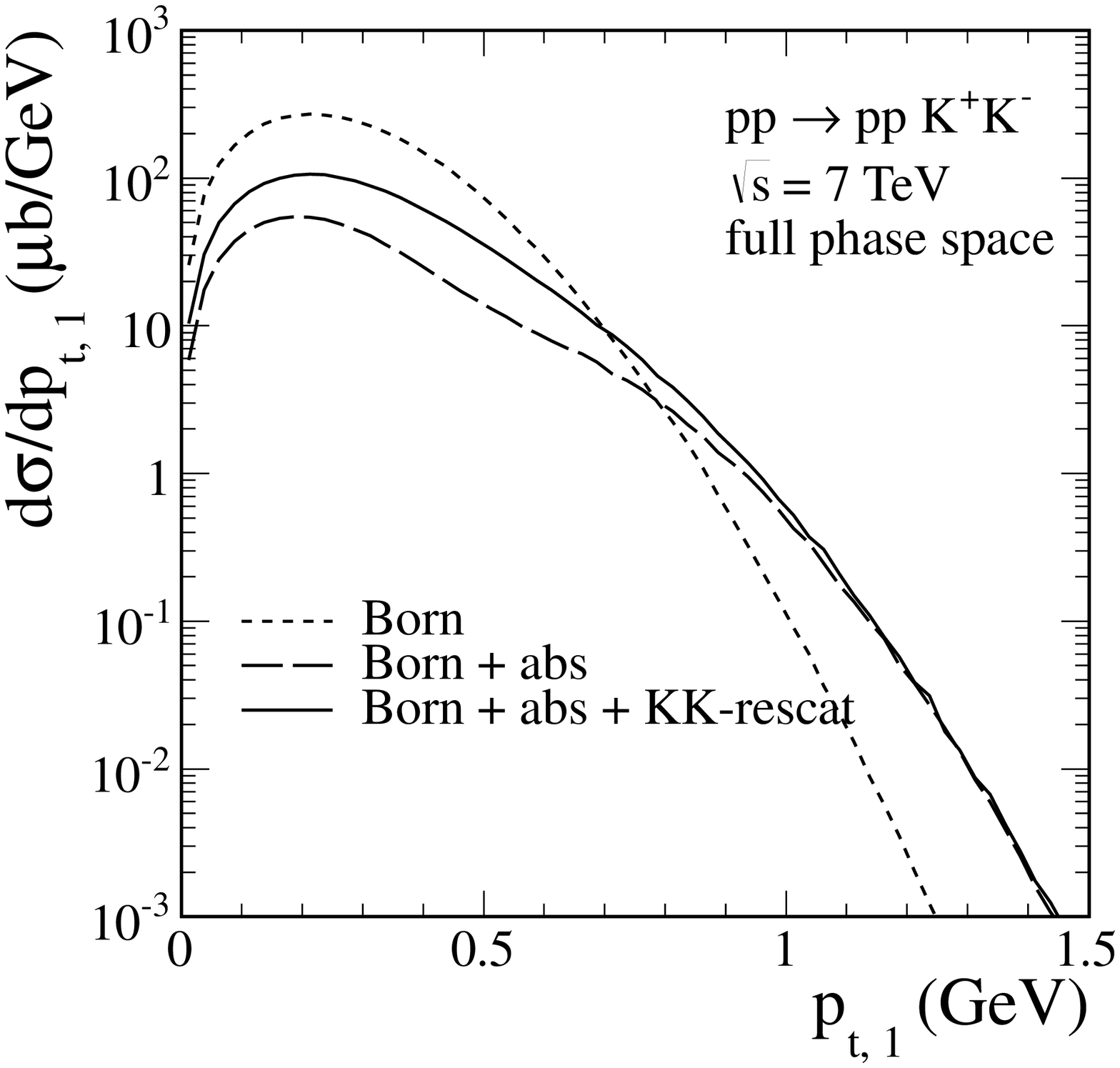}
\includegraphics[width = 0.32\textwidth]{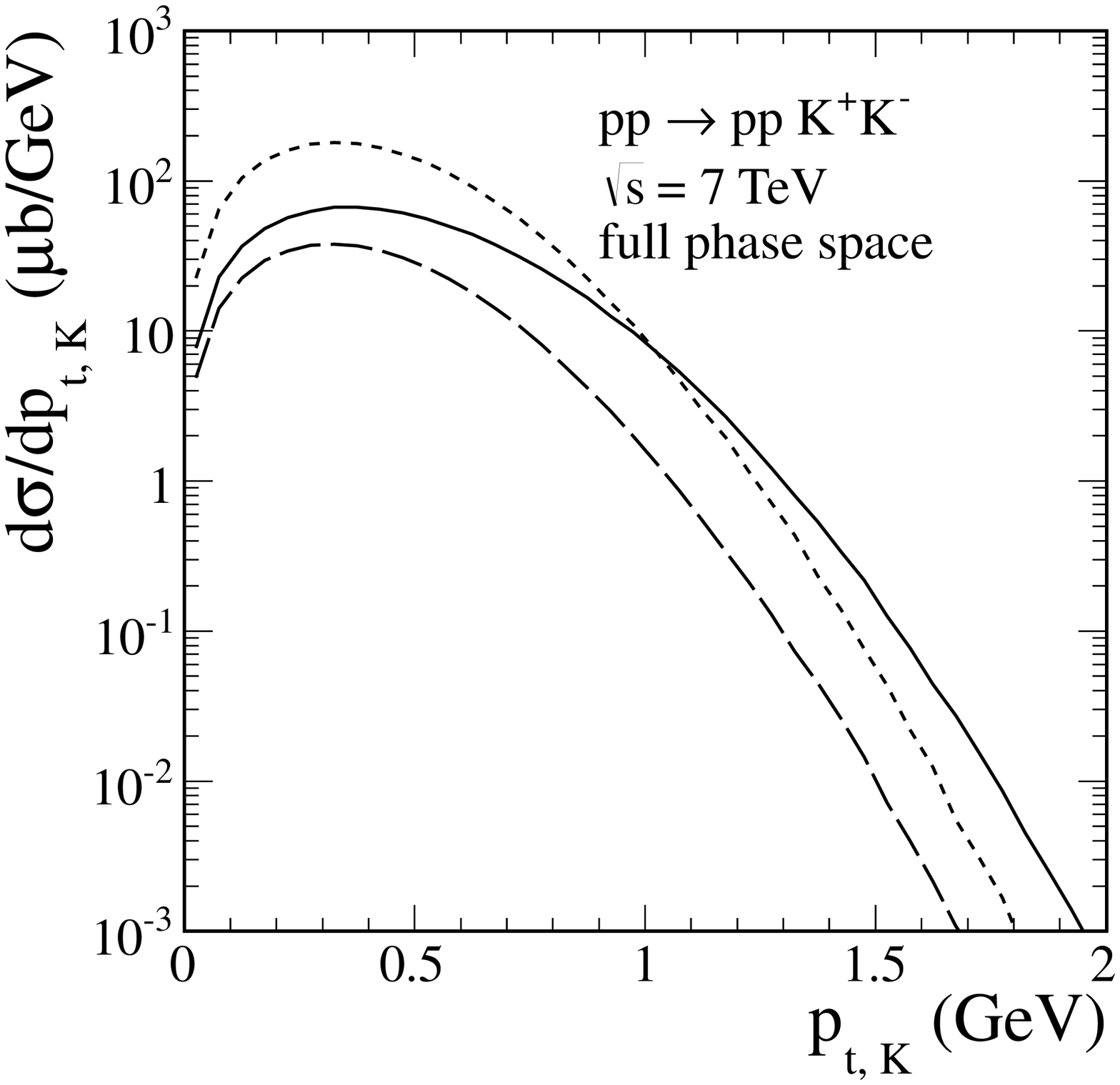}\\
\includegraphics[width = 0.32\textwidth]{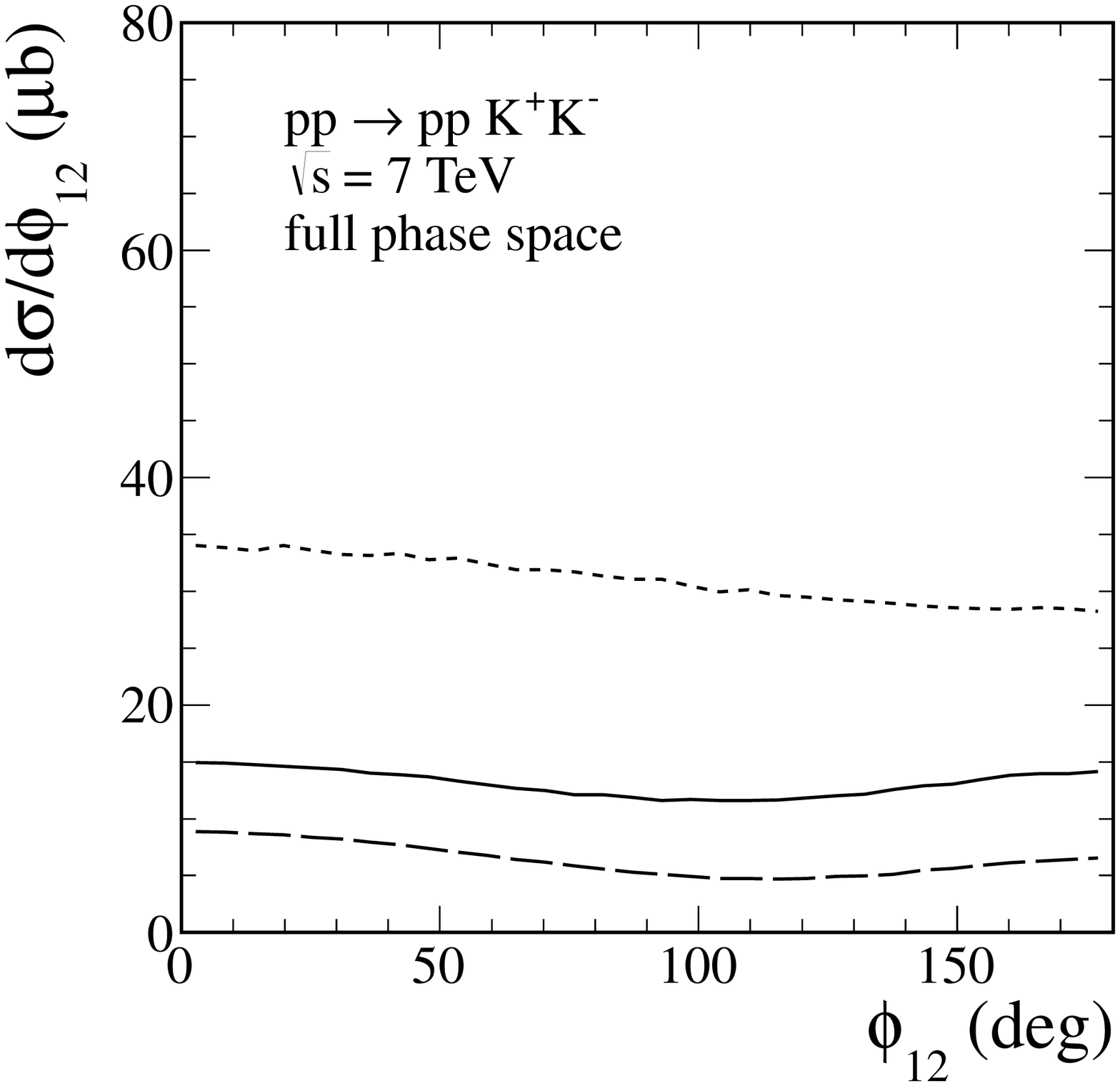}
\includegraphics[width = 0.32\textwidth]{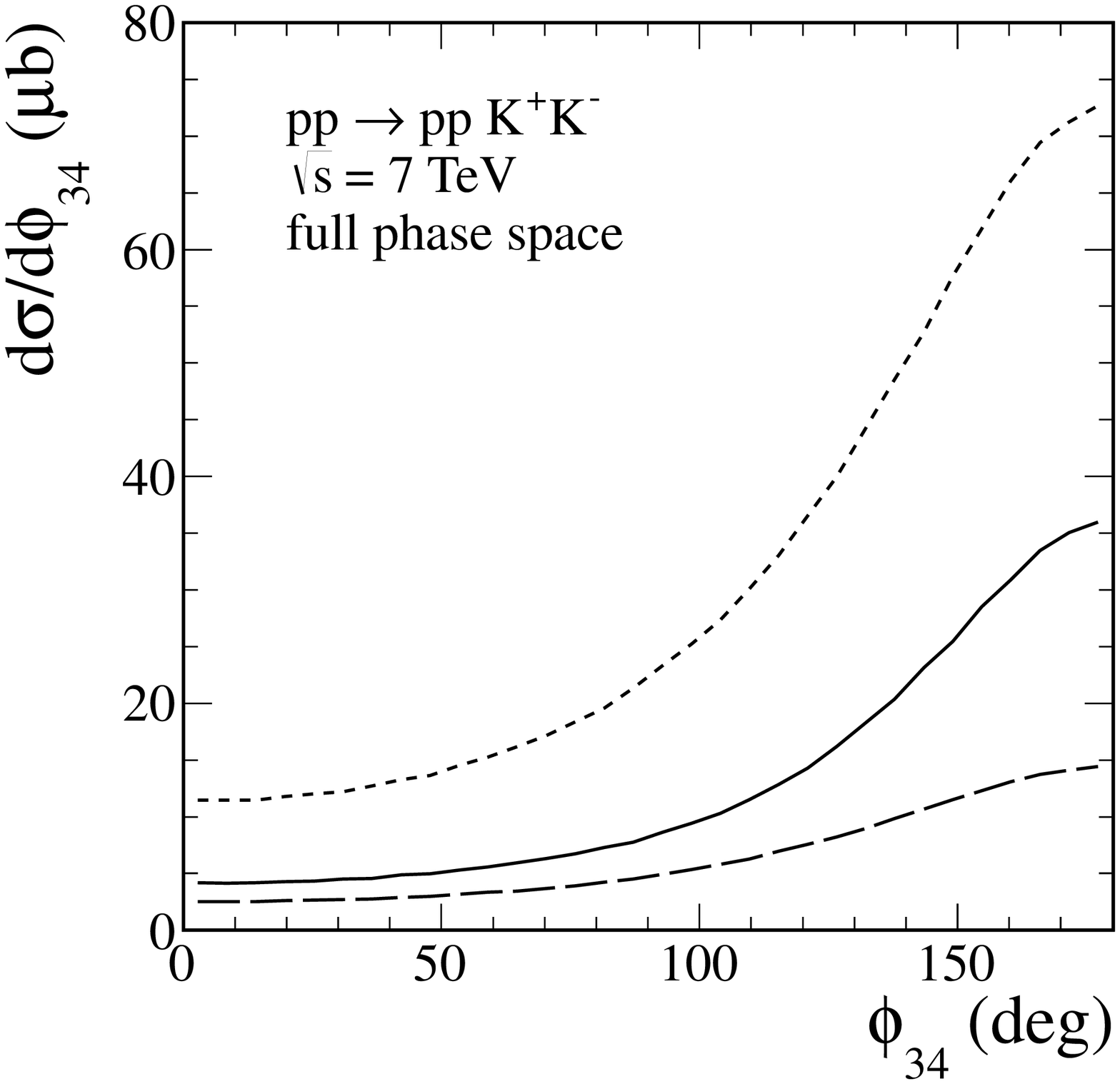}\\
\includegraphics[width = 0.32\textwidth]{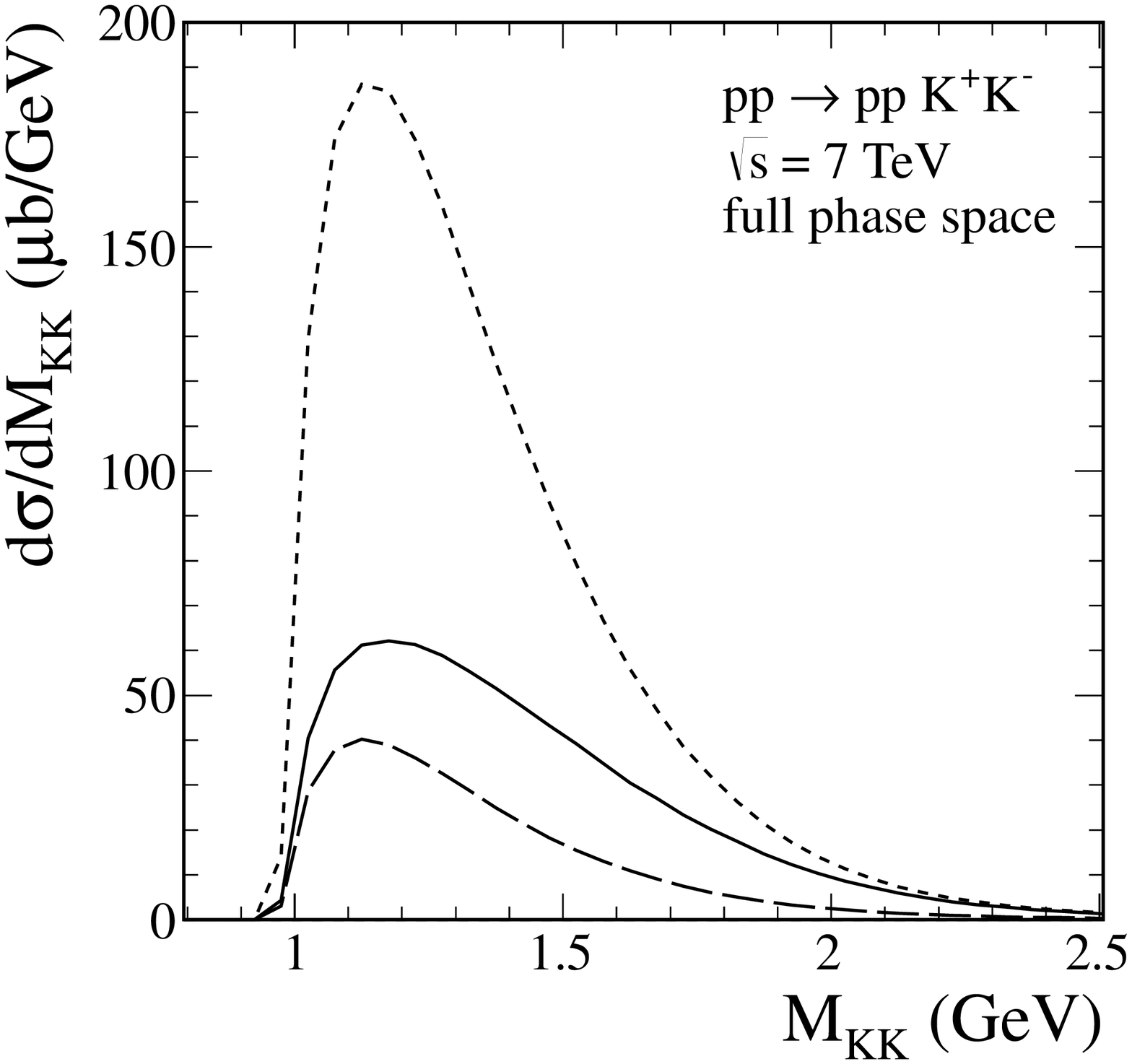}
  \caption{\label{fig:diff_comp}
  \small
Differential cross sections for the $pp \to pp K^{+} K^{-}$ reaction
at $\sqrt{s} = 7$ TeV without (dotted line) and with (solid line) the absorption effects.
These calculations were done with the cut-off parameter $\Lambda_{off}^{2} = 2$ GeV$^{2}$ 
and $\Lambda_{int} = 2$ GeV.}
\end{figure}

%Let us start from presenting various differential cross sections.
In Fig.~\ref{fig:diff_comp} we show differential distributions
for the $pp \to pp K^{+} K^{-}$ reaction
at $\sqrt{s} = 7$ TeV without (dotted line)
and with (solid line) the absorptive corrections.
In most distributions the shape is almost unchanged.
The only exception is the distribution in proton transverse momentum
where we predict a damping of the cross section at small
proton $p_{t}$ and an enhancement of the cross section at large proton $p_{t}$.
%In relative azimuthal angle beetwen outgoing nucleons $(\phi_{12})$
%distribution we observe a dip in the region of $\phi_{12}\sim \pi/2$.
%--------------------------------------------------------
%--------------------------------------------------------

\begin{figure}[!h]
\includegraphics[width = 0.32\textwidth]{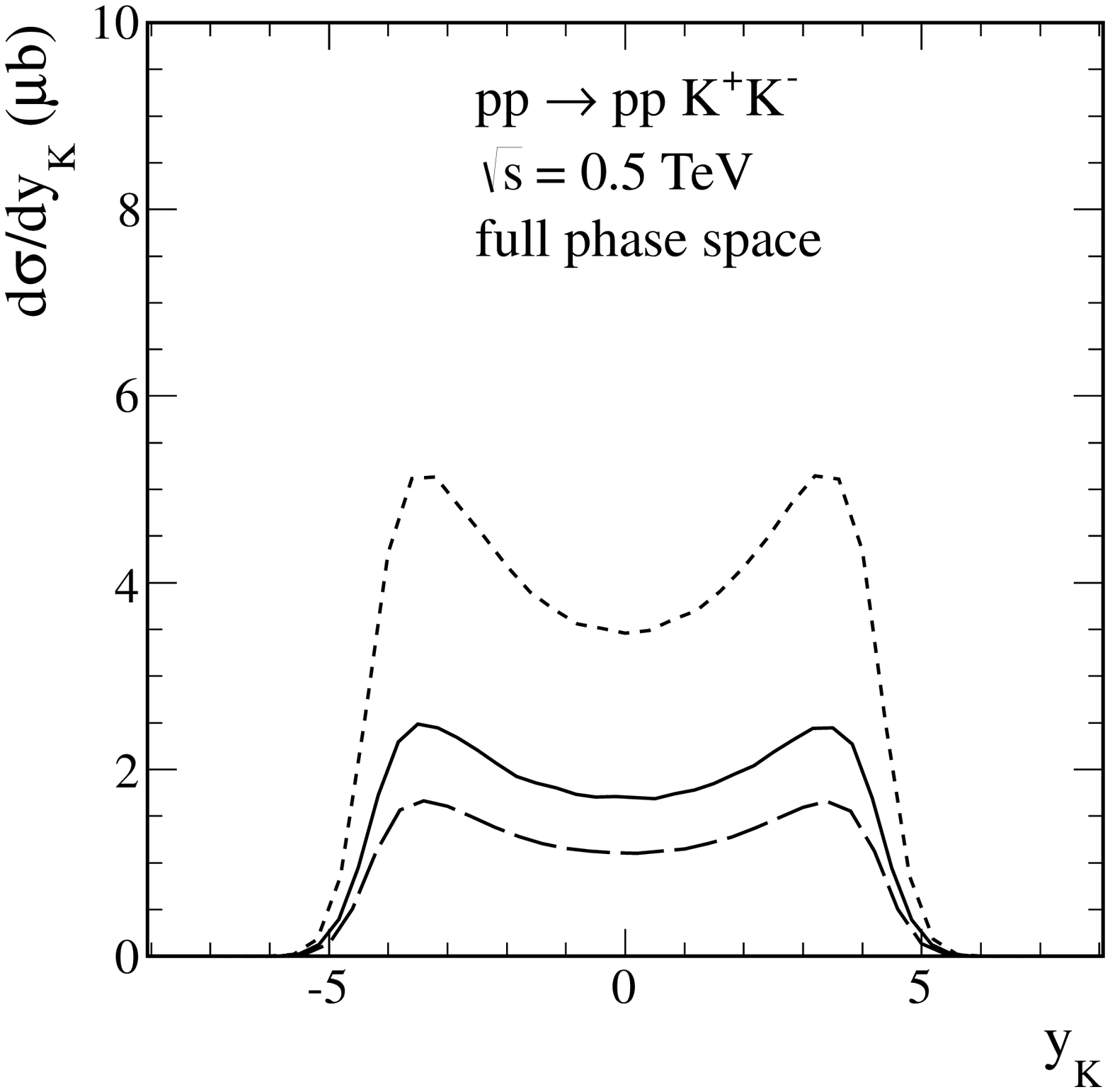}
\includegraphics[width = 0.32\textwidth]{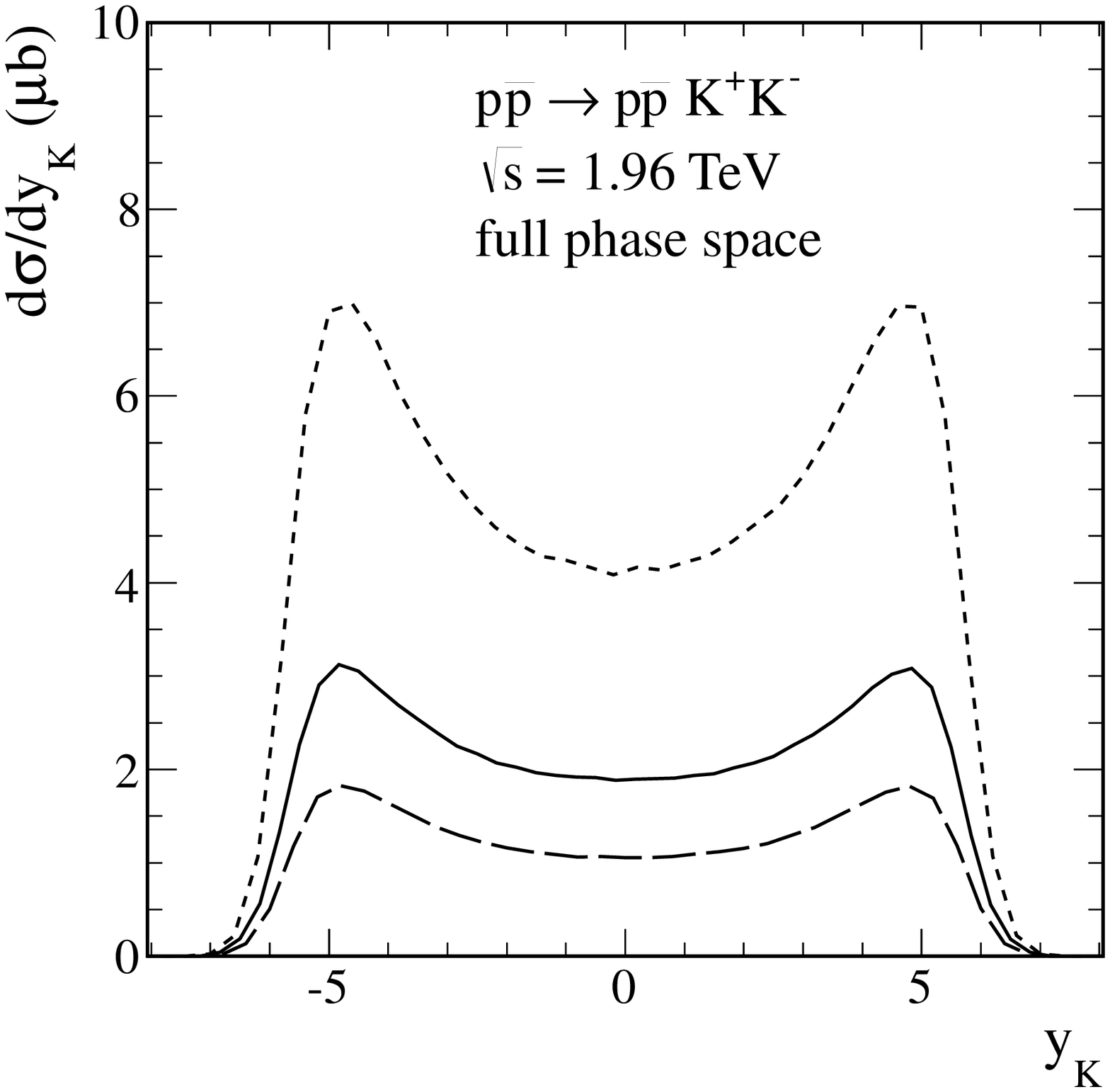}
\includegraphics[width = 0.32\textwidth]{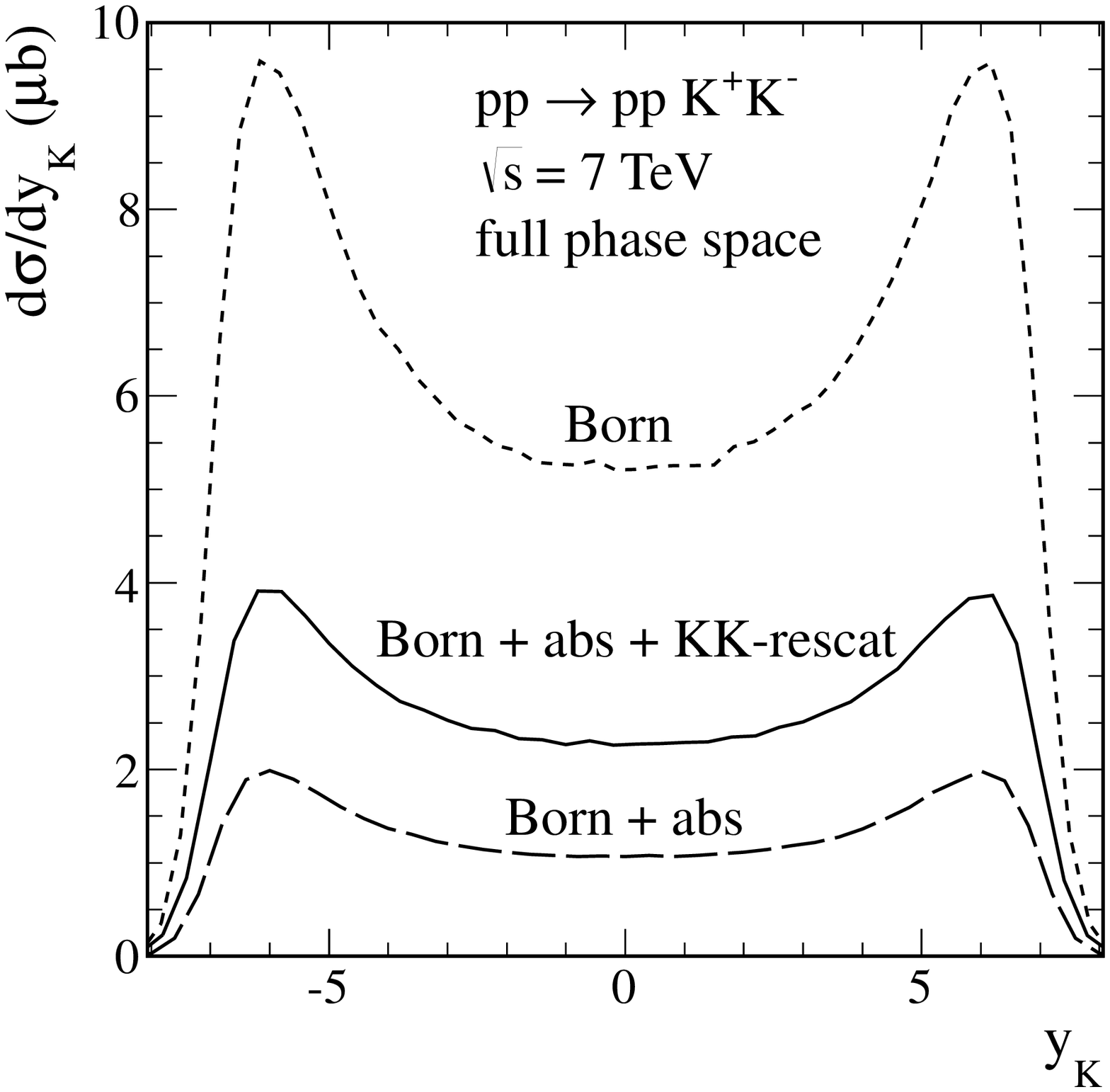}
  \caption{\label{fig:y3}
  \small
Differential cross section $d\sigma/dy_{K}$
for the $pp \to pp K^{+} K^{-}$ reaction
at $\sqrt{s} = 0.5, 1.96, 7$ TeV
with $\Lambda_{off}^{2}=2$ GeV$^{2}$.
The results without (upper lines) and with (bottom lines)
absorption effects due to $pp$-interaction and $KK$-rescattering are shown too.
}
\end{figure}

\begin{figure}[!h]
\includegraphics[width = 0.32\textwidth]{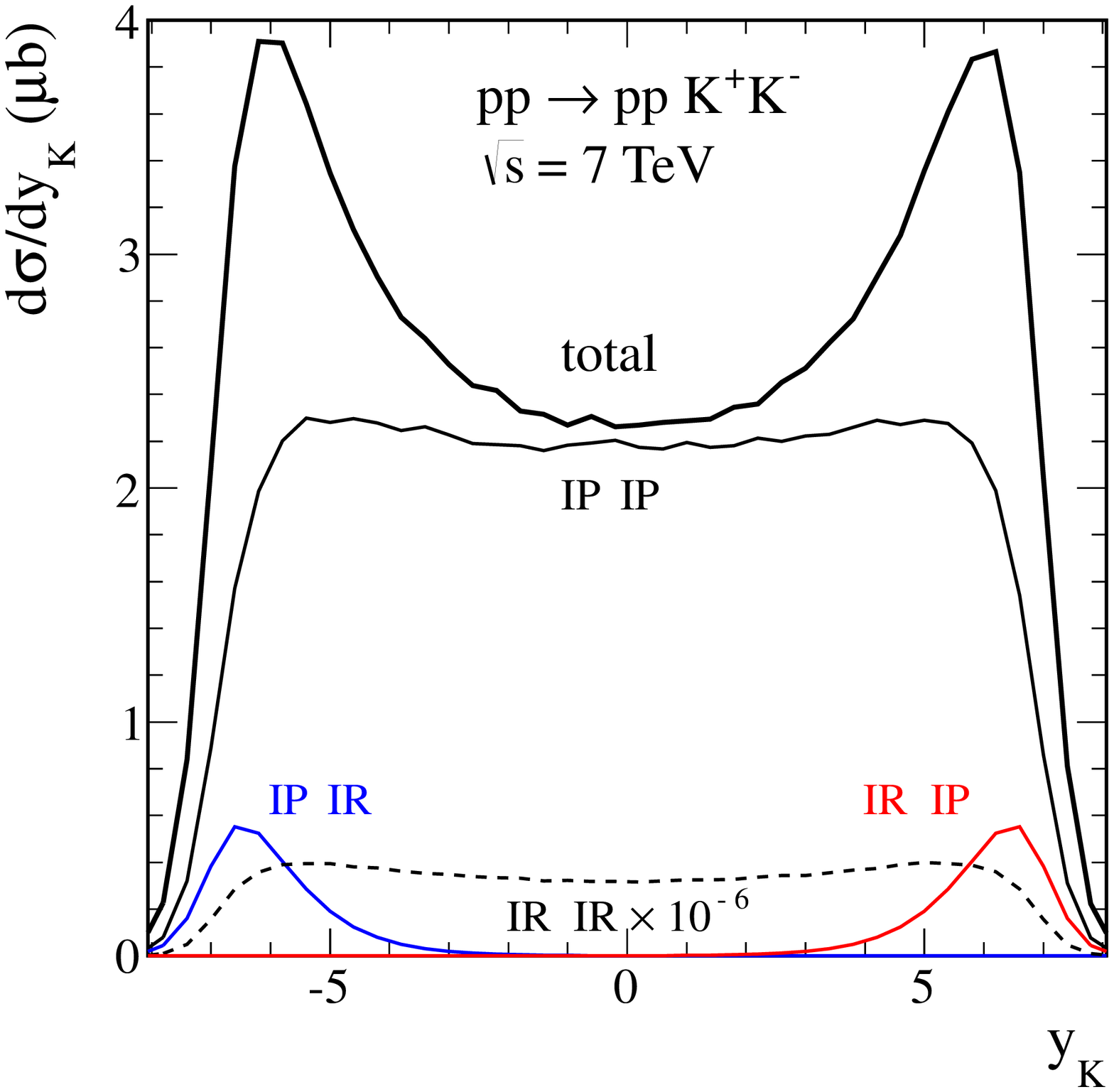}
  \caption{\label{fig:y3_deco}
  \small
Differential cross section $d\sigma/dy_{K}$
for the $pp \to pp K^{+} K^{-}$ reaction at $\sqrt{s} = 7$ TeV 
with $\Lambda_{off}^{2}=2$ GeV$^{2}$.
The different lines corresponds to the situation when all 
and only some components in the amplitude are included.
The details are explained in the main text.
}
\end{figure}

In Fig.~\ref{fig:y3} we show differential distributions in
kaon rapidity $y_{K} = y_3 = y_4$ for the $pp \to pp K^{+} K^{-}$ reaction at
$\sqrt{s} = 0.5, 1.96, 7$ TeV without (upper lines) and with
(bottom lines) absorption effects. 
The integrated cross section slowly rises with incident energy. 
The reader is asked to notice that
the energy dependence of the cross section at $y_{K} \approx 0$
is reversed by the absorption effects which are stronger at higher energies.
In our calculation we include both Pomeron and Reggeon exchanges.
The camel-like shape of the distributions is due to the interference of the components in the amplitude.
In Fig.\ref{fig:y3_deco} we show the distribution in $y_{K} = y_3 = y_4$
for all ingredients included (thick solid line) and when only
Pomeron exchanges are included (solid line),
separately for Pomeron-Reggeon (Reggeon-Pomeron) exchanges
which peaks at backward (forward) kaon rapidities
and in the case when only Reggeon exchanges are included (dashed line).

%--------------------------------------------------------
%--------------------------------------------------------
%--------------------------------------------------------
%--------------------------------------------------------

\begin{figure}[!h]  
\includegraphics[width = 0.3\textwidth]{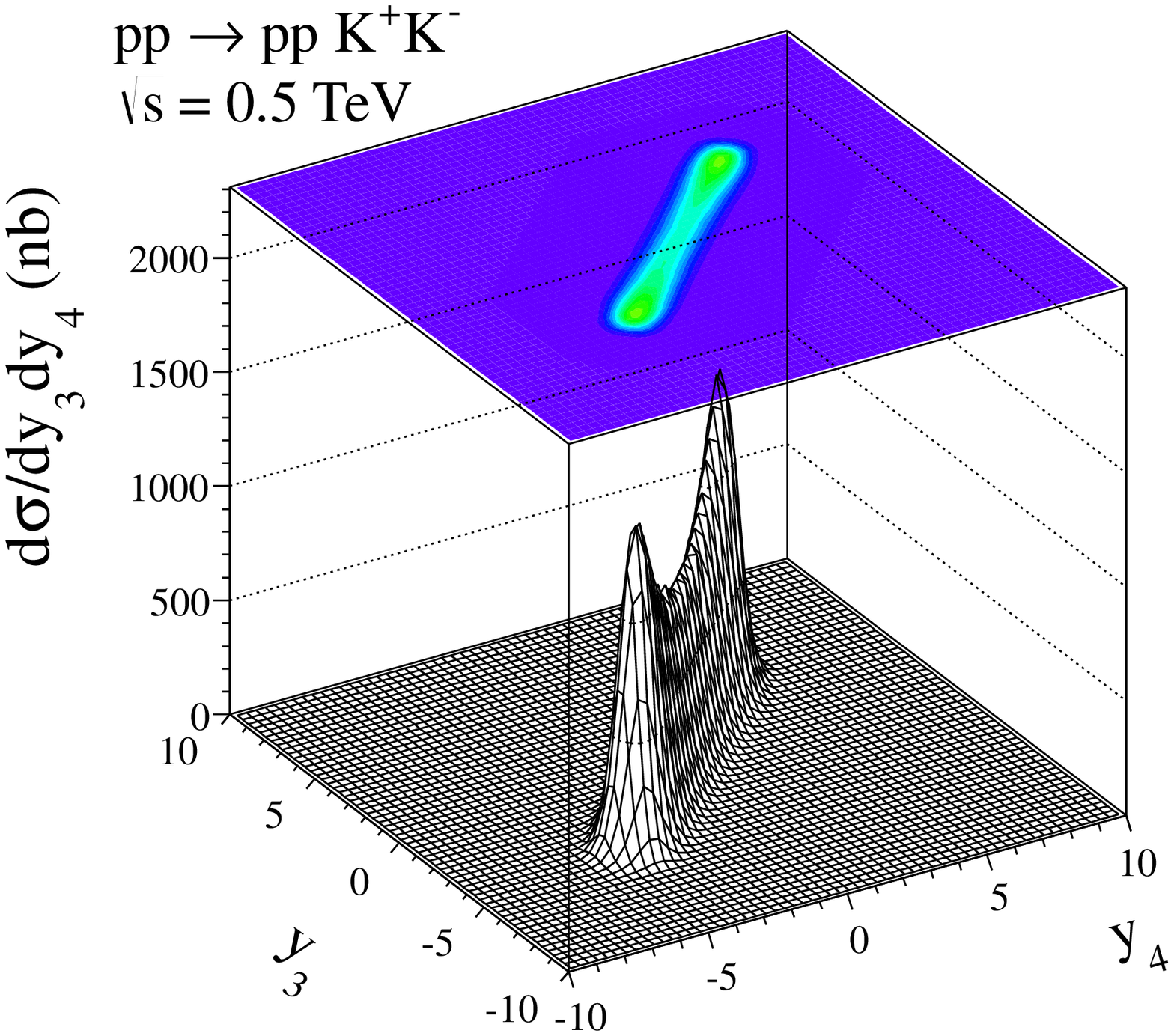}
\includegraphics[width = 0.3\textwidth]{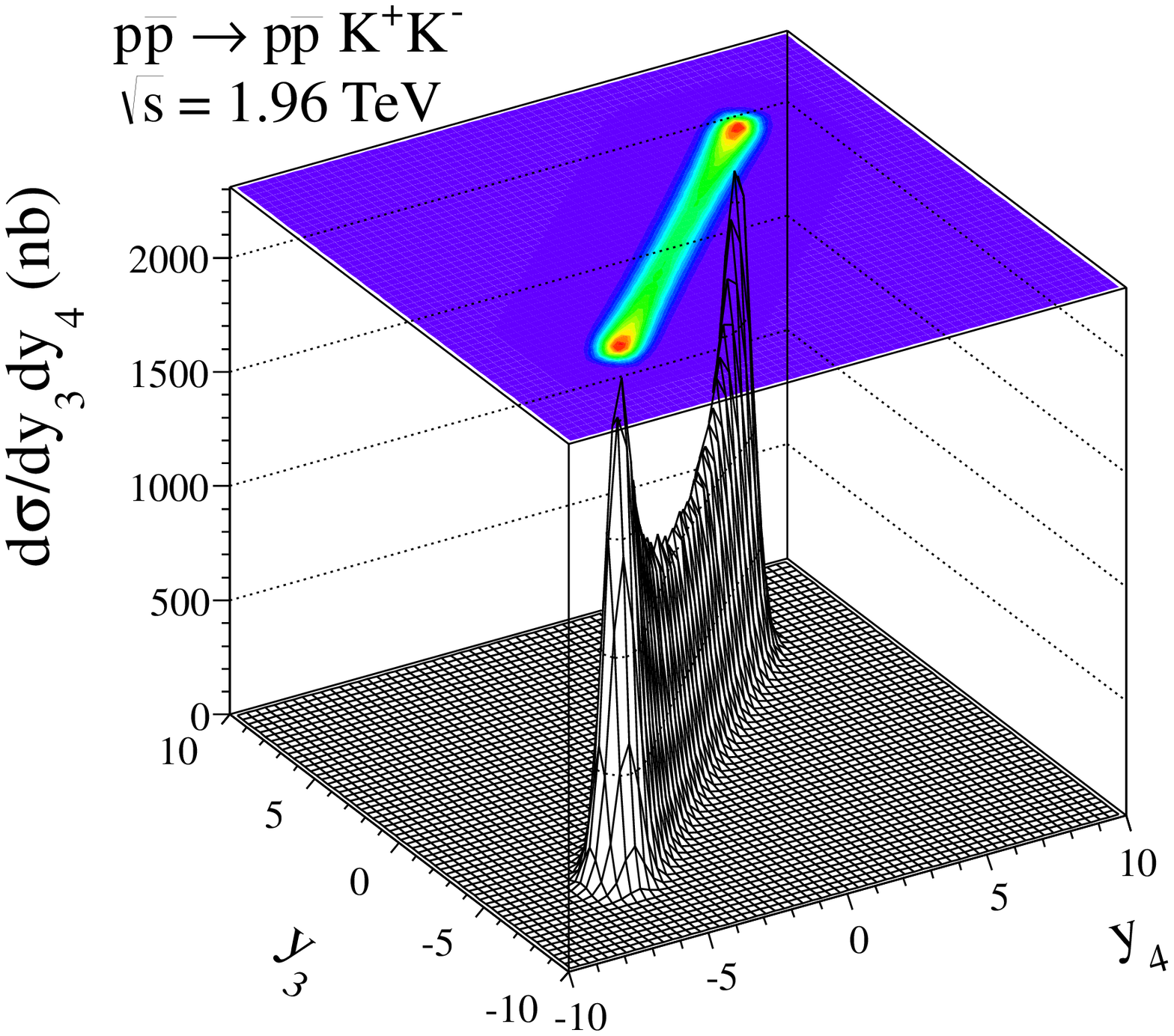}
\includegraphics[width = 0.3\textwidth]{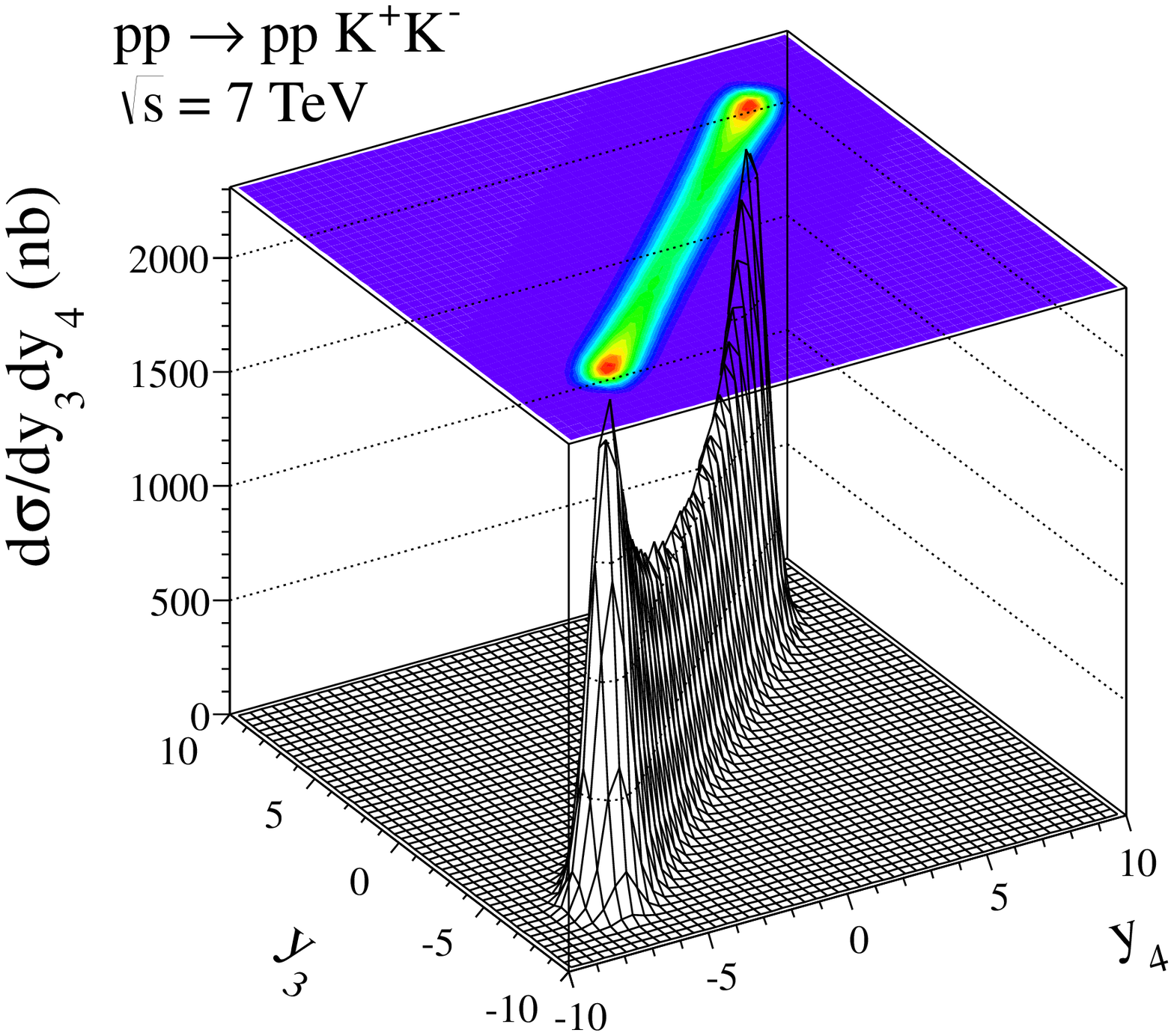}
   \caption{\label{fig:y3y4_kk}
   \small 
Differential cross section in $(y_3,y_4)$ for the central diffractive
contribution for three incident energies $\sqrt{s} = 0.5, 1.96, 7$ TeV.
The absorption effects were included here.
}
\end{figure}

In Fig.\ref{fig:y3y4_kk} we show distributions in the two-dimensional $(y_3, y_4)$ space
at $\sqrt{s} = 0.5, 1.96, 7$ TeV for the  central diffractive contribution.
The cross section grows with $\sqrt{s}$.
At high energies the kaons are emitted preferentially
in the same hemispheres, i.e. $y_3, y_4 >$ 0 or $y_3, y_4 <$ 0.
In this calculation the cut-off parameter $\Lambda^{2}_{off} = 2$ GeV$^{2}$.
%--------------------------------------------------------
%--------------------------------------------------------

\begin{figure}[!h]  
\includegraphics[width = 0.3\textwidth]{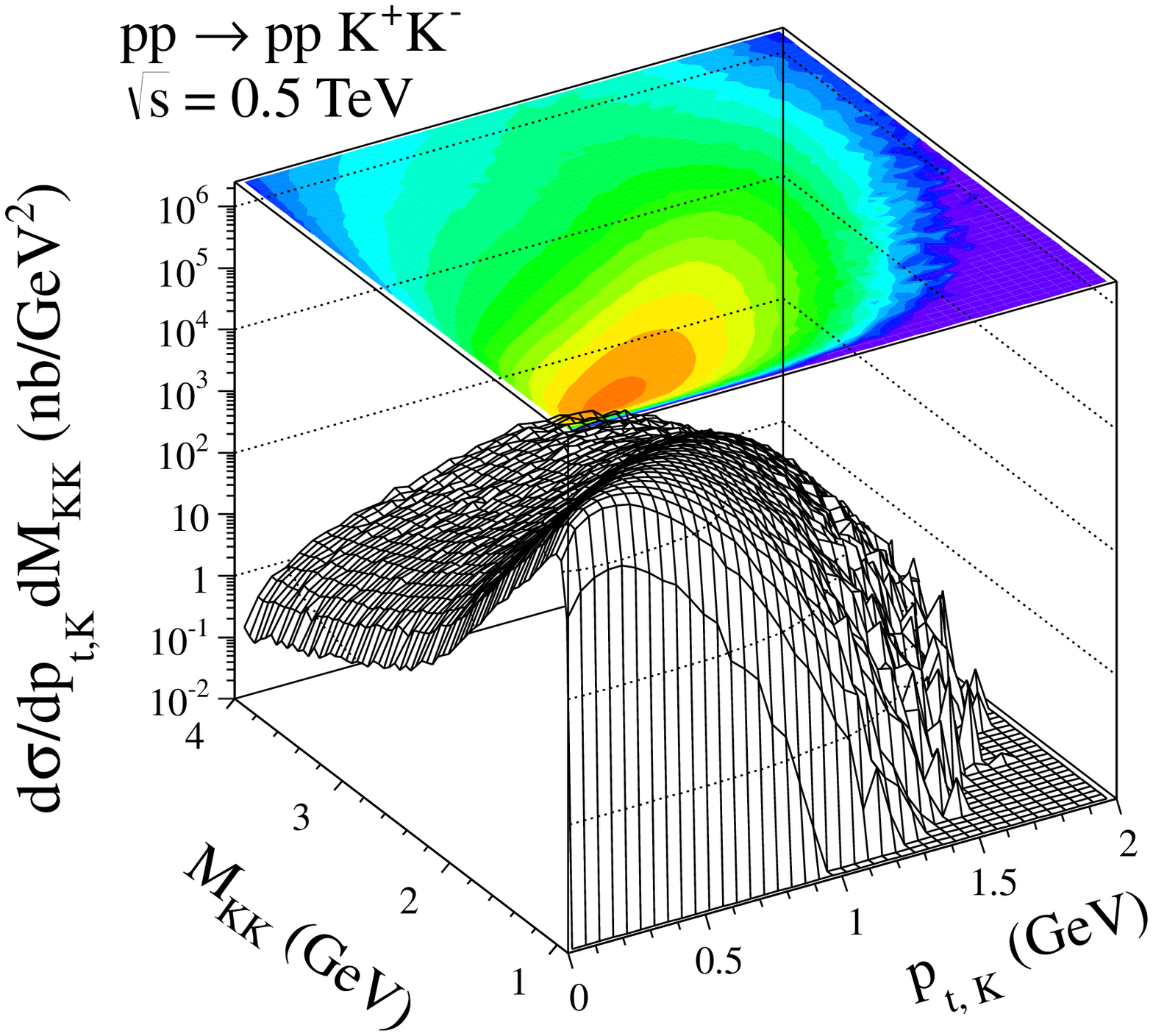}
\includegraphics[width = 0.3\textwidth]{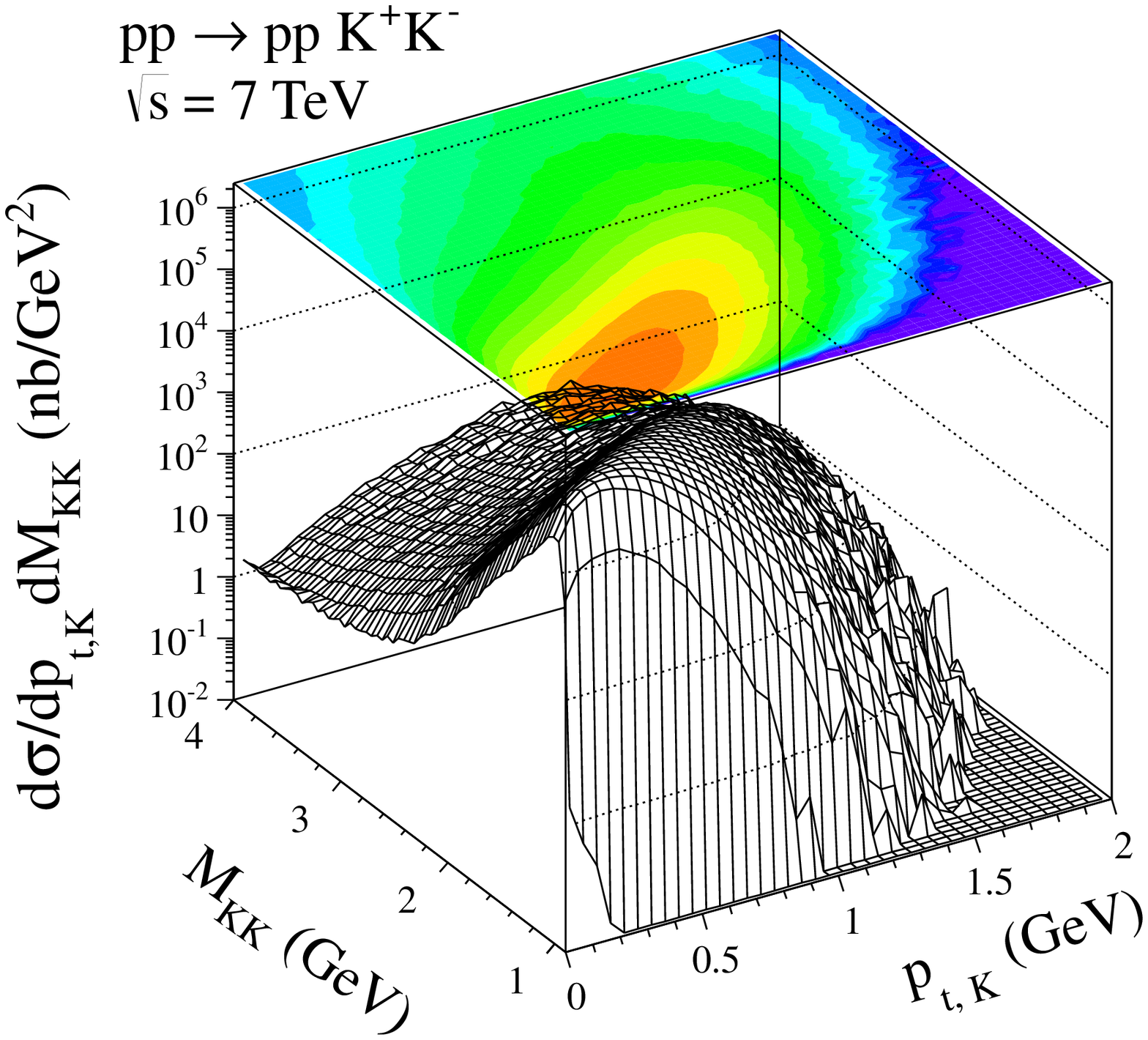}
   \caption{\label{fig:p3tmkk_kk}
   \small 
Differential cross section in $(p_{t,K},M_{KK})$ for the central diffractive
contribution for two incident energies $\sqrt{s} = 0.5, 7$ TeV.
The absorption effects were included here.
}
\end{figure}

In Fig.\ref{fig:p3tmkk_kk} we show distributions in the $(p_{t,K},M_{KK})$ space
at $\sqrt{s} = 0.5, 7$ TeV for the central diffractive contribution.
As expected we observe strong correlation between the two variables.
%--------------------------------------------------------
%--------------------------------------------------------

Now we wish to compare differential distributions of kaon from the
$\chi_{c0}$ decay with those for the continuum kaons.
The amplitude for exclusive central diffractive $\chi_{c0}$ meson
production was calculated within the $k_{t}$-factorization approach
including virtualities of active gluons \cite{PST_chic0}
and the corresponding cross section is calculated with the help of
unintegrated gluon distribution functions (UGDFs) known from the
literature. We apply the following simple procedure.
First we calculate the two-dimensional distribution
$d\sigma(y,p_{t})/dy dp_{t}$, where $y$ is rapidity and $p_{t}$ is 
the transverse momentum of $\chi_{c0}$. The decay of $\chi_{c0} \to K^{+}K^{-}$
is included then in a simple Monte Carlo program assuming isotropic
decay of the scalar $\chi_{c0}$ meson in its rest frame. 
The kinematical variables of kaons are transformed to the overall
center-of-mass frame where extra cuts are imposed. Including the
simple cuts allows us to construct several differential distributions in
different kinematical variables. 

\begin{figure}[!h]
\includegraphics[width = 0.32\textwidth]{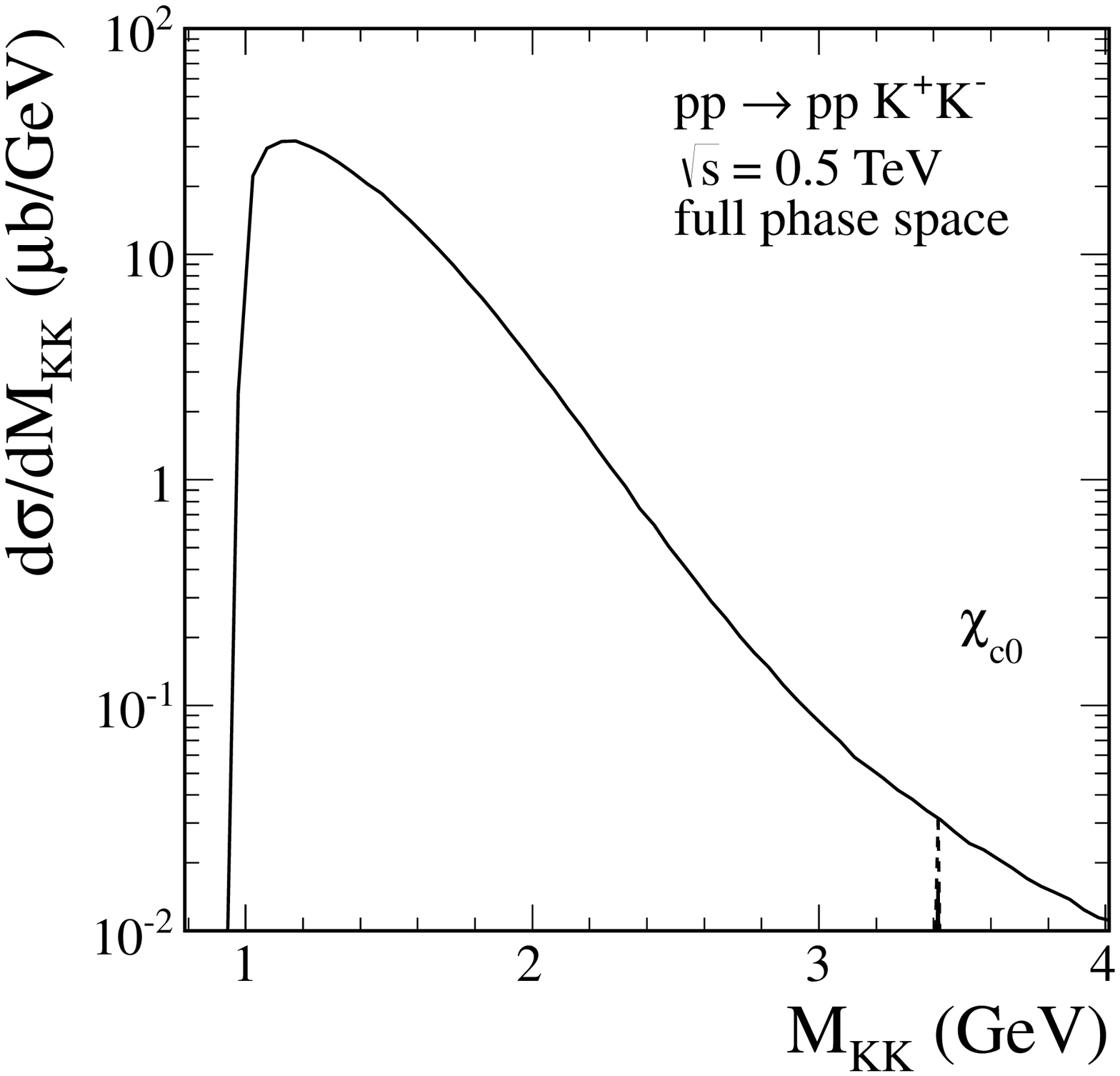}
\includegraphics[width = 0.32\textwidth]{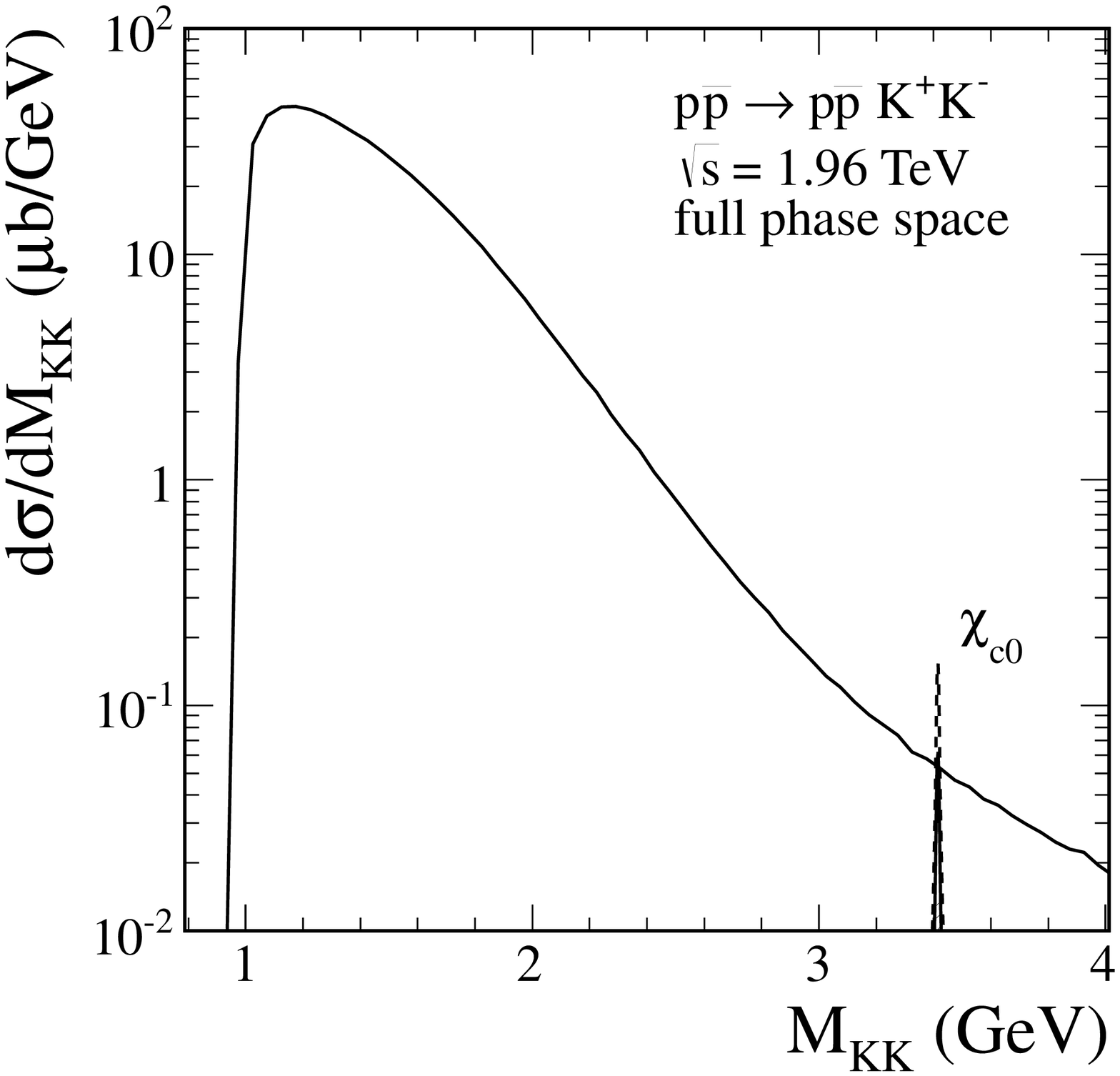}
\includegraphics[width = 0.32\textwidth]{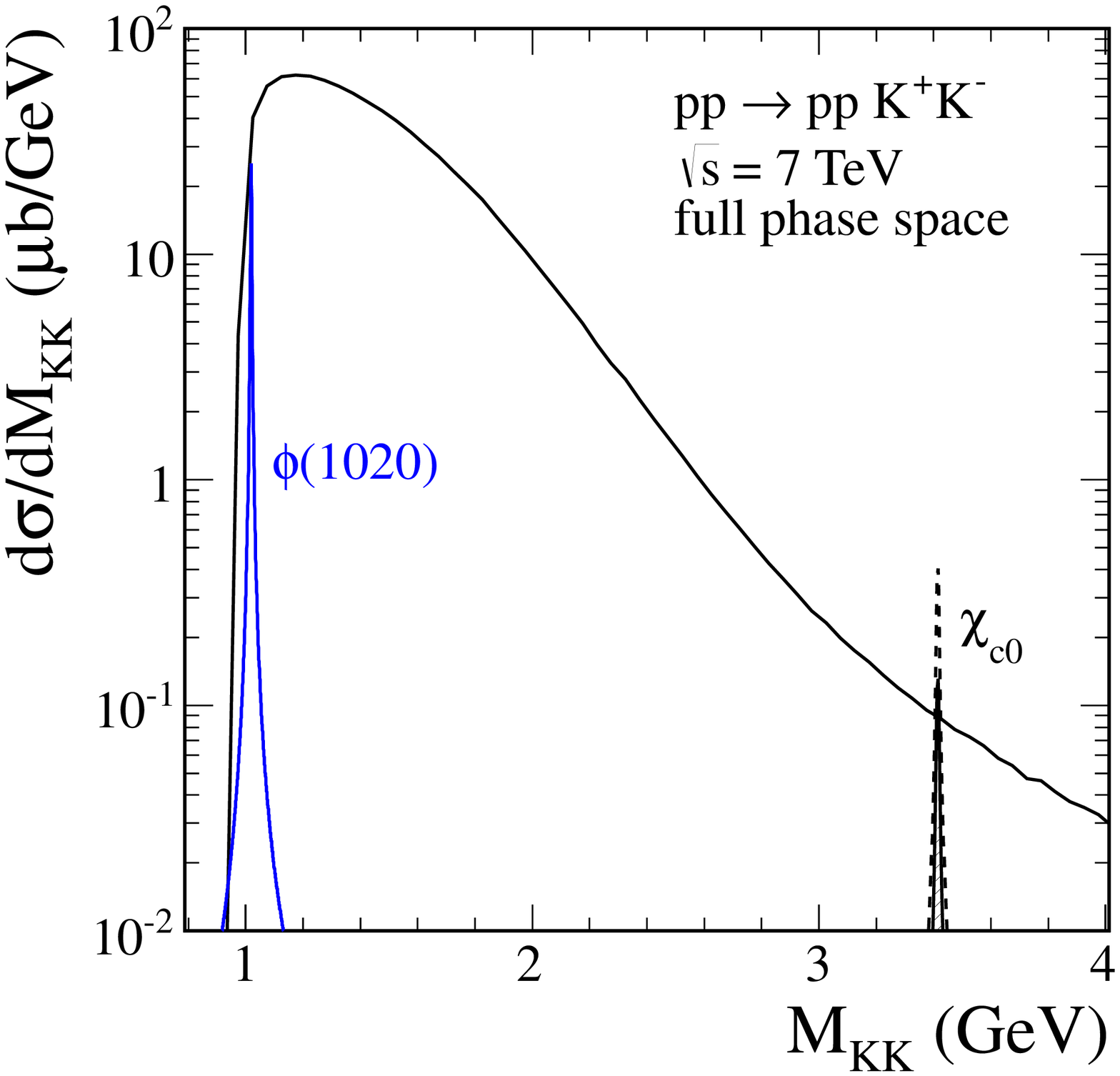}\\
\includegraphics[width = 0.32\textwidth]{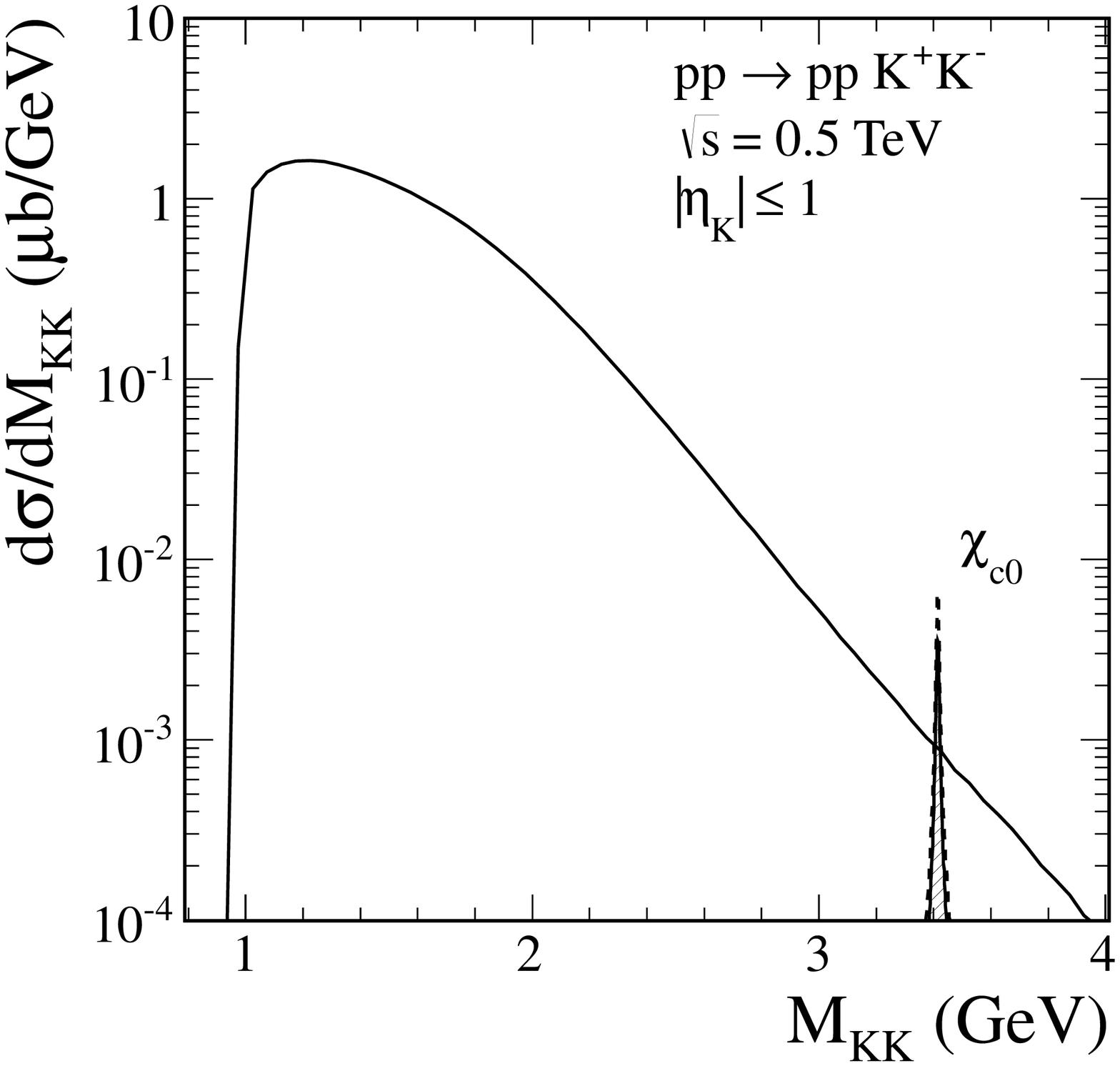}
\includegraphics[width = 0.32\textwidth]{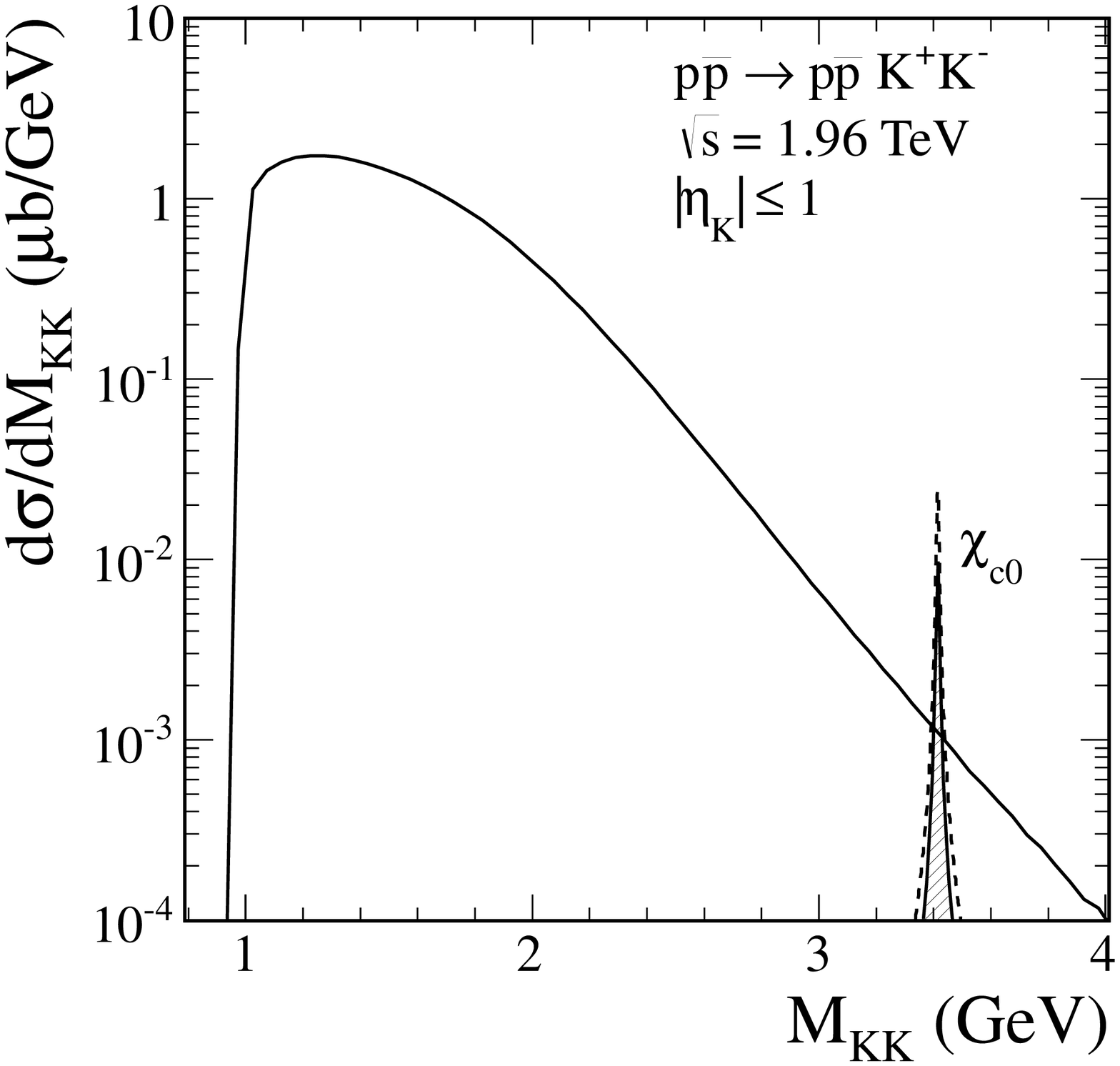}
\includegraphics[width = 0.32\textwidth]{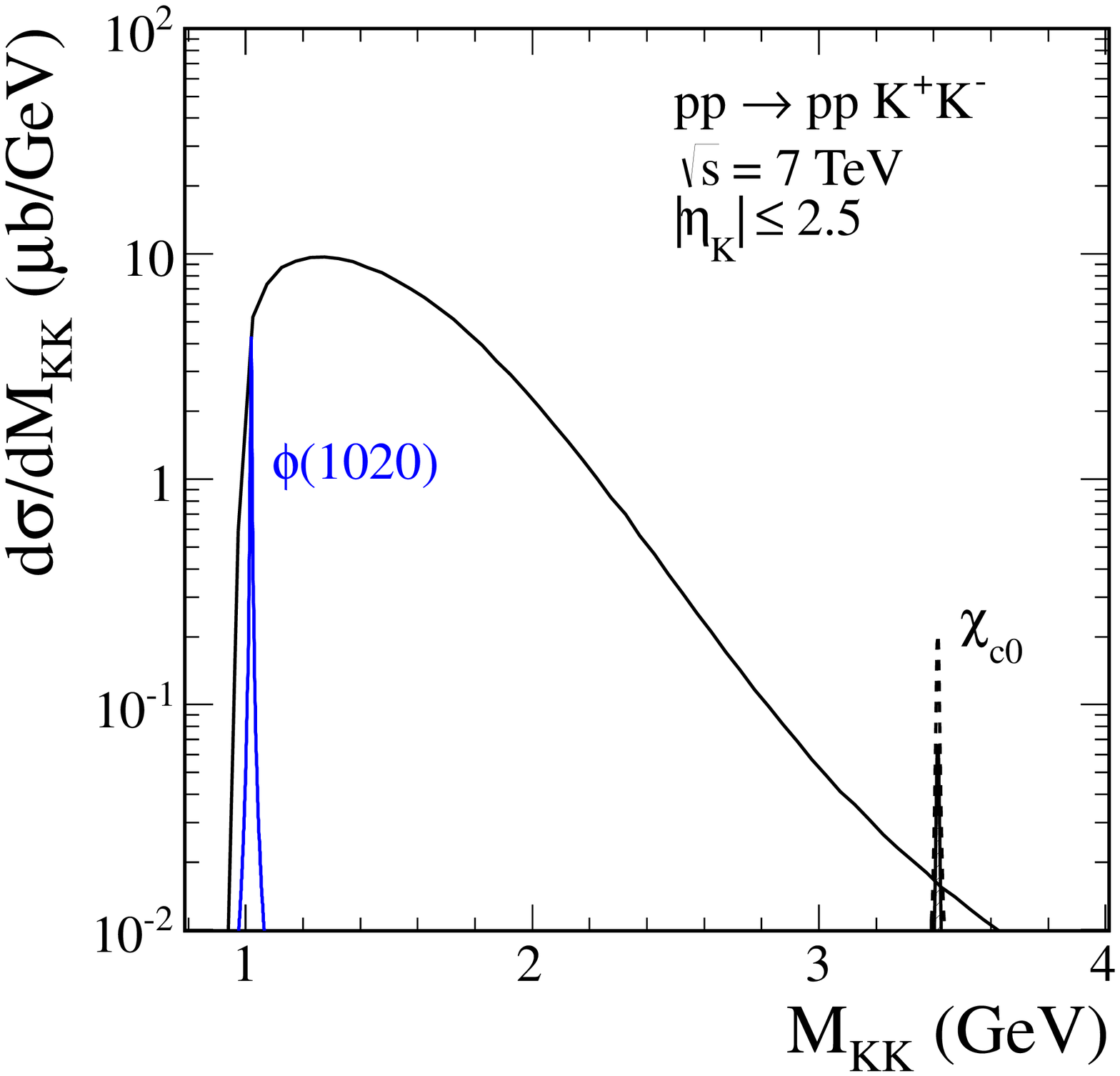}\\
\includegraphics[width = 0.32\textwidth]{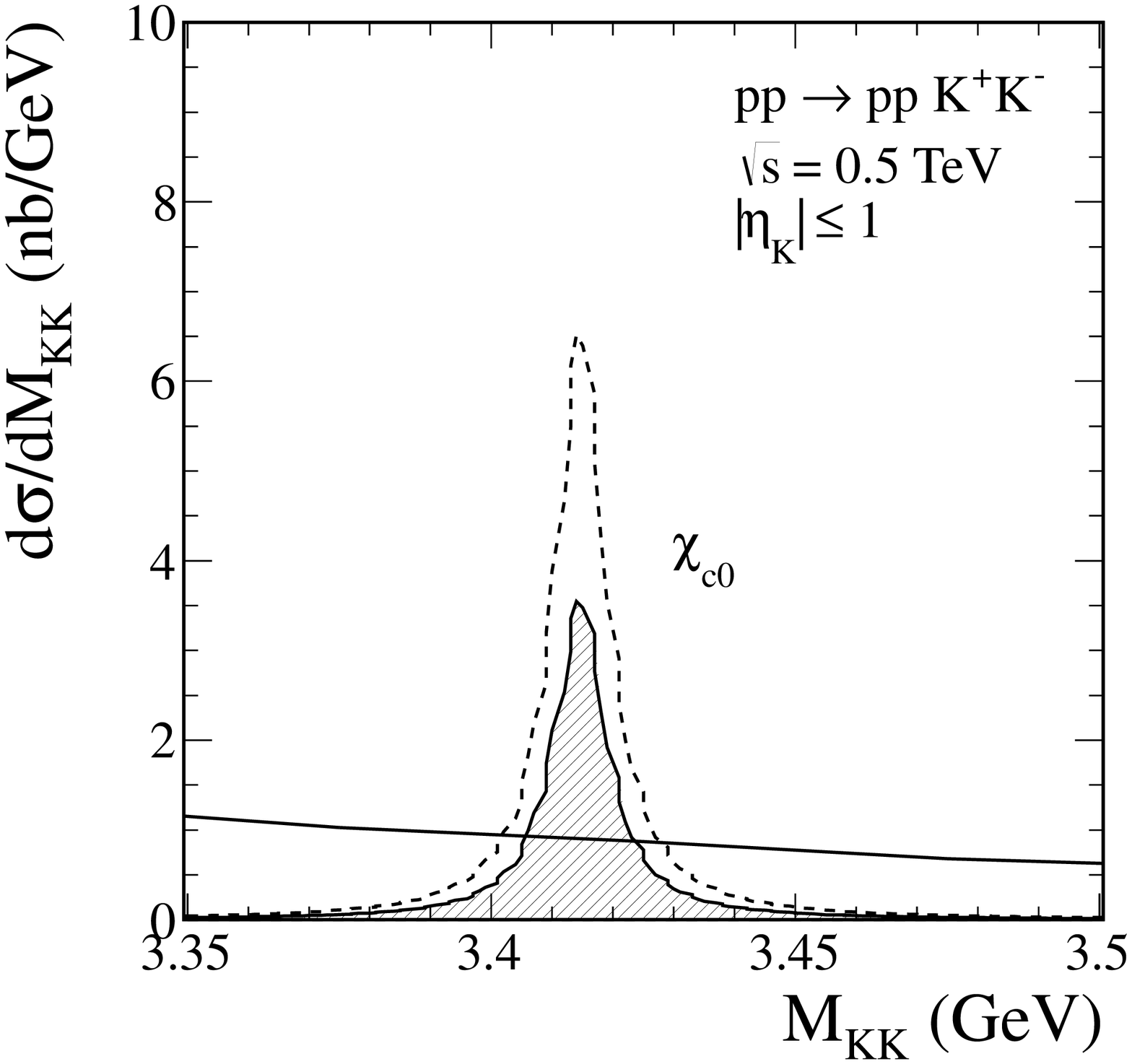}
\includegraphics[width = 0.32\textwidth]{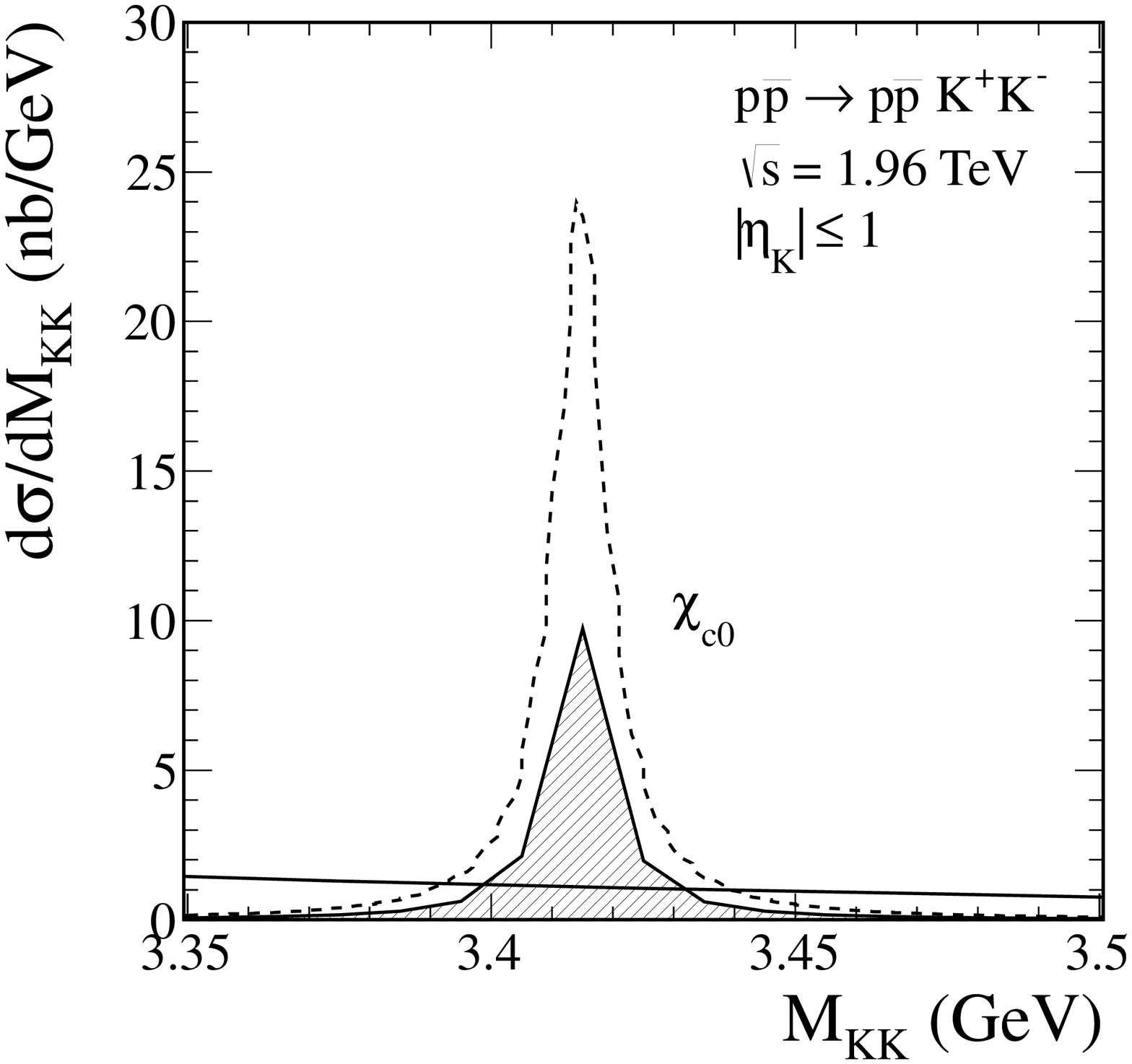}
\includegraphics[width = 0.32\textwidth]{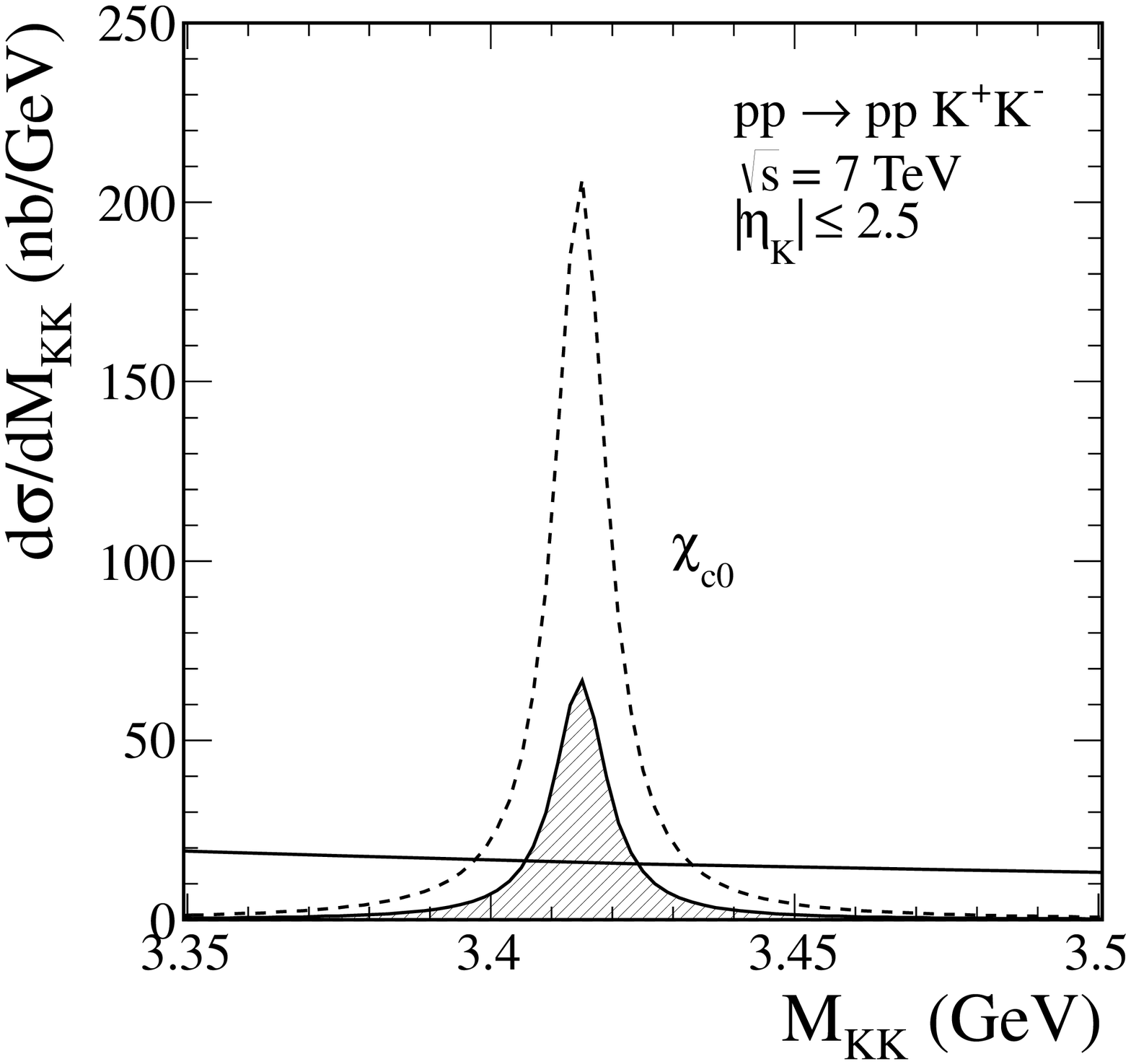}
  \caption{\label{fig:dsig_dmpipi}
  \small
The $K^{+}K^{-}$ invariant mass distribution at $\sqrt{s} = 0.5,1.96, 7$ TeV 
integrated over the full phase space (upper row) and
with the detector limitations in kaon pseudorapidities (lower rows).
The solid lines present the $KK$ continuum
with the cut-off parameters $\Lambda_{off}^{2} = 2$ GeV$^{2}$.
The $\chi_{c0}$ contribution
is calculated with the GRV94 NLO (dotted lines) and GJR08 NLO (filled areas) collinear
gluon distributions. 
The cross section for $\phi$ contribution at $\sqrt{s} = 7$ TeV is calculated as in \cite{CSS10}.
The absorption effects were included in the calculations.
Clear $\chi_{c0}$ signal with relatively small background can be observed.}
\end{figure}

In Fig.~\ref{fig:dsig_dmpipi} we show two-kaon invariant mass
distribution for the central diffractive $KK$ continuum and the
contribution from the decay of the $\chi_{c0}$ meson (see the peak
at $M_{KK} \simeq 3.4$ GeV) and  the contribution from the decay of 
the $\phi$ meson.
The cross section for exclusive production of the $\phi$ meson
has been calculated
within a pQCD $k_{t}$-factorization approach in Ref.\cite{CSS10}.
In these figures the resonant $\mathcal{R} = \phi, \chi_{c0}$ 
distributions was parameterized in the Breit-Wigner form:
\begin{eqnarray}
\frac{d\sigma}{dM_{KK}}=
\mathcal{B}(\mathcal{R} \to K^{+}K^{-})\, \sigma_{pp \to pp \mathcal{R}}\, 2 M_{KK} \,
\frac{1}{\pi}
\frac{M_{KK} \Gamma_{\mathcal{R}}}
     {(M_{KK}^{2}-m_{\mathcal{R}}^{2})^{2} + M_{KK}^{2} \Gamma_{\mathcal{R}}^{2}}\,,
\label{BW_form}
\end{eqnarray}
with parameters according to particle data book \cite{PDG}.
In the calculation of the $\chi_{c0}$ distributions we use 
GRV94 NLO \cite{GRV} and GJR08 NLO \cite{GJR}
collinear gluon distributions. The cross sections for the
$\phi$ and $\chi_{c0}$ production and for the background include absorption effects.
While the upper row shows the cross section integrated over the
full phase space at different energies, the lower rows show results
including the relevant kaon pseudorapidity restrictions 
$-1 < \eta_{K^{+}},\eta_{K^{-}} < 1$ (RHIC and Tevatron) and 
$-2.5 < \eta_{K^{+}},\eta_{K^{-}} < 2.5$ (LHC).
%--------------------------------------------------------
%--------------------------------------------------------
Shown are only purely theoretical predictions. In reality the situation is,
however, somewhat worse as both protons and, in particular,
kaon pairs are measured with a certain precision which leads to an extra
smearing in $M_{KK}$. While the smearing is negligible for the
background, it leads to a modification of the Breit-Wigner peak for
the $\chi_{c0}$ meson
\footnote{An additional experimental resolution not included here can be taken
into account by an extra convolution of the Breit-Wigner shape with
an additional Gaussian function.}.
The results with more modern GJR UGDF are smaller
by about a factor of 2-3 than those for somewhat older GRV UGDF.

\begin{figure}[!h]
\includegraphics[width = 0.32\textwidth]{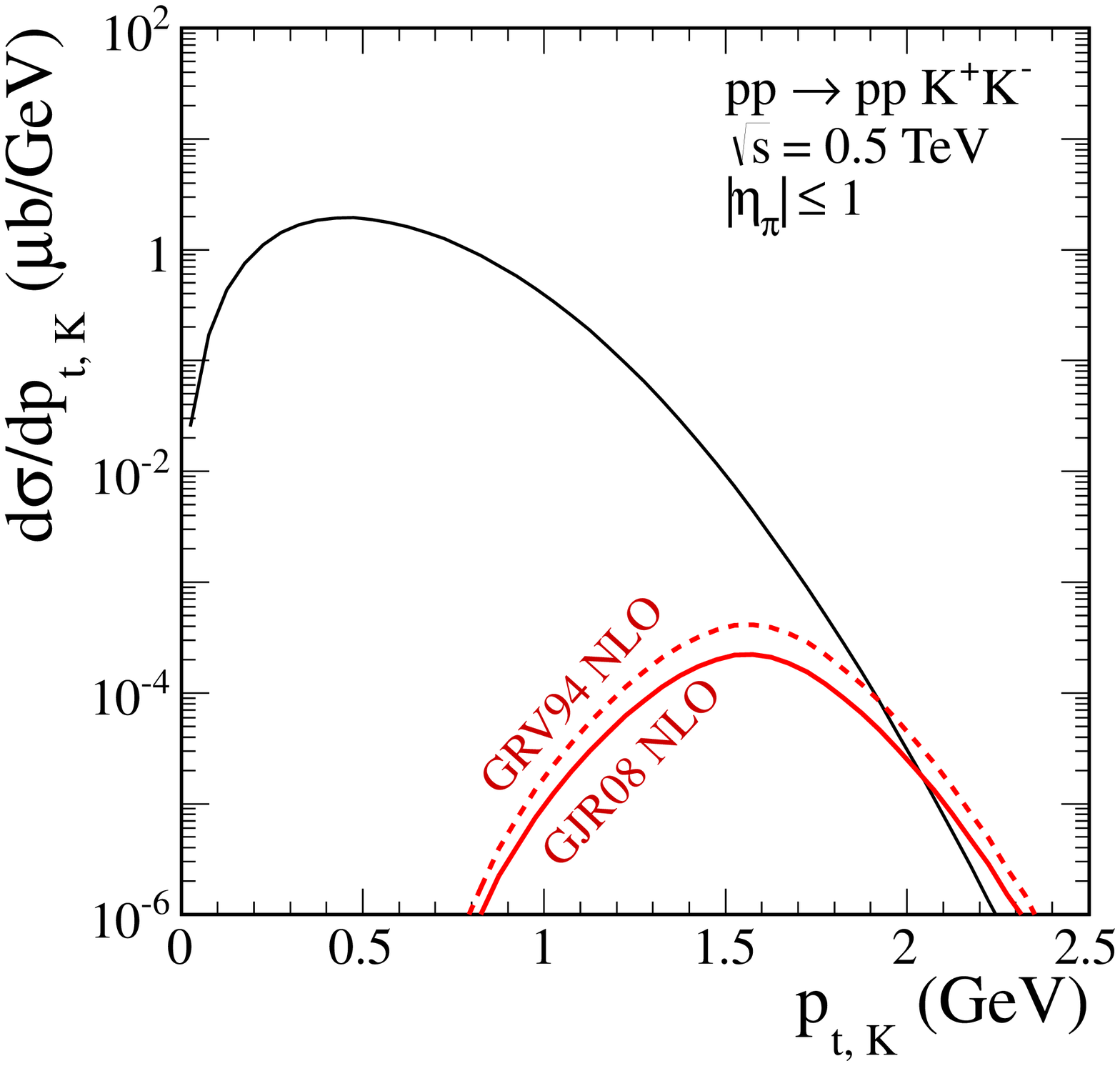}
\includegraphics[width = 0.32\textwidth]{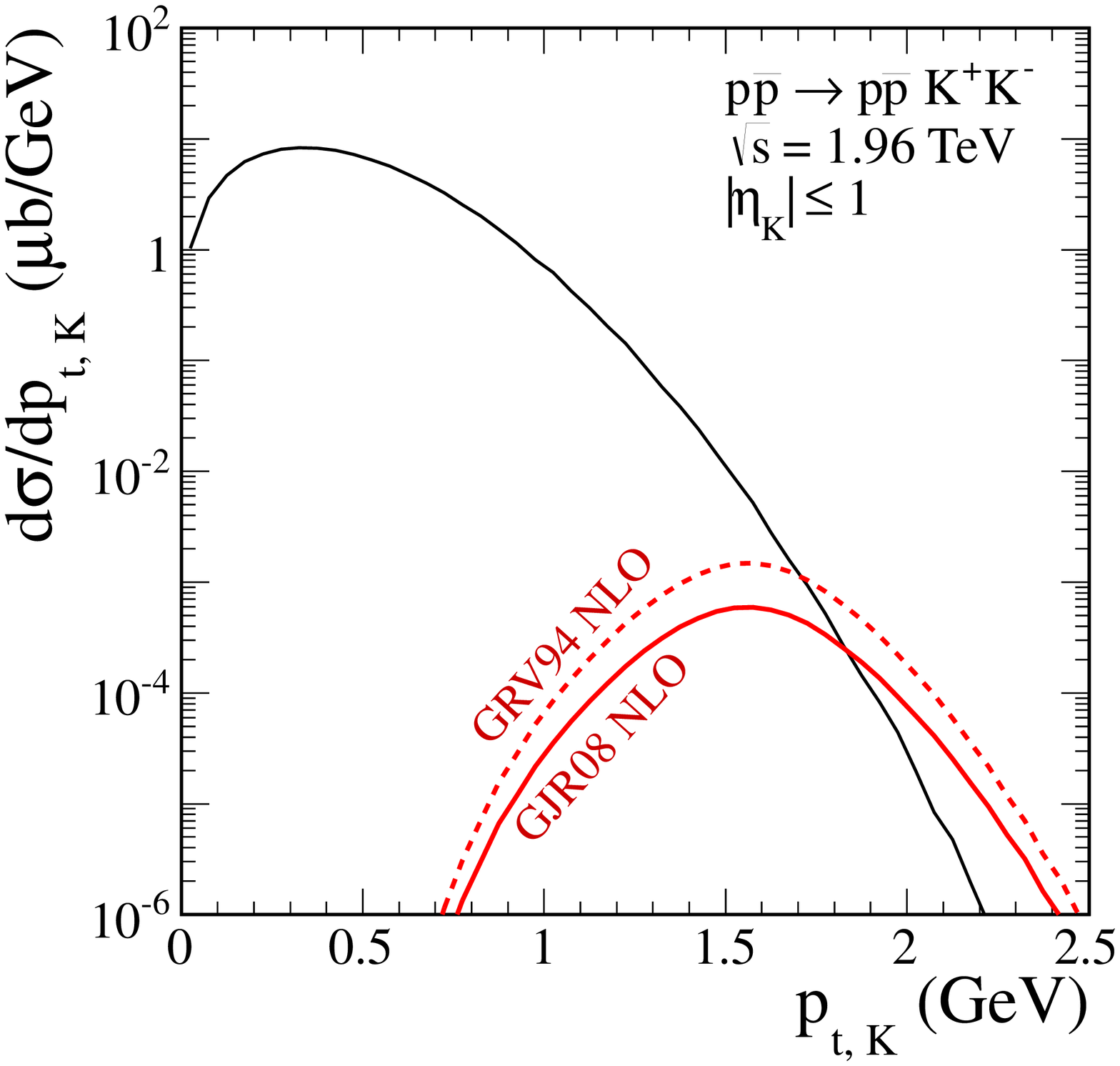}
\includegraphics[width = 0.32\textwidth]{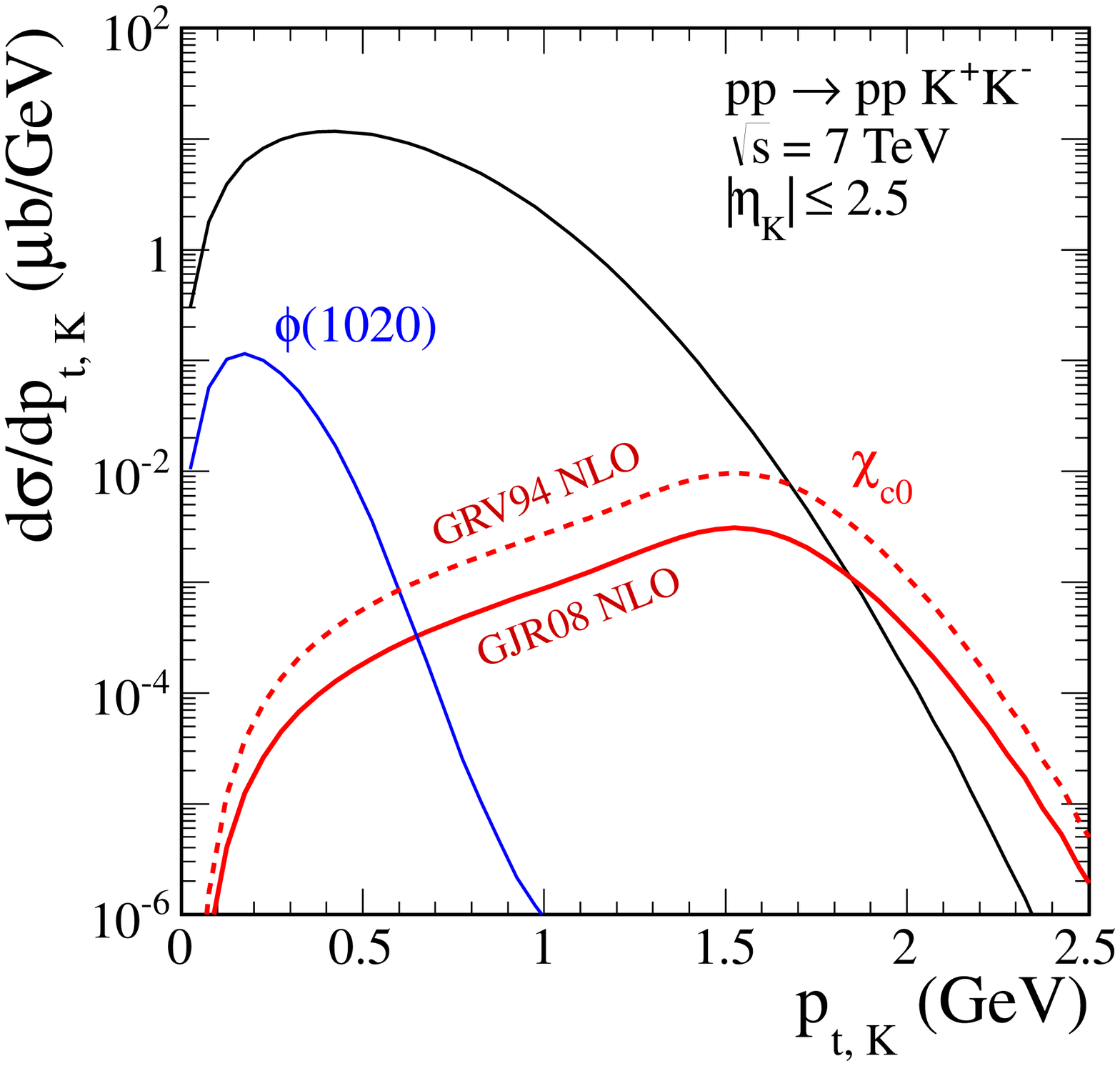}
  \caption{\label{fig:pt_lhc}
  \small
Differential cross section $d\sigma/dp_{t,K}$ at $\sqrt{s} = 0.5,
1.96, 7$ TeV with cuts on the kaon pseudorapidities.
The diffractive background was calculated with the cut-off parameter $\Lambda_{off}^{2}$ = 2 GeV$^{2}$.
Results for the kaons from the decay of the
$\chi_{c0}$ meson including the $K^{+}K^{-}$ branching ratio, 
for the GRV94 NLO (upper lines) and GJR08 NLO (bottom lines) UGDFs, are shown.
In the right panel $\phi$ meson contribution is shown in addition.
The absorption effects were included here.}
\end{figure}

In Fig.~\ref{fig:pt_lhc} we show distributions in kaon transverse momenta.
The kaons from the $\chi_{c0}$ decay are placed at slightly larger $p_{t,K}$.
This can be therefore used to get rid of the bulk of the continuum by imposing
an extra cut on the kaon transverse momenta.
It is not the case for the kaons from the $\phi$ decay which are placed at lower $p_{t,K}$.
%--------------------------------------------------------
%--------------------------------------------------------

In Table~\ref{tab:sig_tot_kk} we have collected 
numerical values of the integrated cross sections 
for exclusive production of $K^{+}K^{-}$ at different energies.
%=============================================================================================
\begin{table}%[h]
\caption{Integrated cross sections in $\mu b$ (with absorption corrections)
for exclusive $K^{+}K^{-}$ production at different energies.
In this calculations we have taken into account
the relevant limitations in the kaon pseudorapidities
$|\eta_{K}| < 1$ at RHIC and Tevatron,
$|\eta_{K}| < 2.5$ at LHC.
%and lower cut on both pion transverse momenta $|p_{t,K}|>$ 1.5 GeV.
}
\label{tab:sig_tot_kk}
\begin{center}
\begin{tabular}{|c|c|c|}
\hline
$\sqrt{s}$ (TeV) & full phase space & with cuts on $\eta_{K}$\\
\hline
0.5  & 18.47 & 1.21 \\
1.96 & 27.96 & 1.37 \\
7    & 41.14 & 7.38 \\
\hline
\end{tabular}
\end{center}
\end{table}
%=======================================================================================
In Table~\ref{tab:sig_tot_chic0} we have collected in addition
numerical values of the integrated cross sections 
(see $\sigma_{pp \to pp \chi_{c0}}$ in Eq.~(\ref{BW_form}))
for exclusive $\chi_{c0}$ production 
for some selected UGDFs at different energies.
%=============================================================================================
\begin{table}[h]
\caption{Integrated cross sections in nb (with absorption corrections)
for exclusive $\chi_{c0}$ production at different energies
with the GRV94 NLO and GJR08 NLO collinear gluon distributions.
In these calculations we have taken into account
the relevant limitations in the kaon pseudorapidities
$|\eta_{K}| < 1$ at RHIC and Tevatron,
$|\eta_{K}| < 2.5$ at LHC
and lower cut on both kaon transverse momenta $|p_{t,K}|>$ 1.5 GeV.}
\label{tab:sig_tot_chic0}
\begin{center}
\begin{tabular}{|c||c|c||c|c||c|c|}
\hline
$\sqrt{s}$&\multicolumn{2}{c||}{full phase space}
          &\multicolumn{2}{c||}{with cuts on $\eta_{K}$}
          &\multicolumn{2}{c|}{with cuts on $\eta_{K}$ and $p_{t,K}$}\\
\cline{2-7}
(TeV)     &GRV   &GJR   &GRV  &GJR  &GRV  &GJR \\
\hline
0.5  & 82.9   & 44.0   & 17.3  & 9.4   & 5.7   & 3.1  \\
1.96 & 406.3  & 165.1  & 63.7  & 25.9  & 20.7  & 8.3  \\
7    & 1076.7 & 347.7  & 548.6 & 177.1 & 114.5 & 36.6 \\
14   & 1566.3 & 449.2  & 735.0 & 210.9 & 152.1 & 43.1 \\
\hline
\end{tabular}
\end{center}
\end{table}

\begin{figure}[!h]  
\includegraphics[width=0.4\textwidth]{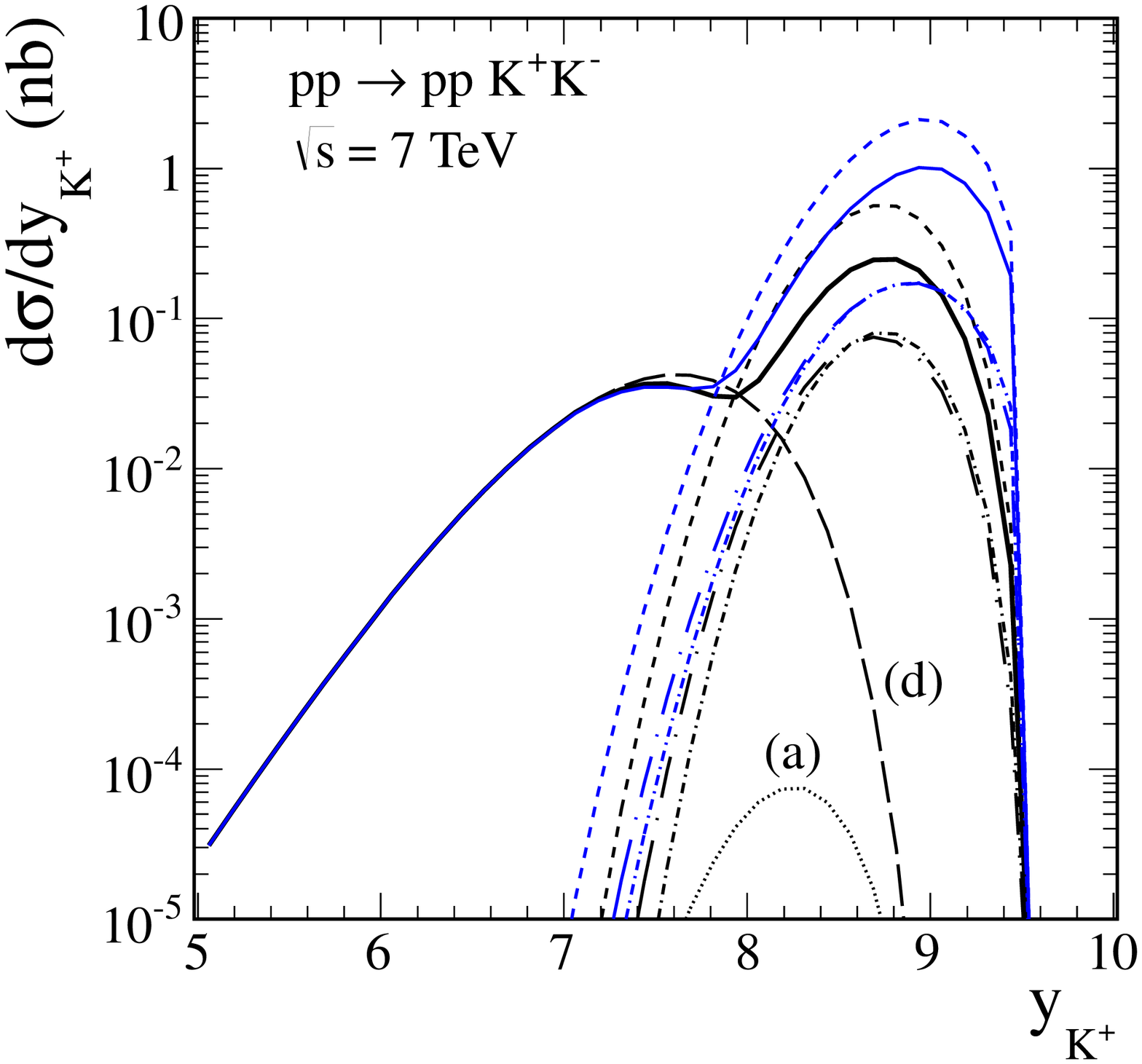}
\includegraphics[width=0.4\textwidth]{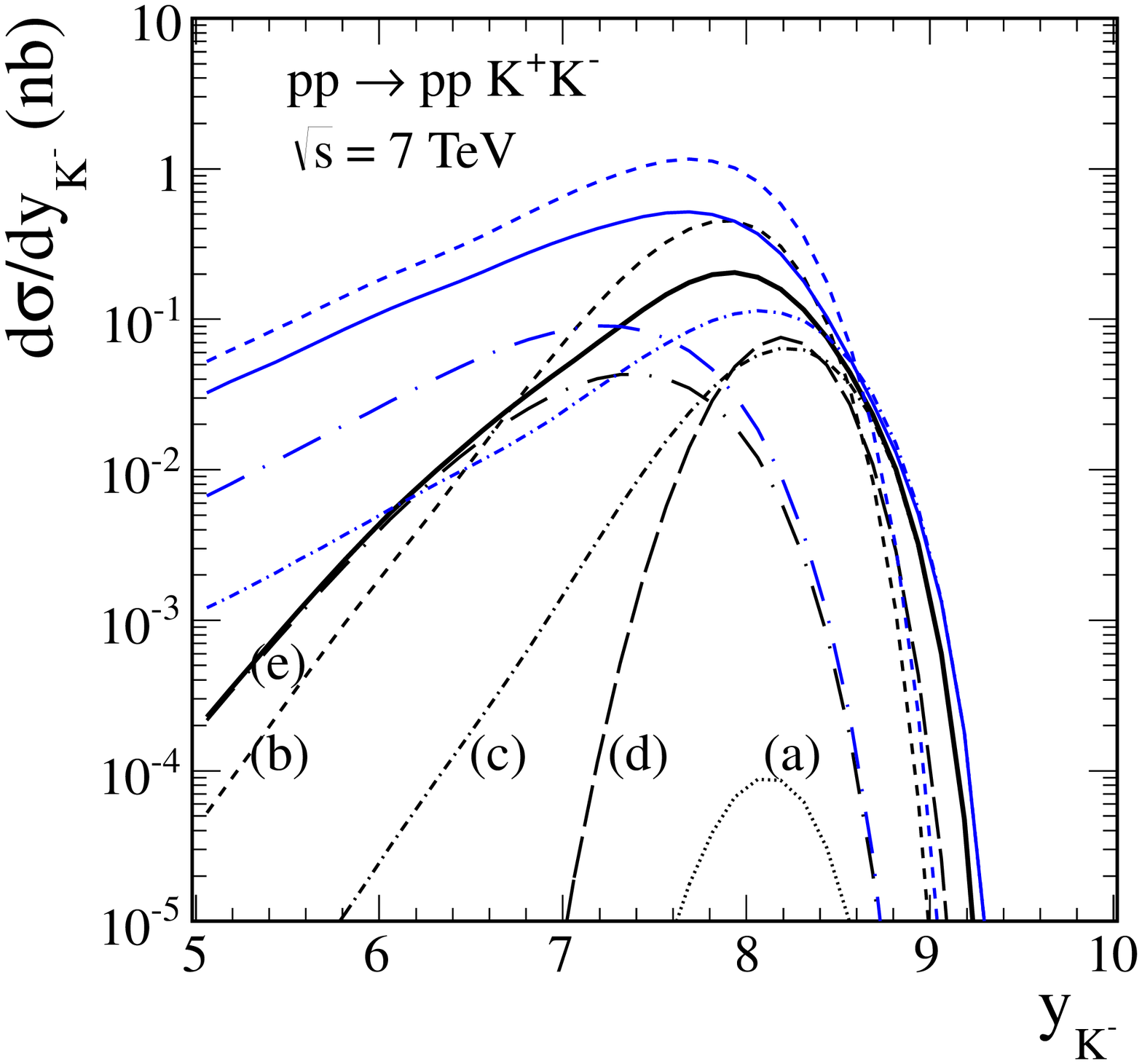}
   \caption{\label{fig:y_deco}
   \small
Differential cross sections $d \sigma / dy_{K^{+}}$ (left panel) and $d \sigma / dy_{K^{-}}$ (right panel)
for the $pp \to pp K^{+} K^{-}$ reaction at $\sqrt{s} = 7$ TeV. 
The solid line represents the coherent sum of all amplitudes.
The dotted, dashed, dash-dotted, long-dashed, long-dash-dotted lines
correspond to contributions from a) - e) diagrams in Fig.\ref{fig:other_diagrams}.
The upper (blue online) lines correspond to contributions without reggezation of $\Lambda$ propagator
in diagrams b), c), e).
}
\end{figure}

In Fig.\ref{fig:y_deco} we present rapidity distribution of $K^{+}$ (left panel)
and rapidity distribution of $K^{-}$ (right panel)
including only diagrams shown in Fig.\ref{fig:other_diagrams}.
The contribution for individual diagrams a) - e) are also shown.
In the discussed here new mechanism not only protons but also kaons are produced dominantly 
in very forward or very background directions forming a large size gap in rapidity.
Please note a very limited range of rapidities shown in the figure.
The reggezation leads to an extra damping of the cross section.
The cross section is much smaller than that for the DPE mechanism discussed above.
It is particularly interesting that the distributions for 
$K^{+}$ and $K^{-}$ have slightly different shape.
%--------------------------------------------------------
%--------------------------------------------------------

\begin{figure}[!h]  
\includegraphics[width=0.4\textwidth]{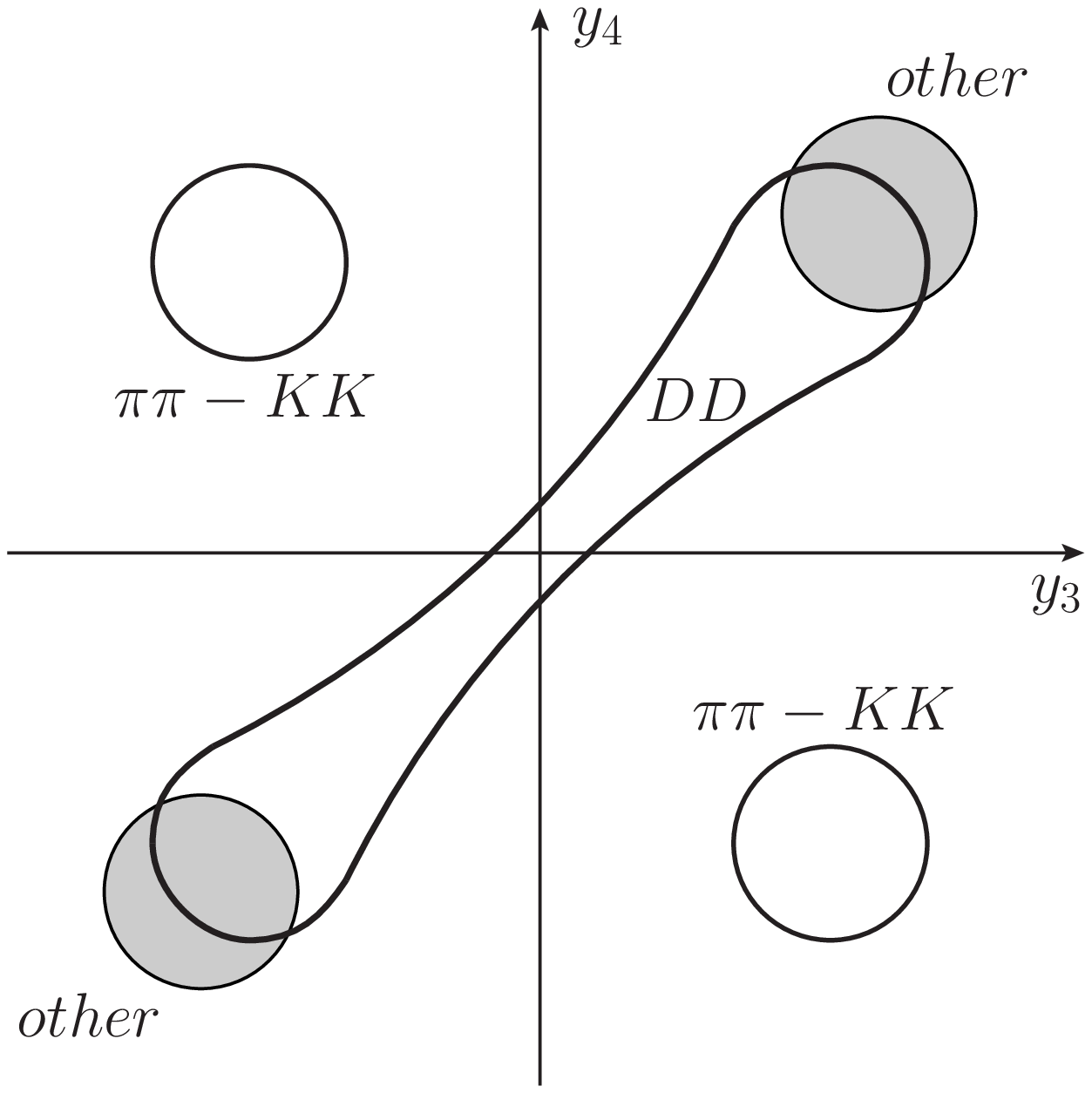}
   \caption{\label{fig:y3y4_mechanisms_localization}
   \small
A schematic localization of different mechanisms for  
the $p p \to p p K^+ K^-$ reaction at high energies.
}
\end{figure}

Finally, the general situation at high energies is sketched
in Fig.\ref{fig:y3y4_mechanisms_localization}.
The discussed in this paper central diffractive (DD) contribution lays along 
the diagonal $y_3 = y_4$ and the classical DPE is placed in the center
$y_3 \approx y_4$.
While the contribution from diagrams in Fig.\ref{fig:other_diagrams}
is predicted at $y_3,y_4 \sim y_{beam}$ or $y_3,y_4 \sim y_{target}$,
the $\pi\pi \to KK$ contribution (see Fig.\ref{fig:pipikk_diagram}) 
is predicted at ($y_3 \sim y_{beam}$ and $y_4 \sim y_{target}$)
or ($y_3 \sim y_{target}$ and $y_4 \sim y_{beam}$), i.e.
well separated from the central diffractive contribution.
The seperation in the $(y_{3},y_{4})$ space can be used
to seperate the two contributions experimentally.

%--------------------------------------------------------
%--------------------------------------------------------

%--------------------------
\section{CONCLUSIONS}
\label{section:V}
%--------------------------

In the present paper we have calculated several differential observables 
for the exclusive $p p \to p p K^+ K^-$ and $p \bar p \to p \bar p K^+ K^-$ reactions.
The full amplitude of central diffractive process
was calculated in a simple model with parameters adjusted to low energy data.
The energy dependence of the amplitudes of the $KN$ subsystems was parametrized in the Regge form
which describes total and elastic cross section for the $KN$ scattering. 
This parametrization includes both leading Pomeron trajectory as well as subleading Reggeon exchanges.
We have predicted large cross sections for RHIC, Tevatron and LHC
which allows to hope that presented by us distributions will be measured.

We have calculated also contributions of several diagrams
where kaons are emitted from the proton lines.
These mechanisms contribute at forward and backward regions
and do not disturb the observation of the central DPE component.

At the Tevatron the measurement of exclusive production of $\chi_{c}$
via decay in the $J/\psi + \gamma$ channel cannot provide production
cross sections for different species of $\chi_{c}$.
In this decay channel the contributions of $\chi_{c}$ mesons
with different spins are similar and experimental resolution is not
sufficient to distinguish them. At LHC situation should be better.

In the present paper we have analyzed a possibility to measure the
exclusive production of $\chi_{c0}$ meson in the proton-(anti)proton
collisions at the LHC, Tevatron and RHIC via $\chi_{c0} \to K^{+}K^{-}$
decay channel.
We have performed detailed studies of several differential
distributions and demonstrated how to impose extra cuts in order to
improve the signal-to-background ratio. 
We have shown that relevant measurements at RHIC, Tevatron and LHC are 
possible.
Since the cross section for exclusive $\chi_{c0}$ production is much
larger than for $\chi_{c(1,2)}$ and the branching fraction to the $KK$ channel
for $\chi_{c0}$ is larger than that for $\chi_{c2}$ ($\chi_{c1}$
does not decay into two kaons) the two-kaon channel should provide
an useful information about the $\chi_{c0}$ exclusive production.

%\vspace{0.5cm}
%--------------------
%{\bf Acknowledgments}
%%--------------------

%This study was partially supported by the Polish grant
%of ......

%---------------------------------------------------------------------

\end{document}